\documentclass[pra,eqsecnum,twocolumn,amsmath,amssymb]{revtex4}
\usepackage{graphicx}% Include figure files
\usepackage{bm}% bold math

\usepackage[usenames,dvipsnames]{color}

\definecolor{altncolor}{rgb}{0,0,0.8}
\usepackage[colorlinks=true, linkcolor=green, anchorcolor=altncolor,
citecolor=altncolor, filecolor=altncolor, menucolor=altncolor,
urlcolor=altncolor]{hyperref}
\begin{document}

\title{Chandrasekhar theory of electromagnetic scattering from strongly
conducting ellipsoidal targets}
% Force line breaks with \\

\author{Peter B. Weichman}

\affiliation{BAE Systems, Advanced Information Technologies, 6 New
England Executive Park, Burlington, MA 01803}

%\affiliation{
%Second institution and/or address \\
%This line break forced with \\ }

%\date{\today}

%\special{papersize=8.5in,11in}

\begin{abstract}

Exactly soluble models in the theory of electromagnetic propagation and
scattering are essentially restricted to horizontally stratified or
spherically symmetric geometries, with results also available for
certain waveguide geometries. However, there are a number of new
problems in remote sensing and classification of buried compact
metallic targets that require a wider class of solutions that, if not
exact, at least support rapid numerical evaluation. Here, the exact
Chandrasekhar theory of the electrostatics of heterogeneously charged
\emph{ellipsoids} is used to develop a ``mean field'' perturbation
theory of low frequency electrodynamics of highly conducting
ellipsoidal targets, in insulating or weakly conducting backgrounds.
The theory is based formally on an expansion in the parameter $\eta_s =
L_s/\delta_s(\omega)$, where $L_s$ is the characteristic linear size of
the scatterer and $\delta_s(\omega)$ is the electromagnetic skin depth.
The theory is then extended to a numerically efficient description of
the intermediate-to-late-time dynamics following an excitation pulse.
As verified via comparisons with experimental data taken using
artificial spheroidal targets, when combined with a previously
developed theory of the high frequency, early-time regime, these
results serve to cover the entire dynamic range encountered in typical
measurements.

\end{abstract}


\maketitle

\section{Introduction}
\label{sec:intro}

There are a number of longstanding economic and humanitarian problems,
such as clearance of unexploded ordnance (UXO) from old practice
ranges, that require remote identification of buried metallic objects.
The most difficult technological issue is not the detection of such
targets, but rather the ability to distinguish between them and
harmless clutter items, such as various sized pieces of exploded
ordnance. Since clutter tends to exist at much higher density, even
modest discrimination ability leads to huge reductions in the economic
cost of remediating such sites.

\subsection{Electromagnetic inverse problems}
\label{sec:em_inverse}

Formally, a successful solution to the electromagnetic (EM)
discrimination problem is a theory or algorithm that allows derivation
of accurate bounds on physical properties of the target scatterer
(its position, shape, orientation, physical composition, etc.)\ from
measurements of the scattered field using a well characterized
experimental apparatus (with known transmitter and receiver coils,
transmitted waveform, and so on). Solution of this \emph{inverse
problem} first requires the ability to generate high-fidelity candidate
solutions to the \emph{forward problem}, namely accurate computations
of the scattered field from a known target in a known subsurface
environment. The general solution to the forward problem requires full
three dimensional numerical solutions to the Maxwell equations, a
difficult and time consuming computational problem. To reduce the
computational burden, it is extremely important to obtain analytic
solutions to as broad an array of exactly soluble model problems as
possible. These solutions may then either be used as crude models of
the target, or as the basis of a perturbation scheme for accurate
modeling of ``nearby'' target geometries.

The only compact targets for which a full analytic solution at any
frequency may be derived are those with spherical symmetry
\cite{Jackson}. These are rather poor approximations to UXO, which tend
to more resemble finite, rounded cylinders with roughly 4:1 aspect
ratio. The approach pursued here is to take advantage of the fact that
\emph{electrostatic} solutions exist for a much broader array of target
geometries, and that these solutions can then be used as the basis for
a controlled perturbation theory, valid at low frequencies. The small
parameter in the theory, $\eta_s \equiv L_s/\delta_s$, is the ratio of
the electromagnetic skin depth $\delta_s(\omega)$ to the linear target
size $L_s$. The theory is dubbed the ``mean field approach,'' since the
smallness of $\eta_s$ means that the expansion is highly nonlocal in
space, with the currents and fields at any given point in the target
being sensitive to their values throughout the target. Although
formally valid only for small $\eta_s$, we will see that the theory may
be extended to higher frequencies, even where $\eta_s$ is significantly
larger than unity, if one generates a sufficient number of terms in the
series \cite{MITrefs}.

The basic zeroth order theory requires one to solve for the
electrostatic field generated by the target in a sequence of background
fields of increasing complexity. The perturbation theory is developed
formally for a general target shape, but even this sequence of simpler
electrostatic problems generally requires a numerical solution.
However, for the case of ellipsoids, such solutions may be computed
analytically via an elegant approach developed by Chandrasekhar
\cite{Chandra,W2011}. Since many targets of interest may be modeled
quite accurately by ellipsoidal or spheroidal shapes, the results of
this paper have an immediately relevant application. The theory will
mainly be illustrated for the case in which both background and
scatterer are nonmagnetic (i.e., the permeability $\mu$ is a uniform
constant), but the extension to permeable targets will be described as
well.

When treating buried targets, the electrodynamics of the soil is also
potentially important. We will assume that the ground is insulating or
sufficiently weakly conducting that its response may be treated as
quasistatic in the frequency range of interest (say, 100 kHz or less).
Specifically, the background EM penetration depth (typically tens of
meters or more at these frequencies) should be large compared to the
measurement domain (typically on the scale of 1 m).

Even with the quasistatic assumption, the total electric field has
a significant, unpredictable variability due to strong variation in
the dielectric function due to varying soil type and inclusions,
surface vegetation, air-ground interface, etc. It transpires, however,
that an induction loop (EMI) measurement (as opposed, say, to a linear
antenna measurement) is effectively sensitive only to the ``magnetic
part'' (curl component) of the electric field, and that the latter is
insensitive to the ground (partially explaining the ubiquity of such
measurements in geophysics), so long as it is nonmagnetic
\cite{foot:magsoils}. In a happy confluence of theory and experiment,
the perturbation theory is most efficiently formulated to isolate and
compute precisely this part of the field.

\subsection{Time-domain measurements}
\label{sec:tdem}

A common experimental probe for metallic targets is the time-domain
electromagnetic (TDEM) measurement, in which one detects the inductive
response of a target following termination of a transmitted pulse. The
pulse generates a characteristic pattern of currents in the target, and
the subsequent dynamics of, say, the electric field may be written as a
superposition of EM eigenmodes:
\begin{equation}
{\bf E}({\bf x},t) = \sum_{n=1}^\infty
A_n {\bf e}^{(n)}({\bf x}) e^{-\lambda_n t},
\label{1.1}
\end{equation}
in which ${\bf e}^{(n)}$ is the mode shape, $\lambda_n$ the decay rate,
and $A_n$ the excitation amplitude. This is analogous to the response
of a drumhead following a strike. However, rather than corresponding to
a set of characteristic eigenfrequencies (with, perhaps, some weak
damping), the dissipative/ohmic dynamics leads here to modes that
exhibit a pure exponential decay in time. The data will typically
consist of the voltage measured in a receiver coil,
\begin{equation}
V(t) = \sum_{n=1}^\infty V_n e^{-\lambda_n t},
\label{1.2}
\end{equation}
which is a superposition of the same exponential decays. The decay
rates and mode shapes are, respectively, eigenvalues and eigenfunctions
of the Maxwell equations at imaginary frequency $\omega_n =
-i\lambda_n$, and may therefore be accessed through the perturbation
technique developed here.

It transpires that in the high-contrast limit there are, in fact, two
classes of excitation with widely separated decay rates. One set
corresponds to electric polarization of the target, the primary example
being an excitation in which a uniform electric field is suddenly
switched off. These ``electric modes'' relax essentially
instantaneously by direct equilibration of the charges induced on the
target surface, and hence have very large, perturbatively inaccessible
$\lambda_n$. On the other hand, the rapid relaxation implies that their
contribution to the signal (\ref{1.2}) disappears almost immediately.

The second, more interesting set of modes, corresponds to magnetic
polarization of the target, the primary example being an excitation in
which a uniform magnetic field is suddenly switched off. In this case
circulating currents are generated. The flows are essentially tangential
at the target boundary, produce no charge polarization, and therefore
take much longer to relax. Being intrinsic to the geometry of the target,
the current patterns associated with these ``magnetic modes'' must vary
on the scale $L_s$ (or even on much smaller scales, as the decay rate
increases and the mode shape becomes more spatially complex), and hence
must lie in the regime $\eta_s(-i\lambda_n) = O(1)$. Once again,
however, these may be accessed via a sufficiently high-order expansion
\cite{WL03}. Moreover, the exponential decay implies that as time
progresses fewer and fewer of even these modes contribute to the signal,
so that such a theory, which accurately computes only a finite set
of the slowest decaying modes, would provide quantitative predictions on
intermediate-to-late-time scales. On the other hand, in the opposite,
early-time regime when a very large number of modes simultaneously
contribute (but not so early that any electric modes still survive), a
complementary ``surface mode'' theory, based on the diffusion of the
initial screening current inward from the target surface, may be
developed \cite{W2003,W2004}. It will be seen that these two regimes
significantly overlap for a sufficiently high order mean field
expansion, enabling a quantitative prediction of the signal over the
full time-domain dynamic range.

\subsection{Outline}
\label{sec:outline}

The outline of the remainder of this paper is as follows. In Sec.\
\ref{sec:mfa} the basic content of the method is introduced, showing
that it reduces to the evaluation of certain integrals of the Coulomb
interaction over the volume of the scatterer. In Sec.\
\ref{sec:chandra_theory} the Chandrasekhar theory of ellipsoidal
electrostatics \cite{Chandra} is reviewed, showing that it too reduces
to this same class of integrals. In Sec.\ \ref{sec:int_compute} this
connection is used to derive explicit expressions for the scattering
integrals. In Secs.\ \ref{sec:solid_ellipsoid} and \ref{sec:degenerate}
these are evaluated explicitly for the case of solid homogeneous
ellipsoids. In Sec.\ \ref{sec:formal_scatt} the solution to the
frequency-domain scattering problem in a known background field is
obtained first formally, and then explicitly to $O(\eta_s^2)$. Electric
and magnetic excitations are identified, the latter being the low
frequency precursors to the freely decaying magnetic modes. In Sec.\
\ref{sec:tdem_response}, the time-domain response is discussed.
Successful comparisons of the theory, extended numerically to
significantly higher order in $\eta_s$, to real data from artificial
spheroidal targets are demonstrated.

Finally, in Sec.\ \ref{sec:generalizations} we conclude by describing
generalizations of the theory (whose detailed applications will be left
to future work). In Sec.\ \ref{sec:inhomo_perm} we consider permeable
targets, $\mu \neq \mu_b$. In Sec.\ \ref{sec:himagcontrast} we consider
simplifications in the high contrast limit $\mu/\mu_b \gg 1$, relevant
to ferrous targets where $\mu/\mu_b = O(10^2)$. In Sec.\
\ref{sec:magmodecomp} we consider the computation of the freely
decaying modes for permeable targets. In Sec.\ \ref{sec:gentarggeom} we
consider more realistic target geometries, including hollow targets and
multiple targets. Finally, in Sec.\ \ref{sec:inhomobgperm} we consider
the effects of background permeability variations.

\section{Mean Field approach to low frequency, high contrast scattering}
\label{sec:mfa}

The mean field approach is based on the Green function formulation of
electromagnetic scattering.  Let $\epsilon_b({\bf x},\omega)$ and
$\epsilon({\bf x},\omega)$ be the space- and frequency-dependent
dielectric constant for background and background plus target,
respectively. At low frequencies these take the form (in Gaussian
units, which are used throughout unless explicitly stated otherwise),
\begin{equation}
\epsilon_b = \epsilon_b' + \frac{4\pi i \sigma_b}{\omega},
\ \ \ \
\epsilon = \epsilon' + \frac{4\pi i \sigma}{\omega},
\label{2.1}
\end{equation}
where $\epsilon_b'({\bf x})$, $\epsilon'({\bf x})$ are static
dielectric constants and $\sigma_b({\bf x})$, $\sigma({\bf x})$ are DC
conductivities. All four quantities are real and frequency independent.
Let $k \equiv \omega/c$ be the vacuum wavenumber, and define
\begin{equation}
\kappa_b^2 = \epsilon_b \mu_b k^2,\ \
\kappa^2 = \epsilon \mu k^2,\ \
Q = \kappa^2 - \kappa_b^2.
\label{2.2}
\end{equation}
For most of this paper, it will be assumed that the permeability $\mu =
\mu_b$ is a uniform constant, the same for both background and target.
Generalization to inhomogeneous $\mu$ will be discussed in Sec.\
\ref{sec:generalizations}. In addition, at the low frequencies of
interest, $\epsilon_b',\epsilon$ are negligible compared to the
conductivity contributions and will usually be dropped
\cite{foot:ebprime}. The high contrast assumption corresponds to
$|\kappa^2/\kappa_b^2| = |\epsilon/\epsilon_b| \sim \sigma/\sigma_b
\gg 1$. Typical metallic conductivities are in the range $\sigma \sim
10^7$ S/m, while typical ground conductivities are in the range
$\sigma_b \sim 0.1$ S/m, so the ratio $\sigma/\sigma_b \sim 10^8$ is
indeed extremely large.

\subsection{Green function formulation}
\label{sec:gf_form}

The background and full electric fields (with uniform $\mu$) satisfy
the frequency domain wave equations
\begin{eqnarray}
\nabla \times \nabla \times {\bf E}_b
- \kappa_b^2 {\bf E}_b &=& {\bf S}
\nonumber \\
\nabla \times \nabla \times {\bf E}
- \kappa^2 {\bf E} &=& {\bf S},
\label{2.3}
\end{eqnarray}
in which ${\bf S}({\bf x}) = (4\pi i \mu k/c) {\bf j}_S({\bf x})$ is
proportional to the source current distribution ${\bf j}_S$. The latter
is generated by a transmitter coil that is assumed to be external to
the target region. The associated source charge distribution is $\rho_S
= (i\omega)^{-1} \nabla \cdot {\bf j}_S$. The magnetic fields follow
from the relation $ik{\bf B} = ik\mu{\bf H} = \nabla \times {\bf E}$.

Let ${\bf \hat G}({\bf x},{\bf x}')$ be the $3 \times 3$ tensor Green
function for the background medium, satisfying,
\begin{equation}
\nabla \times \nabla \times {\bf \hat G}({\bf x},{\bf x}')
- \kappa_b({\bf x})^2 {\bf \hat G}({\bf x},{\bf x}')
= \delta({\bf x}-{\bf x}') \openone.
\label{2.4}
\end{equation}
${\bf \hat G}$ is symmetric, but self adjoint only if $\kappa_b^2$ is
real. For a uniform, homogeneous background medium one obtains the
exact solution
\begin{equation}
{\bf \hat G}({\bf x},{\bf x}') = \left(\openone
+ \frac{1}{\kappa_b^2}\nabla \nabla \right)
g({\bf x},{\bf x}'),
\label{2.5}
\end{equation}
with scalar Green function
\begin{equation}
g({\bf x},{\bf x}')
= \frac{e^{i\kappa_b |{\bf x}-{\bf x}'|}}
{4\pi |{\bf x}-{\bf x}'|},
\label{2.6}
\end{equation}
satisfying the Helmholtz equation, $-(\nabla^2 + \kappa_b^2)g =
\delta({\bf x}-{\bf x}')$. The background electric field may then be
expressed in the form
\begin{equation}
{\bf E}_b({\bf x}) = \int d^3x'
{\bf \hat G}({\bf x},{\bf x}') \cdot {\bf S}({\bf x}'),
\label{2.7}
\end{equation}
The operator acting on $g$ in (\ref{2.5}) essentially enforces
transverse polarization via the divergence condition $\nabla \cdot
(\epsilon_b {\bf E}_b) = 4\pi \rho_S$. Note that quite generally, as
indicated by (\ref{2.5}) and (\ref{2.6}), ${\bf \hat G}$ has a $|{\bf
x}-{\bf x}'|^{-3}$ divergence at small separation, hence there is a
potential logarithmic singularity in (\ref{2.7}). The integral over the
anisotropic dipole-like angular dependence regularizes this
singularity, but careful limiting procedures must still be used when
dealing with Green function integrals of this type.

By taking the difference of the two lines of (\ref{2.3}), the source
${\bf S}$ drops out and one obtains
\begin{equation}
\nabla \times \nabla \times ({\bf E}-{\bf E}_b)
- \kappa_b^2 ({\bf E}-{\bf E}_b) = Q {\bf E}.
\label{2.8}
\end{equation}
Applying the Green function to both sides, the full field satisfies the
integral equation,
\begin{eqnarray}
{\bf E}({\bf x}) &=& {\bf E}_b({\bf x})
+ \int_{V_s} d^3x'  Q({\bf x}')
{\bf \hat G}({\bf x},{\bf x}') \cdot {\bf E}({\bf x}')
\nonumber \\
&=& {\bf E}_b({\bf x}) + \left(\openone + \frac{1}{\kappa_b^2}
\nabla \nabla \right) \cdot \int_{V_s} d^3x'
g({\bf x},{\bf x}')
\nonumber \\
&& \hskip1.5in \times\ Q({\bf x}') {\bf E}({\bf x}'),
\ \ \ \ \
\label{2.9}
\end{eqnarray}
where second line is valid for the case of a homogeneous background.
The ability to factor the double gradient outside of the integral
renders the latter explicitly convergent in this case. The utility of
this formulation is that $Q({\bf x}')$ vanishes outside of the scatterer
volume, $V_s$. Hence if ${\bf x}$ is restricted to $V_s$, (\ref{2.9})
becomes a closed equation for the field internal to the scatterer. When
${\bf x}$ lies outside of $V_s$, the external field then follows by
simple integration of the internal field.

\subsection{High contrast limit}
\label{sec:highconlim}

As alluded to earlier, even though $\epsilon_b$ is assumed very small,
it can still vary by many orders of magnitude (e.g., between weakly
conducting ground and insulating air), generating highly variable
contributions to ${\bf \hat G}$ and to the background electric field.
However, we will now see that this component does not contribute to a
magnetic field or EMI measurement (though it will contribute, e.g., to
a linear antenna electric field measurement), and can be removed from
the computation at the outset. This is a key result because the
non-inductive part of the field, due to this intrinsic variability, is
essentially unpredictable.

The fact that the $\kappa_b \to 0$ limit of (\ref{2.8}) is singular is
evident from the fact that $Q {\bf E}$ is not divergence free (at
minimum, there is a delta-function contribution due to the target
boundary discontinuity); this is also evident from the $1/\kappa_b^2$
term on the right hand side of (\ref{2.9}). Therefore (\ref{2.8}) has
no solution if the $\kappa_b^2$ term on the left is simply dropped. To
account for this, decompose ${\bf E}$ and ${\bf E}_b$ into divergence
free and curl free parts,
\begin{eqnarray}
{\bf E}_b &=& ik {\bf A}_b - \nabla \Phi_b
\nonumber \\
{\bf E} &=& ik {\bf A} - \nabla \Phi.
\label{2.10}
\end{eqnarray}
Here the vector potentials generate the magnetic field via $\nabla
\times {\bf A} = {\bf B}$, $\nabla \times {\bf A}_b = {\bf B}_b$ and,
for later convenience, we choose the Coulomb gauge
\begin{equation}
\nabla \cdot {\bf A} = 0, \ \ \nabla \cdot {\bf A}_b = 0.
\label{2.11}
\end{equation}
It will be seen that a consistent $\kappa_b$-independent solution for
${\bf A}, {\bf A}_b$ exists when $\kappa_b \to 0$, while $\Phi, \Phi_b$
continue to depend strongly on $\kappa_b$ in this limit. However, the
integral of a gradient around any closed curve vanishes, and this
variability then indeed disappears in an EMI measurement.

\subsubsection{Background field}
\label{subsec:Ebgnd}

Consider first the background field. Using (\ref{2.11}), the first line
of (\ref{2.3}) may be written the form \cite{foot:ebprime}
\begin{equation}
-\nabla^2 {\bf A}_b = \frac{4\pi \mu}{c}({\bf j}_S
+ \sigma_b {\bf E}_b),
\label{2.12}
\end{equation}
with formal solution
\begin{equation}
{\bf A}_b({\bf x}) = \frac{\mu}{c}
\int d^3x' \frac{{\bf j}_S({\bf x}')
+ \sigma_b({\bf x}'){\bf E}_b({\bf x}')}
{|{\bf x}-{\bf x}'|}.
\label{2.13}
\end{equation}
This result is equivalent to the Biot-Savart law for the magnetic field
arising from the combination of source/transmitter current ${\bf j}_S$
and the induced background currents $\sigma_b {\bf E}_b$. For an
insulating or weakly conducting background the latter term is expected
to be negligible, and the formal limit $\kappa_b \to 0$ yields
\begin{equation}
{\bf A}_b({\bf x}) = \frac{\mu}{c}
\int d^3x' \frac{{\bf j}_S({\bf x}')}{|{\bf x}-{\bf x}'|},
\label{2.14}
\end{equation}
which is indeed entirely independent of the background details.

To establish conditions for consistency of this conclusion, an equation
for $\Phi_b$ is obtained by taking the divergence of both sides of
(\ref{2.12}):
\begin{equation}
\nabla \cdot (\sigma_b \nabla \Phi_b)
= -i\omega \rho_S + ik (\nabla \sigma_b) \cdot {\bf A}_b.
\label{2.15}
\end{equation}
The first term on the right generates the usual static Coulomb field
(distorted by the nonuniform $\epsilon_b$). If $\rho_S$ is nonzero,
then $\sigma_b {\bf E}_b$ could indeed be of the same order as ${\bf
j}_S$, and (\ref{2.14}) is invalid (the $\sigma_b {\bf E}_b \approx
-\sigma_b \nabla \Phi_b$ term contains a potentially large quasistatic
Coulomb contribution). However if, as is typical, the transmitter is
purely inductive, i.e., does not generate any free charges, $\rho_S =
i\omega \nabla \cdot {\bf j}_S = 0$, then this term is absent,
(\ref{2.14}) is valid, and (\ref{2.15}) may put in the form
\begin{equation}
\frac{1}{\sigma_b} \nabla \cdot (\sigma_b \nabla \Phi_b)
= \frac{ik}{\sigma_b} (\nabla \sigma_b) \cdot {\bf A}_b.
\label{2.16}
\end{equation}
Both sides of this (generalized Poisson) equation for $\Phi_b$ depend
explicitly on the ``shape'' of $\kappa_b({\bf x})$, but not its overall
magnitude. Substituting (\ref{2.14}) into its right hand side, the
formal $\kappa_b \to 0$ limit then produces a finite value of $\Phi_b$,
but with, as alluded to earlier, a complicated spatial dependence that
depends on the detailed geometry of the ground conductivity. The
physical interpretation of this result is that the oscillating magnetic
field generated by (\ref{2.14}), in combination with the nonuniform
conductivity, induces a small background charge density, which then
produces a finite quasistatic electric field. However, the contribution
$-\sigma_b \nabla \Phi_b$ to the background current formally vanishes
when $\kappa_b \to 0$, producing (\ref{2.14}).

Note that, comparing (\ref{2.5}) and (\ref{2.6}), this limit also
implies a near-field approximation in which propagating wave effects
are neglected. Quantitatively, this requires that $|\kappa_b| R \ll 1$
where $R$ is the length scale of the measurement domain (e.g.,
transmitter-target separation). This is equivalent to $R/\xi \ll 1$,
where $\xi = c/\sqrt{2\pi \mu \sigma_b \omega}$ is a characteristic
background skin depth.  In MKS units one obtains
\begin{equation}
\frac{\xi}{10\ \mathrm m} = \frac{5}{\pi}
\left(\frac{\mu_0}{\mu} \right)^{1/2}
\left(\frac{0.1\ \mathrm{S/m}}{\sigma_b} \right)^{1/2}
\left(\frac{10\ \mathrm{kHz}}{f} \right)^{1/2},
\label{2.17}
\end{equation}
which confirms that in the parameter domain of interest the
approximations we have used will certainly be valid for $R$ on the
scale of a few meters or less.

We emphasize again that the contribution of ${\bf E}_b$ to an induction
measurement is only through ${\bf A}_b$, which is insensitive to the
background conductivity (and, more generally, to $\epsilon_b'$
\cite{foot:ebprime}) as claimed.

\subsubsection{Full field}
\label{subsec:Efull}

We now proceed in a similar fashion to obtain a convenient formulation
of equation (\ref{2.9}) for the full electric field in the high
contrast limit. Analogous to (\ref{2.12}), substituting (\ref{2.10})
into (\ref{2.8}) yields
\begin{equation}
{\bf A}({\bf x}) - {\bf A}_b({\bf x})
= \frac{\mu}{c} \int d^3x'
\frac{\sigma({\bf x}') {\bf E}({\bf x}')
- \sigma_b({\bf x}') {\bf E}_b({\bf x}')}
{|{\bf x}-{\bf x}'|}.
\label{2.18}
\end{equation}
Consistently, from (\ref{2.3}), the divergence of the right hand side
vanishes. One expects the scattered and background fields to be of the
same order in the measurement region, and so in the limit
$\kappa_b/\kappa \to 0$ one obtains, analogous to (\ref{2.14}),
\begin{eqnarray}
ik[{\bf A}({\bf x}) - {\bf A}_b({\bf x})]
&=& {\bf E}({\bf x}) - {\bf E}_b({\bf x})
+ \nabla [\Phi({\bf x}) - \Phi_b({\bf x})]
\nonumber \\
&=& \frac{ik\mu}{c} \int_{V_s} d^3x'
\frac{\sigma({\bf x}') {\bf E}({\bf x}')}
{|{\bf x}-{\bf x}'|}.
\label{2.19}
\end{eqnarray}
Since $\sigma = \sigma_b$ outside $V_s$, correct to the same order, we
have also dropped the exterior contribution from the $\sigma {\bf E}$
term and restricted the integral to $V_s$.

A generalized Poisson equation for $\Phi$ is obtained from the
divergence of the second line of (\ref{2.3}):
\begin{equation}
\frac{1}{\sigma} \nabla \cdot (\sigma \nabla \Phi)
= \frac{ik}{\sigma} (\nabla \sigma) \cdot {\bf A}.
\label{2.20}
\end{equation}
A purely inductive transmitter, $\rho_S = 0$, has again been assumed.
The solution to this equation again depends on the detailed geometry of
$\sigma = \sigma_b$ outside $V_s$. For a homogeneous target and
background, the right hand side is supported at the discontinuity of
$\sigma$ at the target boundary.

\subsection{Solution strategy}
\label{sec:solnstrat}

The remainder of this paper will be focused on solving (\ref{2.19}) for
${\bf A}$ [with ${\bf A}_b$ already determined by (\ref{2.14})], while
entirely avoiding the computation of $\Phi,\Phi_b$. The basic strategy
is to first solve for the full electric field ${\bf E}_\mathrm{int}$
inside $V_s$, and then insert this into the left hand side of
(\ref{2.19}) to compute ${\bf A}$ outside $V_s$. A measurement of the
magnetic field ${\bf B} = \nabla \times {\bf A}$, or of the induced
voltage
\begin{equation}
V = \oint_{C_R} {\bf E} \cdot d{\bf l}
= ik \oint_{C_R} {\bf A} \cdot d{\bf l}
\label{2.21}
\end{equation}
in a receiver loop $C_R$, are then independent of $\Phi$.

The key to solving for ${\bf E}_\mathrm{int}$ is to project
(\ref{2.19}) onto the appropriate vector function space, namely the
space of functions ${\bf E}$ restricted to the domain $V_s$ and
satisfying
\begin{equation}
\left\{\begin{array}{ll}
\nabla \cdot (\sigma {\bf E}) = 0, & {\bf x} \in V_s \\
{\bf \hat n} \cdot {\bf E} = 0, & {\bf x} \in \partial V_s.
\end{array} \right.
\label{2.22}
\end{equation}
The first condition is equivalent, in the high contrast limit, to
$\nabla \cdot (\kappa^2 {\bf E}) = 0$, while the second, in the same
limit, follows from the usual EM boundary condition, namely continuity
of $\epsilon {\bf \hat n} \cdot {\bf E}$ across the boundary $\partial
V_s$, where ${\bf \hat n}$ is the local surface normal. The latter
implies that ${\bf \hat n} \cdot {\bf E}_\mathrm{int} =
(\epsilon_b/\epsilon) {\bf \hat n} \cdot {\bf E}_\mathrm{ext} \to 0$.
Physically this follows from the fact that since background currents
are much smaller than those in the target, the latter must flow
parallel to the boundary in order to maintain charge conservation.

Next, the identity
\begin{widetext}
\begin{equation}
\int_{V_s} d^3 x \sigma({\bf x})
{\bf E}({\bf x}) \cdot \nabla f({\bf x})
= -\int_{V_s} d^3 x f({\bf x}) \nabla
\cdot [\sigma({\bf x}) {\bf E}({\bf x})]
+ \int_{\partial V_s} dA \sigma({\bf x}) f({\bf x})
{\bf \hat n}({\bf x}) \cdot {\bf E}({\bf x}) = 0,
\label{2.23}
\end{equation}
\end{widetext}
which follows by applying Green's theorem, and is valid for any scalar
function $f$, shows that the function space (\ref{2.22}) is orthogonal
to the space of gradients, in the sense of the above inner product with
kernel $\sigma$. Denoting the orthogonal projection onto the space
(\ref{2.22}) by $\hat {\cal P}_\sigma$, equation (\ref{2.19}), when
restricted to the scatterer volume, takes the form
\begin{equation}
{\bf E}({\bf x}) - \hat {\cal P}_\sigma{\bf E}_b({\bf x})
= \frac{ik\mu}{c} \hat {\cal P}_\sigma \int_{V_s} d^3x'
\frac{\sigma({\bf x}') {\bf E}({\bf x}')}{|{\bf x}-{\bf x}'|}.
\label{2.24}
\end{equation}
This is a closed equation for the internal electric field [restricted
to the space (\ref{2.22})], and is explicitly independent of $\Phi_b$
and of the form of $\Phi$ outside $V_s$.

\subsubsection{Basis function expansion}
\label{subsec:basisfn}

A more explicit form of this equation, which is the foundation for the
mean field perturbation scheme to be developed, is obtained by
expanding
\begin{equation}
\sigma({\bf x}) {\bf E}({\bf x}) = \sum_M a_M {\bf Z}_M({\bf x})
\label{2.25}
\end{equation}
in terms of a complete (though not necessarily orthogonal) set of basis
functions ${\bf Z}_M$ consistent with (\ref{2.22}):
\begin{equation}
\left\{\begin{array}{ll}
\nabla \cdot {\bf Z}_M = 0, & {\bf x} \in V_s \\
{\bf \hat n} \cdot {\bf Z}_M = 0, & {\bf x} \in \partial V_s.
\end{array} \right.
\label{2.26}
\end{equation}
By inserting this expansion into (\ref{2.24}), and taking the inner
product on the left with ${\bf Z}_l^*$, one obtains a formal matrix
equation for the coefficients
\begin{equation}
\sum_M O_{LM} a_M = a_{b,L}
+ \frac{ik \mu}{c} \sum_M H_{LM} a_M
\label{2.27}
\end{equation}
in which
\begin{equation}
O_{LM} = \int_{V_s} d^3x \frac{{\bf Z}_L({\bf x})^*
\cdot {\bf Z}_M({\bf x})}{\sigma({\bf x})}
\label{2.28}
\end{equation}
are basis function inner products,
\begin{equation}
H_{LM} = \int_{V_s} d^3x \int_{V_s} d^3x'
\frac{{\bf Z}_L({\bf x})^* \cdot
{\bf Z}_M({\bf x}')}{|{\bf x}-{\bf x}'|}
\label{2.29}
\end{equation}
represents the projection of the Coulomb integral onto the basis
function, and
\begin{equation}
a_{b,L} = \int_{V_s} d^3x {\bf Z}_L({\bf x})^* \cdot {\bf E}_b({\bf x})
\label{2.30}
\end{equation}
represents a similar projection of the background field. The $\nabla
(\Phi - \Phi_b)$ term drops out via (\ref{2.23}). The formal inverse
\begin{equation}
{\bf a} = \left({\bf O} - \frac{ik \mu}{c} {\bf H} \right)^{-1} {\bf a}_b
\label{2.31}
\end{equation}
determines the internal field expansion coefficients in terms of those
for the background field. Since $\sigma$ is real, the matrices ${\bf O}$,
${\bf H}$ are both self-adjoint.

\subsubsection{Basis functions for ellipsoids}
\label{subsec:ellpsdbasis}

For general targets, explicit forms for the ${\bf Z}_M$ (or,
equivalently, for the projection operator ${\cal P}_\sigma$) may be
hard to come by, but for ellipsoids they may be constructed directly
from known forms for the sphere. Specifically, if ${\bf Z}^{(S)}_M({\bf
x})$ are basis functions obeying (\ref{2.26}) on the unit sphere, then
\begin{equation}
{\bf Z}^{({\bf a})}_M({\bf x})
= \sum_{\alpha = 1}^3 a_\alpha {\bf \hat e}_\alpha
Z_{M,\alpha}^{(S)}(x_1/a_1,x_2/a_2,x_3/a_3),
\label{2.32}
\end{equation}
are basis functions for the ellipsoid with principal radii ${\bf a} =
(a_1,a_2,a_3)$, where ${\bf \hat e}_\alpha$ is the unit vector along
principle axis $\alpha$. The unit surface normal is given by
\begin{equation}
{\bf \hat n}({\bf x}) = \frac{\sum_\alpha
{\bf \hat e}_\alpha x_\alpha/a_\alpha^2}
{\sqrt{\sum_\alpha x_\alpha^2/a_\alpha^4}},\
{\bf x} \in \partial V_s,
\label{2.33}
\end{equation}
where the boundary is defined by
\begin{equation}
\sum_\alpha \frac{x_\alpha^2}{a_\alpha^2} = 1,\
{\bf x} \in \partial V_s.
\label{2.34}
\end{equation}

A convenient complete set of basis functions for the sphere may be
constructed from the vector spherical harmonics \cite{Jackson}
\begin{widetext}
\begin{eqnarray}
{\bf X}_{lm}(\theta,\phi) &=& \frac{1}{\sqrt{l(l+1)}} {\bf \hat L} Y_{lm}
\nonumber \\
&=& \frac{1}{\sqrt{l(l+1)}} \left(\frac{1}{2} c_{lm} Y_{l,m+1}
+ \frac{1}{2} c_{l,-m} Y_{l,m-1},\ \
\frac{1}{2i} c_{lm} Y_{l,m+1} - \frac{1}{2i} c_{l,-m} Y_{l,m-1},\ \
m Y_{lm} \right),
\label{2.35}
\end{eqnarray}
\end{widetext}
in which ${\bf \hat L} = -i {\bf x} \times \nabla$ is the angular
momentum operator and $c_{lm} = \sqrt{(l-m)(l+m+1)}$. The key
properties are that the vector harmonics are tangent everywhere to the
sphere surface, and divergence free,
\begin{eqnarray}
{\bf \hat x} \cdot {\bf X}_{lm} &=& 0
\nonumber \\
\nabla \cdot {\bf X}_{lm} &=& 0,
\label{2.36}
\end{eqnarray}
and the functions ${\bf X}_{lm}, {\bf \hat x} \times {\bf X}_{lm}$, $l
\geq 1$, $-l \leq m \leq l$, form a complete basis for the tangent
vector fields, obeying the orthonormality relations
\begin{eqnarray}
\int d\Omega {\bf X}_{lm}^* \cdot {\bf X}_{l'm'}
&=& \delta_{ll'} \delta_{mm'}
\nonumber \\
\int d\Omega {\bf X}_{lm}^* \cdot
({\bf \hat x} \times {\bf X}_{l'm'}) &=& 0.
\label{2.37}
\end{eqnarray}
Furthermore, since $x^l Y_{lm}$ are polynomials of degree $l$ in
$x_1,x_2,x_3$, so are the components of $x^l {\bf X}_{lm}$. Similarly,
$x^{l+1} {\bf \hat x} \times {\bf X}_{lm}$ is a vector polynomial of
degree $l+1$. Finally, note the identity
\begin{equation}
\nabla \times [f(x) {\bf X}_{lm}]
= i {\bf \hat x} \frac{f(x)}{x} \sqrt{l(l+1)} Y_{lm}
+ \frac{[xf(x)]'}{x} {\bf \hat x} \times {\bf X}_{lm}
\label{2.38}
\end{equation}
which shows that the choice $f(x) = (1-x^2)x^l$ produces a divergence
free polynomial of degree $l+1$ that is tangent everywhere to the unit
sphere surface, $x=1$.

From the completeness property it follows that a complete set of basis
functions ${\bf Z}_{lmp}^{(i)}$ for the sphere, in the form of
divergence free vector polynomials that are tangent at the sphere
surface, and in which $M = (l,m,p,i)$ is now a composite index, may
be taken as
\begin{eqnarray}
{\bf Z}_{lmp}^{(1)} &=& x^{l+2p} {\bf X}_{lm}
\nonumber \\
{\bf Z}_{lmp}^{(2)} &=& \nabla \times
\left[\left(1 - x^2 \right) x^{l+2p} {\bf X}_{lm} \right]
\label{2.39} \\
&=& \left(1 - x^2 \right)
\nabla \times [x^{l+2p} {\bf X}_{lm}]
- 2 x^{l+2p+1} {\bf \hat x} \times {\bf X}_{lm}
\nonumber
\end{eqnarray}
for $p=0,1,2,\ldots$. The $i=1$ basis functions are of degree $n = l+2p
\geq 1$, while those for $i=2$ are of degree $n = l+2p+1 \geq 2$. At a
given degree $n \geq 1$, there are $(n+1)(n+2)/2$ polynomials of the
first type for $n$ odd, and $n(n+3)/2$ for $n$ even; and there are
$n(n+1)/2$ polynomials of the second type for $n$ even, and
$(n-1)(n+2)/2$ for $n$ odd. Including both types, there are $P(n) =
n(n+2)$ basis functions of degree $n$ for both even and odd $n$.

The transformation (\ref{2.32}) produces the corresponding basis
functions ${\bf Z}_{lmp}^{({\bf a};i)}$ for the ellipsoid, which are
then polynomials of identical degree, but with coefficients rescaled by
appropriate powers of the $a_\alpha$.

\subsection{Time domain eigenvalue equation}
\label{sec:td_eigen}

Equation (\ref{2.31}) describes the target response to an external
source at fixed frequency. In time-domain measurements one is instead
interested in the free evolution (\ref{1.1}) of the system following
pulse termination. Although the excitation coefficients $A_n$ will
depend on the details of the pulse (and their computation will be dealt
with in Sec.\ \ref{sec:tdem_response}), the mode shapes ${\bf e}^{(n)}$
and decay rates $\lambda_n$ do not. The eigenvalue equation these
satisfy corresponds to the second of equations (\ref{2.3}) with the
replacement $\omega = -i\lambda$ and with \emph{vanishing source,}
${\bf S} \equiv 0$:
\begin{equation}
\nabla \times \nabla \times {\bf e}^{(n)}
- \kappa^2(-i\lambda_n) {\bf e}^{(n)} = 0.
\label{2.40}
\end{equation}
Correspondingly, these modes satisfy the \emph{homogeneous} form
of the integral equation (\ref{2.19}) or (\ref{2.24}) in the
\emph{absence} of the background field:
\begin{equation}
{\bf e}^{(n)}({\bf x})
= \frac{\lambda_n \mu}{c^2} \hat {\cal P}_\sigma \int_{V_s} d^3x'
\frac{\sigma({\bf x}') {\bf e}^{(n)}({\bf x}')}{|{\bf x}-{\bf x}'|}.
\label{2.41}
\end{equation}
The corresponding basis function expansion of the modes
\begin{equation}
\sigma({\bf x}) {\bf e}^{(n)}({\bf x})
= \sum_M a_M^{(n)} {\bf Z}_M({\bf x})
\label{2.42}
\end{equation}
then produces the self-adjoint \emph{generalized eigenvalue equation}
\begin{equation}
{\bf O} {\bf a} = \lambda \frac{\mu}{c^2} {\bf H} {\bf a}
\label{2.43}
\end{equation}
The expansion coefficients may be normalized to obey the
orthonormality condition
\begin{equation}
{\bf a}^{(m) \dagger} {\bf O} {\bf a}^{(n)} = \delta_{mn}
\label{2.44}
\end{equation}
It is precisely this form of the equations that will be analyzed in
Sec.\ \ref{sec:tdem_response}. The normalization condition is
equivalent to
\begin{equation}
\int d^3x \sigma({\bf x}) {\bf e}^{(m)*}({\bf x})
\cdot {\bf e}^{(n)}({\bf x}) = \delta_{mn},
\label{2.45}
\end{equation}
which follows directly from self-adjointness of the double curl
operator in (\ref{2.40}). In the high contrast limit, the integral may
be restricted to $V_s$, with errors of order $\sigma_b/\sigma$.

\subsection{Role of the Chandrasekhar theory}
\label{sec:chandrarole}

The vector harmonics allow one to diagonalize the system (\ref{2.31})
for the sphere, but no such simplification occurs for more general
target geometries. One is therefore forced to develop approximate
solutions based on computation of the array elements
(\ref{2.28})--(\ref{2.30}) for a truncated set of basis functions.

The key observation, however, is that for homogeneous ellipsoids the
polynomial character of the basis functions allows one to perform the
Coulomb integrals (\ref{2.29}) analytically. Specifically, one requires
integrals of the form
\begin{equation}
{\cal D}_{\bf m}({\bf x}) \equiv \int_{V_s} d^3 x'
\frac{{x'_1}^{m_1} {x'_2}^{m_2} {x'_3}^{m_3}}
{|{\bf x} - {\bf x}'|}.
\label{2.46}
\end{equation}
in which the domain of the ${\bf x}'$ integral is restricted to the
ellipsoid interior, but ${\bf x}$ may be either inside or outside. The
problem of computing ${\cal D}_{\bf m}({\bf x})$ is isomorphic to that
of computing the electrostatic potential due to an ellipsoid with
static charge density $\rho({\bf x}) = x_1^{m_1} x_2^{m_2} x_3^{m_3}$.
In his book \cite{Chandra}, Chandrasekhar presents an elegant formalism
for computing potentials of precisely this type. Moreover, for ${\bf x}
\in V_s$, ${\cal D}_{\bf m}({\bf x})$ is also a polynomial (of degree
$m_1+m_2+m_3+2$). The ${\bf x}$-integral in (\ref{2.29}) therefore has
a pure polynomial integrand and is trivial to perform. In Sec.\
\ref{sec:chandra_theory} an overview of his method is presented.
Following that, the results of these evaluations will be applied to the
solution of the scattering problem.

In numerical applications, the matrix equation (\ref{2.31}) will be
truncated at finite order. As the index $M$ increases [more
specifically, as the indices $l,p$ in (\ref{2.39}) increase], the
spatial complexity of the basis functions increase, so this truncation
works best for smoother (typically, lower frequency) field
distributions. Correspondingly, mode complexity increases as the decay
rate $\lambda_n$ increases, and the truncated eigenvalue equation
(\ref{2.43}) will be accurate only for a finite set of more slowly
decaying modes.

\section{The Chandrasekhar theory of ellipsoidal electrostatics}
\label{sec:chandra_theory}

We will now develop a theory for the analytic calculation of
Coulomb integrals of the form
\begin{equation}
{\cal D}_\rho({\bf x}) = \int_{V_s} d^3x'
\frac{\rho({\bf x}')}{|{\bf x}-{\bf x}'|},
\label{3.1}
\end{equation}
for a certain class of charged densities $\rho$, which include the
monomial forms (\ref{2.45}), for the case in which $V_s$ is an
ellipsoid, defined by its principal axes ${\bf a} \equiv
(a_1,a_2,a_3)$.

\subsection{Homoeoids}
\label{sec:homoeoids}

For any given ellipsoid one may define a family of similar concentric
ellipsoids by
\begin{equation}
\sum_\alpha \frac{x_\alpha^2}{a_\alpha^2} = \mu,
\label{3.2}
\end{equation}
with $\mu \geq 0$.  Equation (\ref{3.2}) clearly defines an ellipsoid
with principal axes $\sqrt{\mu} a_1, \sqrt{\mu} a_2, \sqrt{\mu} a_3$.
A \emph{homoeoid} is defined to be a shell bounded two similar
concentric ellipsoids, i.e., the set of points ${\bf x}$ such that
\begin{equation}
\mu_1 \leq \sum_\alpha \frac{x_\alpha^2}{a_\alpha^2} \leq \mu_2,
\label{3.3}
\end{equation}
for some $0 \leq \mu_1 \leq \mu_2$.  An homogeneous homoeoid is one
that has a uniform charge density over its interior.  An infinitesimal
homoeoid is the 2D surface that emerges when $\mu_2-\mu_1 \to 0$. A
homogeneous infinitesimal homoeoid will be called a \emph{hi-homoeoid}.
Initially a special kind of heterogeneous homoeoid will be considered,
in which surfaces of constant charge density are concentric, similar
hi-homeoids: $\rho = \rho(\mu)$ only. These will be called
\emph{s-homoeoids} (or \emph{s-ellipsoids} if there is no inner
bounding surface). In particular, $\rho$ in (\ref{3.1}) will be
taken to be a function of $\mu$ alone. Since one may view a s-homoeoid
as a superposition of hi-homoeoids, the potential due to a s-homoeoid
may be written in the form
\begin{equation}
\phi({\bf x}) = \int_{\mu_1}^{\mu_2}
\rho(\mu) \phi(\mu;{\bf x}) d\mu
\label{3.4}
\end{equation}
in which $\phi(\mu;{\bf x}) d\mu$ is the potential due to a hi-homoeoid
of uniform unit charge density bounded by $\mu$ and $\mu+d\mu$.
Moreover, by simple homogeneity of the Coulomb integral one obtains
\begin{eqnarray}
\phi(\mu;{\bf x}) &\equiv&
\int \frac{d^3x'}{|{\bf x} - {\bf x}'|}
\delta\left(\mu - \sum_\alpha
\frac{x_\alpha^{\prime 2}}{a_\alpha^2} \right)
\nonumber \\
&=& \int \frac{d^3y'}
{|\mu^{-1/2}{\bf x} - {\bf y}'|}
\delta\left(1 - \sum_\alpha
\frac{y_\alpha^{\prime 2}}{a_\alpha^2} \right)
\nonumber \\
&=& \phi(1;{\bf x}/\sqrt{\mu}),
\label{3.5}
\end{eqnarray}
where the change of variable ${\bf x}' = \sqrt{\mu}{\bf y}'$ has been
used. Thus, all results for s-homoeoids may be obtained by considering
only the potential due to a hi-homoeoid with parameter $\mu=1$.

\subsection{Potential in the interior of a hi-homoeoid}
\label{sec:hh_int}

Consider first the potential \emph{interior} to the hi-homoeoid.  Note
that (by the usual rules for manipulating arguments of $\delta$-functions)
the surface charge density implied by (\ref{3.5}) is nonuniform, even
though the original volume charge density is uniform. Using spherical
coordinates with origin at the observation point ${\bf x}$, one obtains
\begin{equation}
\phi(1;{\bf x}) = \int d\Omega \int_0^\infty r dr \delta[f(r)]
= \int d\Omega \frac{r(\Omega)}{|f'[r(\Omega)]|},
\label{3.6}
\end{equation}
with
\begin{equation}
f(r) \equiv \sum_\alpha
\frac{(x_\alpha + r \hat n_\alpha)^2}{a_\alpha^2} - 1,
\label{3.7}
\end{equation}
where ${\bf \hat n}(\Omega)$ is the unit vector in the direction
$\Omega$, $r(\Omega)$ is the positive solution to $f(r) = 0$, and the
$1/|f'|$ factor follows from the rules for integrating delta-functions
with nontrivial arguments. Recalling that $\sum_\alpha \hat
n_\alpha^2/a_\alpha^2 = 1/R^2$, where $R(\Omega)$ is the radius of the
shell in direction ${\bf \hat n}$ measured from the natural origin at
the ellipsoid center, one obtains a quadratic equation for $r(\Omega)$:
\begin{equation}
0 = r^2 + 2 R^2 r \sum_\alpha
\frac{\hat n_\alpha x_\alpha}{a_\alpha^2}
+ R^2 \left(\sum_\alpha
\frac{x_\alpha^2}{a_\alpha^2} - 1 \right),
\label{3.8}
\end{equation}
with solutions
\begin{widetext}
\begin{equation}
r_\pm = -R^2 \sum_\alpha
\frac{\hat n_\alpha x_\alpha}{a_\alpha^2} \pm
\sqrt{R^4\left(\sum_\alpha
\frac{\hat n_\alpha x_\alpha}{a_\alpha^2} \right)^2
+ R^2\left(1 - \sum_\alpha
\frac{x_\alpha^2}{a_\alpha^2}\right)}.
\label{3.9}
\end{equation}
Since $\sum_\alpha x_\alpha^2/a_\alpha^2 < 1$ (${\bf x}$ is interior to
the shell), these solutions exist for any ${\bf \hat n}$.  One may now
write $f(r) = (r-r_+)(r-r_-)/R^2$, and hence $f'(r_+) = -f'(r_-) =
(r_+ - r_-)/R^2$. The positive solution is $r_+$, while $r_-(\Omega) =
-r_+(-\Omega)$ corresponds to the point on the ellipsoid in the direction
$-{\bf \hat n}$. Averaging these two contributions and using $R(\Omega)
= R(-\Omega)$, one obtains the remarkably simple result,
\begin{equation}
\frac{1}{2}\left\{\frac{r_+(\Omega)}{R(\Omega)^{-2}
[r_+(\Omega) - r_-(\Omega)]}
+ \frac{r_+(-\Omega)}{R(-\Omega)^{-2}
[r_+(-\Omega) - r_-(-\Omega)]}\right\}
= \frac{1}{2} R(\Omega)^2
\frac{r_+(\Omega) - r_-(\Omega)}{r_+(\Omega) - r_-(\Omega)}
= \frac{1}{2} R(\Omega)^2,
\label{3.10}
\end{equation}
\end{widetext}
One obtains therefore,
\begin{equation}
\phi(1;{\bf x}) = \frac{1}{2} \int d\Omega R(\Omega)^2,
\label{3.11}
\end{equation}
independent of the the observation point ${\bf x}$: the potential
interior to a hi-homoeoid is constant. In a more Newtonian style,
Chandrasekhar \cite{Chandra} demonstrates this geometrically by showing
that the electrostatic force due to infinitesimal elements of charge at
$\pm {\bf \hat n}$ precisely cancel one another. Using (\ref{3.4}) and
(\ref{3.5}), the potential in the interior of any s-homoeoid is also
constant and given by
\begin{equation}
\phi({\bf x}) = \phi(1;{\bf x})
\int_{\mu_1}^{\mu_2} d\mu \rho(\mu)
= \phi(1,{\bf x})[\psi(\mu_2)-\psi(\mu_1)],
\label{3.12}
\end{equation}
in which, for later convenience, the integrated charge density
function,
\begin{equation}
\psi(\mu) = \int_0^\mu \rho(\mu') d\mu',
\label{3.13}
\end{equation}
has been defined, so that $\rho = \partial \psi/\partial \mu$.  For a
finite homogeneous homoeoid, bounded by $\mu = \mu_0$ and $\mu = 1$ and
with uniform charge density $\rho_0$, one obtains $\psi(\mu_2)
-\psi(\mu_1) = (1 - \mu_0) \rho_0$ in (\ref{3.12}).

\subsection{Representation of the basic integral in terms of
elliptic functions}
\label{sec:elliptic_fns}

One may reduce the angular integral (\ref{3.11}) to a more convenient
form as follows. First, choose the usual spherical angular coordinates
$\theta$ and $\phi$ to be centered at the origin.  In order to adapt the
present notation to that which is standard for elliptic integrals, one
uses $a_1$ as the $z$-axis, $a_2$ as the $y$-axis, and $a_3$ as the
$x$-axis. Chandrasekhar \cite{Chandra} makes a different choice, but
the final answers must clearly be fully symmetric in the components of
${\bf a}$. One obtains
\begin{equation}
\frac{1}{R(\Omega)^2} = \frac{\cos^2(\theta)}{a_1^2}
+ \sin^2(\theta) \left[\frac{\cos^2(\phi)}{a_3^2}
+ \frac{\sin^2(\phi)}{a_2^2} \right].
\label{3.14}
\end{equation}
Substituting this form into $\phi(1;{\bf x})$, and using $d\Omega =
\sin(\theta) d\theta d\phi$, one first performs the $\phi$
integration [using the substitution $u = \tan(\phi)$] to obtain
\begin{widetext}
\begin{equation}
\phi(1;{\bf x}) = 2\pi a_1^2 a_2 a_3
\int_0^{\pi/2} \frac{\sec^2(\theta) \sin(\theta) d\theta}
{\sqrt{a_2^2 + a_1^2 \tan^2(\theta)}
\sqrt{a_3^2 + a_1^2 \tan^2(\theta)}}.
\label{3.15}
\end{equation}
\end{widetext}
The substitution $t = a_1^2 \tan^2(\theta)$ now simplifies this to
\begin{equation}
\phi(1;{\bf x}) = v({\bf a}) A({\bf a}),
\label{3.16}
\end{equation}
in which
\begin{eqnarray}
v({\bf a}) &\equiv& \pi a_1 a_2 a_3
\nonumber \\
A({\bf a}) &\equiv& \int_0^\infty \frac{dt}{\Delta({\bf a},t)}
\nonumber \\
\Delta({\bf a},t) &\equiv& \sqrt{(a_1^2 + t) (a_2^2 + t) (a_3^2 + t)}.
\label{3.17}
\end{eqnarray}

From Ref.\ \cite{GR}, p.\ 220, with the convention $a_1 \geq a_2 \geq
a_3$, $A({\bf a})$ may be expressed as:
\begin{equation}
A({\bf a}) = \frac{2}{\sqrt{a_1^2 - a_3^2}} F(\varphi,k)
\label{3.18}
\end{equation}
in which the elliptic integral $F(\varphi,k)$ is defined by
\begin{eqnarray}
F(\varphi,k) &\equiv& \int_0^\varphi
\frac{d\alpha}{\sqrt{1-k^2 \sin^2(\alpha)}}
\nonumber \\
&=& \int_0^{\sin(\varphi)}
\frac{dx}{\sqrt{(1-x^2)(1-k^2x^2)}},
\label{3.19}
\end{eqnarray}
with parameters
\begin{eqnarray}
\sin^2(\varphi) &=& 1-a_3^2/a_1^2
\nonumber \\
k^2 &=& \frac{a_1^2 - a_2^2}{a_1^2 - a_3^2}, \ \
k^{\prime 2} = 1-k^2 = \frac{a_2^2 - a_3^2}{a_1^2 - a_3^2}.
\ \ \ \ \
\label{3.20}
\end{eqnarray}
The result (\ref{3.18}) follows directly from the substitution $x =
\sqrt{(a_1^2 - a_3^2)/(a_1^2 + t)}$ in (\ref{3.17}), or equivalently
the substitution $x = \sqrt{1-a_3^2/a_1^2} \cos(\theta)$ in
(\ref{3.15}). For later purposes, we define also the auxiliary elliptic
function
\begin{eqnarray}
E(\varphi,k) &\equiv& \int_0^\varphi
d\alpha \sqrt{1-k^2 \sin^2(\alpha)}
\nonumber \\
&=& \frac{1}{2} \sqrt{a_1^2-a_3^2}
\int_0^\infty \frac{dt}{\Delta({\bf a},t)}
\frac{a_2^2+t}{a_1^2+t}.
\label{3.21}
\end{eqnarray}
The two functions obey
\begin{eqnarray}
\frac{\partial E}{\partial \varphi} &=& \frac{a_2^2}{a_1^2},\ \
\frac{\partial F}{\partial \varphi} = \frac{a_1^2}{a_2^2}
\nonumber \\
\frac{\partial E}{\partial k} &=& \frac{E-F}{k}
\nonumber \\
\frac{\partial F}{\partial k} &=&  \frac{1}{k^{\prime 2}}
\left[\frac{E - k^{\prime 2}F}{k} - \frac{a_3}{a_2}
\sqrt{\frac{a_1^2}{a_2^2} - 1} \right]
\label{3.22}
\end{eqnarray}
Thus if one has available numerical algorithms to evaluate $E$ and $F$,
then all derivatives of $E$ and $F$ follow by algebraic manipulations
alone. This will be used below to generate an iterative procedure for
computing all required integrals.

\subsection{Potential exterior to an hi-homoeoid}
\label{sec:hh_ext}

The remarkable cancelation that produces (\ref{3.10}) fails when the
observation point ${\bf x}$ is external to the hi-homoeoid, which will
be denoted $E_1$.  To make progress one must use a different approach.
The key idea is to seek the equipotential surfaces of $\phi(1,{\bf
x})$. One may reasonably guess that such surfaces form a family of
ellipsoidal surfaces, but it is not obvious what the corresponding
family of principal axes should be. For a given ${\bf x}$, let ${\bf
a}' = (a'_1,a'_2,a'_3)$ label a concentric hi-homoeoid $E_2$ passing
through ${\bf x}$: $\sum_\alpha x_\alpha^2/ a_\alpha^{\prime 2} = 1$.
One seeks conditions on ${\bf a}'$ such that this ellipsoidal surface
is an equipotential.

For each point ${\bf x}_1$ on $E_1$, let a corresponding point ${\bf
x}_2$ on $E_2$ be defined by the relation $x_{2,\alpha}/a'_\alpha =
x_{1,\alpha}/a_\alpha$. This correspondence may now be used to map
integrals over the surface $E_1$ into integrals over the surface $E_2$.
Volume elements then translate as $d^3x_2 = (a'_1 a'_2 a'_3/a_1 a_2
a_3) d^3x_1$.  The potential at ${\bf x}$ may now be manipulated as
follows:
\begin{widetext}
\begin{equation}
\phi(E_1,{\bf x}) = \int_{E_1}
\frac{d^3x_1}{|{\bf x}_1 - {\bf x}|}
\delta\left(1 - \sum_\alpha
\frac{x_{1,\alpha}^2}{a_\alpha^2} \right)
= \frac{a_1 a_2 a_3}{a'_1 a'_2 a'_3}
\int_{E_2} \frac{d^3x_2}{\sqrt{\sum_\alpha
\left(\frac{a'_\alpha}{a_\alpha} x'_\alpha
- \frac{a_\alpha}{a'_\alpha}x_{2,\alpha} \right)^2}}
\delta\left(1 - \sum_\alpha \frac{x_{2,\alpha}^2}
{a_\alpha^{\prime 2}} \right)
\label{3.23}
\end{equation}
where ${\bf x}'$ is the point on $E_1$ corresponding to ${\bf x}$:
$x'_\alpha/a_\alpha = x_\alpha/a'_\alpha$.  If it were not for the
altered Coulomb factor in the denominator, the second line would
correspond precisely to the potential at a point on $E_1$ due to a
charged homoeoid $E_2$.  The idea now is to choose ${\bf a}'$ in such a
way as to restore the Coulomb factor to the form $|{\bf x}' - {\bf
x}_2|$.  One therefore seeks a solution to the equation,
\begin{equation}
0 = \sum_\alpha \left(\frac{a'_\alpha}{a_\alpha} x'_\alpha
- \frac{a_\alpha^{\phantom\prime}}
{a'_\alpha} x_{2,\alpha}\right)^2
- \sum_\alpha (x'_\alpha - x_{2,\alpha})^2
= \sum_\alpha (a_\alpha^{\prime 2} - a_\alpha^2)
\left(\frac{x_\alpha^{\prime 2}}{a_\alpha^2}
- \frac{x_{2,\alpha}^2}{a_\alpha^{\prime 2}} \right).
\label{3.24}
\end{equation}
\end{widetext}
If $a_\alpha^{\prime 2} - a_\alpha^2 = \lambda$ is independent of
$\alpha$, the ellipsoidal conditions $\sum_\alpha x_\alpha^{\prime
2}/a_\alpha^2 = 1 = \sum_\alpha x_{2,\alpha}^2/a_\alpha^{\prime 2}$
make the final sum vanish identically for any choice of the two points
${\bf x}'$ and ${\bf x}_2$ (or, equivalently, ${\bf x}$ and ${\bf
x}_1$).  Two ellipsoids related by this condition are called
\emph{confocal}, and the corresponding ``covariance'' of the Coulomb
factor is known as Ivory's Lemma \cite{Chandra}.

One obtains therefore the following remarkable result: if $\lambda
> 0$ labels the unique ellipsoidal surface $E_2$ confocal to $E_1$
and passing through the point ${\bf x}$, i.e.,
\begin{equation}
\sum_\alpha \frac{x_\alpha^2}{a_\alpha^2 + \lambda} = 1
\label{3.25}
\end{equation}
then
\begin{equation}
\phi(E_1,{\bf x}) = \frac{a_1 a_2 a_3}
{\sqrt{(a_1^2 + \lambda)(a_2^2 + \lambda)(a_3^2 + \lambda)}}
\phi(E_2,{\bf x}').
\label{3.26}
\end{equation}
The prefactor on the right hand side may be stated geometrically as a
condition that $E_1$ and $E_2$ carry the same total charge. To
conclude the argument, note that since $\phi(E_2,{\bf x}')$ is the
potential in the interior of a homogeneous homoeoid, which by the
previous results is a constant independent of ${\bf x}'$, the original
potential $\phi(E_1,{\bf x})$ will be a constant, independent of the
point ${\bf x}$ on the confocal ellipsoidal surface $E_2$.  This
verifies that the family of confocal ellipsoids, which sweep out all of
the space external to the hi-homoeoid as $\lambda$ varies over the
range $0 \leq \lambda < \infty$, characterize completely the
equipotential surfaces.  The completeness of this family of surfaces
also demonstrates that there can be no further independent solutions to
(\ref{3.24}). Furthermore, (\ref{3.26}), together with (\ref{3.17}),
yields the explicit formula
\begin{widetext}
\begin{equation}
\phi(E_1,{\bf x}) = v({\bf a})
\int_0^\infty \frac{dt}
{\sqrt{(a_1^2 + \lambda + t)(a_2^2 + \lambda + t)
(a_3^2 + \lambda + t)}} = v({\bf a})
\int_\lambda^\infty \frac{dt}{\Delta}
\equiv v({\bf a}) A({\bf a},\lambda).
\label{3.27}
\end{equation}
\end{widetext}
From (\ref{3.17}) and (\ref{3.18}) one then obtains the general
elliptic integral representation, valid both exterior and interior to
the hi-homoeoid:
\begin{equation}
\phi(E_1;{\bf x}) = \frac{v({\bf a})}{\sqrt{a_1^2 - a_3^2}}
F[\varphi(\lambda),k],
\label{3.28}
\end{equation}
in which $k$ remains as defined in (\ref{3.20}), while
$\varphi(\lambda)$ is now obtained from
\begin{equation}
\sin^2[\varphi(\lambda)]
= 1 - \frac{a_3^2 + \lambda}{a_1^2 + \lambda}
= \frac{a_1^2 - a_3^2}{a_1^2 + \lambda}.
\label{3.29}
\end{equation}

\subsection{Potential in the interior and exterior of an s-ellipsoid}
\label{sec:s_pot}

One may finally use the results (\ref{3.17}) and (\ref{3.27}) for the
potentials inside and outside a hi-homoeoid to construct the full
potential due to any s-ellipsoid, $E$. From (\ref{3.4}) and (\ref{3.5})
one obtains for the \emph{exterior} potential
\begin{equation}
\phi({\bf x}) = v({\bf a}) \int_0^1 d\mu \rho(\mu)
\int_{\lambda(\mu)}^\infty \frac{dt}{\Delta},
\ \ {\bf x} \in E^c
\label{3.30}
\end{equation}
in which $E^c$ is the complement of $E$, and $\lambda(\mu)$ is the
solution to the equation $\sum_\alpha x_\alpha^2/[a_\alpha^2 +
\lambda(\mu)] = \mu$, and therefore parameterizes the confocal
ellipsoidal surface passing through the point ${\bf x}/\sqrt{\mu}$.
This formula includes the case of the general s-homoeoid which simply
corresponds to a vanishing $\rho(\mu)$ for $\mu$ smaller than some
$\mu_0 > 0$. Interchanging the order of the integrations, and noting
that $\lambda(\mu) \to \infty$ as $\mu \to 0$, one has
\begin{eqnarray}
\phi({\bf x}) &=& v({\bf a})
\int_\lambda^\infty \frac{dt}{\Delta}
\int_{\mu(t)}^1 \rho(\mu) d\mu
\nonumber \\
&=& v({\bf a}) \int_\lambda^\infty
\frac{dt}{\Delta} \left\{\psi(1) - \psi[\mu(t)] \right\},
\ \ {\bf x} \in E^c \ \ \ \ \
\label{3.31}
\end{eqnarray}
in which, as in (\ref{3.27}), $\lambda \equiv \lambda(1)$ labels the
confocal ellipsoidal surface passing through ${\bf x}$ itself,
$\psi(\mu)$ was defined in (\ref{3.13}), and $\mu(t) \equiv \sum_\alpha
x_\alpha^2/(a_\alpha^2 + t)$.

To calculate the interior potential, one divides the contributions into
two parts: the potential $\phi_{\rm in}$ from the similar s-ellipsoid
whose surface passes through ${\bf x}$, and $\phi_{\rm out}$ from the
complement s-homoeoid whose inner surface passes through ${\bf x}$.
From (\ref{3.31}) one obtains
\begin{eqnarray}
\phi_{\rm in}({\bf x}) &=& v({\bf a}) \int_0^{\mu(0)} d\mu \rho(\mu)
\int_{\lambda(\mu)}^\infty \frac{dt}{\Delta}
\nonumber \\
&=& v({\bf a}) \int_0^\infty \frac{dt}{\Delta}
\left\{\psi[\mu(0)] - \psi[\mu(t)] \right\},
\label{3.32}
\end{eqnarray}
where $\mu(0) = \sum_\alpha x_\alpha^2/a_\alpha^2$ labels the surface
of the interior s-ellipsoid. The lower limit on the $t$ integral
vanishes because ${\bf x}$ lies right on this surface. On the other
hand, from (\ref{3.12}) one obtains
\begin{eqnarray}
\phi_{\rm out}({\bf x}) &=& v({\bf a})
\int_{\mu(0)}^1 d\mu \rho(\mu)
\int_0^\infty \frac{dt}{\Delta}
\nonumber \\
&=& \{\psi(1) - \psi[\mu(0)]\}
\int_0^\infty \frac{dt}{\Delta}.
\label{3.33}
\end{eqnarray}
The total interior potential is then given by
\begin{equation}
\phi({\bf x}) = v({\bf a}) \int_0^\infty \frac{dt}{\Delta}
\left\{\psi(1) - \psi[\mu(t)] \right\}, \ \ {\bf x} \in E.
\label{3.34}
\end{equation}
The only difference between (\ref{3.31}) and (\ref{3.34}) is the
lower limit $\lambda$ on the $t$ integral.

As a simple application of these formulas, for a uniformly charged
ellipsoid, $\rho(\mu) \equiv \rho_0$, one has $\psi(1) - \psi[\mu(t)] =
\rho_0[1 - \sum_\alpha x_\alpha^2/(a_\alpha^2 + t)]$, and therefore
\begin{equation}
\phi({\bf x}) = \rho_0 v({\bf a}) \left[A(\lambda)
- \sum_\alpha x_\alpha^2 A^{(1)}_\alpha(\lambda)\right]
\label{3.35}
\end{equation}
in which
\begin{equation}
A^{(1)}_\alpha(\lambda) = \int_\lambda^\infty
\frac{dt}{\Delta (a_\alpha^2 + t)}.
\label{3.36}
\end{equation}
For interior points one simply sets $\lambda = 0$.

\subsection{Generalization to certain classes of non-s-ellipsoids}
\label{sec:non_s}

The monomial charge density $\rho({\bf x}) = x_{\alpha_1} \ldots
x_{\alpha_n}$ corresponding to (\ref{2.45}) clearly does not fall into
the s-ellipsoid category. The following trick, however, may be used to
generalize the results to include this form. Given any solution to the
Poisson equation $-\nabla^2 \phi({\bf x}) = 4\pi\rho({\bf x})$, by
taking derivatives of both sides it is a trivial observation that
$\partial_\alpha \phi({\bf x})$ is the solution for charge density
$\partial_\alpha \rho({\bf x})$. Charge densities of the form $\rho
= \rho(\mu)$ with $\mu = \sum_\alpha x_\alpha^2/a_\alpha^2$ have so far
been considered. One obtains therefore
\begin{eqnarray}
\partial_\beta \rho &=& \rho'(\mu)
\frac{2x_\beta}{a_\beta^2}
\nonumber \\
\partial_\beta \partial_\gamma \rho &=& \rho''(\mu)
\frac{4x_\beta x_\gamma}{a_\beta^2 a_\gamma^2} + \rho'(\mu)
\frac{2}{a_\beta^2} \delta_{\beta \gamma},
\label{3.37}
\end{eqnarray}
and so on.  Repeated applications of the derivative trick therefore
produce s-ellipsoid charge densities multiplied by polynomials in the
components of ${\bf x}$. Clearly (\ref{2.45}) is a special case of this.

\section{Computation of the basic integrals}
\label{sec:int_compute}

All the necessary tools to compute the potentials ${\cal D}_{\bf
m}({\bf x})$, equation (\ref{3.1}), are now at hand. Motivated
by (\ref{3.37}), before specializing to these, we consider the
more general class of integrals
\begin{equation}
{\cal D}^{(q)}_{\bf m}({\bf x}) = \int_{V_s} d^3x'
\frac{q({\bf x}'){x'_1}^{m_1} {x'_2}^{m_2} {x'_3}^{m_3}}
{|{\bf x} - {\bf x}'|},
\label{4.1}
\end{equation}
in which $q = q(\mu)$ only. These reduce to (\ref{2.45}) when
$q \equiv 1$. The formalism developed in this section allows analytic
treatment of all integrals of the form (\ref{4.1}) by selecting of an
appropriate $\rho(\mu)$ upon which to apply the derivative trick. The
result for ${\cal D}_{\bf 0}({\bf x})$ is contained already in
(\ref{3.31}) and (\ref{3.34}):
\begin{eqnarray}
{\cal D}_{\bf 0}({\bf x}) &=& \int d^3x'
\frac{q({\bf x}')}{|{\bf x}-{\bf x}'|}
\nonumber \\
&=& v({\bf a}) \int_\lambda^\infty
\frac{dt}{\Delta} \rho_1[\mu(t)]
\label{4.2}
\end{eqnarray}
in which $\lambda({\bf x}) \equiv 0$ for ${\bf x}$ interior to the
ellipsoid, and is determined by (\ref{3.25}) for ${\bf x}$ exterior to
the ellipsoid, and
\begin{equation}
\rho_1(\mu) = \int_\mu^1 q(\mu') d\mu'.
\label{4.3}
\end{equation}
Thus $q = -d\rho_1/d\mu$ and both $q$ and $\rho_1(\mu)$ vanish for $\mu
> 1$.  For future reference, the sequence of charge densities
$\rho_n(\mu)$ are defined iteratively via
\begin{equation}
\rho_{n+1}(\mu) = \int_\mu^1 \rho_n(\mu') d\mu',\ \
n=1,2,3,\ldots,
\label{4.4}
\end{equation}
so that $\rho_n = -d\rho_{n+1}/d\mu$, and all $\rho_n$ vanish for $\mu
\geq 1$.

Next, from the first line of (\ref{3.37}) one has $q x_\alpha =
-(a_\alpha^2/2) \partial_\alpha \rho_1$, and hence
\begin{eqnarray}
{\cal D}^{(q)}_{{\bf \hat e}_\alpha}({\bf x}) &=& \int d^3x'
\frac{q({\bf x}') x'_\alpha}{|{\bf x}-{\bf x}'|}
\nonumber \\
&=& -\frac{1}{2} a_\alpha^2 \partial_\alpha \int d^3x'
\frac{\rho_1({\bf x}')}{|{\bf x}-{\bf x}'|}
\nonumber \\
&=& -\frac{1}{2} a_\alpha^2 v({\bf a})
\partial_\alpha \int_\lambda^\infty
\frac{dt}{\Delta} \rho_2[\mu(t)]
\nonumber \\
&=& a_\alpha^2 v({\bf a}) x_\alpha \int_\lambda^\infty
\frac{dt}{\Delta (a_\alpha^2 + t)} \rho_1[\mu(t)],
\label{4.5}
\end{eqnarray}
in which ${\bf \hat e}_1 = (1,0,0)$, etc., and a term coming from the
${\bf x}$-dependence of the lower limit of integration $\lambda$
vanishes since $\mu(\lambda) = 1$, and hence $\rho_2[\mu(\lambda)] =
\rho_2(1) = 0$.

A similar procedure, using the identity
\begin{equation}
q x_\alpha x_\beta = \frac{1}{4} a_\alpha^2 a_\beta^2
\partial_\alpha \partial_\beta \rho_2
+ \frac{1}{2} a_\alpha^2 \rho_1 \delta_{\alpha \beta},
\label{4.6}
\end{equation}
yields
\begin{widetext}
\begin{eqnarray}
{\cal D}_{{\bf \hat e}_\alpha + {\bf \hat e}_\beta}({\bf x})
&=& \int d^3x' \frac{q({\bf x}') x'_\alpha x'_\beta}{|{\bf x}-{\bf x}'|}
\nonumber \\
&=& \frac{1}{4} a_\alpha^2 a_\beta^2 v({\bf a})
x_\alpha x_\beta \partial_\alpha \partial_\beta
\int_\lambda^\infty \frac{dt}{\Delta} \rho_3[\mu(t)]
+ \frac{1}{2} a_\alpha^2 v({\bf a}) \delta_{\alpha\beta}
\int_\lambda^\infty \frac{dt}{\Delta} \rho_2[\mu(t)]
\nonumber \\
&=& v({\bf a}) a_\alpha^2 a_\beta^2
x_\alpha x_\beta \int_\lambda^\infty
\frac{dt}{\Delta (a_\alpha^2 + t) (a_\beta^2 + t)} \rho_1[\mu(t)]
+ \frac{1}{2} v({\bf a}) a_\alpha^2 \delta_{\alpha \beta}
\int_\lambda^\infty \frac{t dt}
{\Delta (a_\alpha^2 + t)} \rho_2[\mu(t)].
\label{4.7}
\end{eqnarray}
Similarly, the identity
\begin{equation}
x_\alpha x_\beta x_\gamma q
= -\frac{1}{8} a_\alpha^2 a_\beta^2 a_\gamma^2
\partial_\alpha \partial_\beta \partial_\gamma \rho_3
- \frac{1}{4} \left[a_\alpha^2 a_\gamma^2 \delta_{\alpha \beta}
\partial_\gamma \rho_2
+ (\gamma \leftrightarrow \beta)
+ (\gamma \leftrightarrow \alpha) \right],
\label{4.8}
\end{equation}
yields
\begin{eqnarray}
{\cal D}_{{\bf \hat e}_\alpha + {\bf \hat e}_\beta
+ {\bf \hat e}_\gamma}({\bf x})
&=& \int d^3x' \frac{q({\bf x}') x'_\alpha x'_\beta x'_\gamma}
{|{\bf x}-{\bf x}'|}
\nonumber \\
&=& v({\bf a}) a_\alpha^2 a_\beta^2 a_\gamma^2
x_\alpha x_\beta x_\gamma
\int_\lambda^\infty \frac{dt}
{\Delta (a_\alpha^2 + t)(a_\beta^2 + t)(a_\gamma^2 + t)}
\rho_1[\mu(t)]
\nonumber \\
&&+\ \frac{1}{2} v({\bf a})
\left\{a_\alpha^2 a_\gamma^2 x_\gamma \delta_{\alpha \beta}
\int_\lambda^\infty \frac{t dt}
{\Delta (a_\alpha^2 + t)(a_\gamma^2 + t)}
\rho_2[\mu(t)] + (\gamma \leftrightarrow \beta)
+ (\gamma \leftrightarrow \alpha) \right\}.
\label{4.9}
\end{eqnarray}
\end{widetext}
In all cases, terms arising from derivatives acting on the lower limit
$\lambda$ do not contribute since $\rho_n(1) = 0$ for $n \geq 1$.  It
is clear that calculations become substantially more tedious as $n$
increases.  General cases beyond $n=3$ will not be treated explicitly
in the present work. Simplifications for solid homogeneous ellipsoids
that allow general evaluation of ${\cal D}_{\bf m}({\bf x})$ will be
treated in Sec.\ \ref{sec:solid_ellipsoid}.

\section{Explicit evaluations for the case of a solid homogeneous
ellipsoid}
\label{sec:solid_ellipsoid}

For the special case that the scatterer is a solid, homogeneous
ellipsoid, $q({\bf x}) \equiv 1$, which will be the focus of all
explicit computations in this and later sections, one obtains
$\rho_n(\mu) = (1-\mu)^n/n!$ for $0 \leq \mu \leq 1$, $\rho_n(\mu)
\equiv 0$ for $\mu \geq 1$. From (\ref{3.31}), the potential due to
$\rho_n$ is given by
\begin{eqnarray}
\phi_n({\bf x}) &=& \frac{1}{n!}
\int_{V_s} \frac{d^3x'}{|{\bf x}-{\bf x}'|}
\left[1 - \sum_\alpha \left(\frac{x_\alpha'}{a_\alpha}
\right)^2 \right]^n
\nonumber \\
&=& \frac{v({\bf a})}{(n+1)!} \int_\lambda^\infty \frac{dt}{\Delta}
\left[1 - \sum_\alpha \frac{x_\alpha^2}{a_\alpha^2+t} \right]^{n+1}
\label{5.1} \\
&=& v({\bf a}) {\sum_{\bf k}}'
\frac{(-1)^{|{\bf k}|} x_1^{2k_1} x_2^{2k_2} x_3^{2k_3}}
{k_1! k_2! k_3! (n + 1 - |{\bf k}|)!}
A_{\bf k}({\bf a},\lambda) ,
\nonumber
\end{eqnarray}
in which we define $|{\bf k}| = \sum_\alpha k_\alpha$, and the
prime on the last sum indicates that it is restricted by $|{\bf k}|
\leq n+1$, and we have defined the basic integrals
\begin{equation}
A_{\bf k}({\bf a},\lambda)
= \int_\lambda^\infty \frac{dt}{\Delta({\bf a},t)
\prod_\alpha (a_{\alpha}^2 + t)^{k_\alpha}},
\label{5.2}
\end{equation}
with $A_{\bf 0} = A$ [see (\ref{3.17})]. These obey the simple
iterative relation:
\begin{equation}
\frac{\partial}{\partial a_\alpha^2} A_{\bf k}
= -(k_\alpha + 1/2) A_{{\bf k} + {\bf e}_\alpha},
\label{5.3}
\end{equation}
in which all derivatives are at fixed $\lambda$.  Thus,
\begin{eqnarray}
A_{\bf k}({\bf a},\lambda)
&=& \frac{(-1)^{|{\bf k}|} \Gamma(1/2)^3}
{\Gamma(k_1 + 1/2)\Gamma(k_2 + 1/2)\Gamma(k_3 + 1/2)}
\nonumber \\
&&\times\ \frac{\partial^{|{\bf k}|}}{\partial (a_1^2)^{k_1}
\partial (a_2^2)^{k_2} \partial (a_3^2)^{k_3}} A({\bf a},\lambda).
\label{5.4}
\end{eqnarray}
It is apparent that interior to the ellipsoid, where $\lambda = 0$,
$\phi_n$ is an even polynomial of degree $2(n+1)$.

\subsection{Monomial charge densities: even cases}
\label{sec:mono_even}

One may now use (\ref{5.1}) to derive formulas for the potential due
to monomial charge densities, extending the results of Sec.\
\ref{sec:int_compute} to general index values. From the first line of
(\ref{5.1}), one obtains for \emph{even} monomials
\begin{equation}
{\cal D}_{2{\bf l}}({\bf x}) \equiv \int_{V_s}
\frac{d^3x'}{|{\bf x}-{\bf x}'|}
\prod_\alpha (x_\alpha')^{2l_\alpha}
= \partial^{\bf l}_{\bf a} \phi_{|{\bf l}|}({\bf x};{\bf a})
\label{5.5}
\end{equation}
where, again, $|{\bf l}| = \sum_\alpha l_\alpha$, and the operator
acting on the ${\bf a}$-dependence is defined by:
\begin{equation}
\partial^{\bf l}_{\bf a} \equiv (-1)^{|{\bf l}|}
\frac{\partial^{|{\bf l}|}}
{\partial (a_1^{-2})^{l_1} \partial (a_2^{-2})^{l_2}
\partial (a_3^{-2})^{l_3}}.
\label{5.6}
\end{equation}
The derivatives eliminate all monomials of order lower than
$|{\bf l}|$, and there are no contributions from the
${\bf a}$-dependence of the integration region $V_s$ because the
integrand of $\phi_n$, along with its first $n-1$ derivatives,
vanish on the boundary. This observation allows us to apply the same
derivatives to the last line of (\ref{5.1}) at constant $\lambda$:
terms arising from derivatives of the ${\bf a}$-dependence of
$\lambda$ must, via the second line of (\ref{5.1}), all cancel. One
obtains, therefore,
\begin{equation}
{\cal D}_{2{\bf l}}({\bf x}) = {\sum_{\bf k}}'
{\cal D}_{2{\bf k}}^{(2{\bf l})}(\lambda)
x_1^{2k_1} x_2^{2k_2} x_3^{2k_3},
\label{5.7}
\end{equation}
in which the prime on the sum indicates the constraint $|{\bf k}| \leq
|{\bf l}|+ 1$, and the coefficients are given by
\begin{equation}
{\cal D}_{2{\bf k}}^{(2{\bf l})}(\lambda)
= \frac{(-1)^{|{\bf k}|}}
{k_1! k_2! k_3! (|{\bf l}|-|{\bf k}|+1)!}
\partial^{\bf l}_{\bf a}
[v({\bf a}) A_{\bf k}({\bf a},\lambda)].
\label{5.8}
\end{equation}
Internal to the scatterer, ${\cal D}_{2{\bf k}}^{(2{\bf l})}(\lambda)$
is a polynomial of degree $2(|{\bf l}| + 1)$.

By iterating the relation
\begin{widetext}
\begin{equation}
\frac{\partial}{\partial (a^{-2})}
\frac{a^{2m+1}}{(a^2+t)^{k+m+\frac{1}{2}}}
= \left(k + m + \frac{1}{2} \right)
\frac{a^{2m+5}}{(a^2+t)^{k+m+\frac{3}{2}}}
- \left(m + \frac{1}{2} \right)
\frac{a^{2m+3}}{(a^2+t)^{k+m+\frac{1}{2}}},
\label{5.9}
\end{equation}
one obtains
\begin{equation}
(-1)^l \frac{\partial}{\partial (a^{-2})^l}
\frac{a^{2m+1}}{(a^2+t)^{k+\frac{1}{2}}}
= (-1)^k k! \sum_{p=0}^l C^{(klm)}_p
\frac{a^{2(m+l+p)+1}}{(a^2+t)^{k+m+p+\frac{1}{2}}},
\label{5.10}
\end{equation}
in which we have defined the coefficients
\begin{equation}
C^{(klm)}_p = \frac{(-1)^{k+l}}{k!}
\left(\begin{array}{c} l \\ p \end{array} \right)
\frac{\Gamma(\frac{1}{2} - m - p)}
{\Gamma(\frac{1}{2} - m - l)}
\frac{\Gamma(k + m + p + \frac{1}{2})}
{\Gamma(k + m + \frac{1}{2})}.
\label{5.11}
\end{equation}
This produces the explicit form
\begin{equation}
{\cal D}_{2{\bf k}}^{(2{\bf l})}(\lambda)
= \frac{v({\bf a})}{(|{\bf l}|-|{\bf k}|+1)!}
{\sum_{\bf p}}' A_{{\bf k}+{\bf p}}({\bf a},\lambda)
\prod_\alpha C^{(k_\alpha l_\alpha 0)}_{p_\alpha}
a_\alpha^{2 (l_\alpha + p_\alpha)},
\label{5.12}
\end{equation}
where the prime indicates that the sum is limited to the range $0
\leq p_\beta \leq l_\beta$. Inserted in to (\ref{5.7}), the result
(\ref{5.12}) explicitly exhibits the monomial integral (\ref{5.5})
to a linear combination of the basic integrals (\ref{5.2}).

If one wishes to compute the coefficients ${\cal D}_{2{\bf
k}}^{(2{\bf l})}$ for all (even) monomials of degree at most $N$, since
$0 \leq |{\bf l}| \leq [N/2]$, and hence $0 \leq |{\bf k}| \leq
[N/2]+1$, one therefore needs to compute all basic integrals $A_{\bf n}$
with indices constrained by $0 \leq |{\bf n}| \leq 2[N/2]+1$.

\subsection{Monomial charge densities: odd cases}
\label{sec:mono_odd}

Consider now cases where the monomial is not even. The desired results
follow immediately from the identities
\begin{eqnarray}
-\frac{a_\alpha^2}{2} \partial_\alpha \phi_{n+1}({\bf x})
&=& \frac{1}{n!} \int \frac{d^3x'}{|{\bf x}-{\bf x}'|} x'_\alpha
\left[1 - \sum_\gamma \left(\frac{x'_\gamma}{a_\gamma^2} \right)^2 \right]^n
\nonumber \\
\frac{a_\alpha^2 a_\beta^2}{4} \partial_\alpha \partial_\beta \phi_{n+2}({\bf x})
&=& \frac{1}{n!} \int \frac{d^3x'}{|{\bf x}-{\bf x}'|} x'_\alpha x'_\beta
\left[1 - \sum_\gamma \left(\frac{x'_\gamma}{a_\gamma^2} \right)^2 \right]^n,
\ \ \ \ \alpha \neq \beta
\nonumber \\
-\frac{a_1^2 a_2^2 a_3^2}{8} \partial_1 \partial_2 \partial_3 \phi_{n+3}({\bf x})
&=& \frac{1}{n!} \int \frac{d^3x'}{|{\bf x}-{\bf x}'|} x'_1 x'_2 x'_3
\left[1 - \sum_\gamma \left(\frac{x'_\gamma}{a_\gamma^2} \right)^2 \right]^n.
\label{5.13}
\end{eqnarray}
\end{widetext}
By applying the operator $\partial_{\bf a}^{\bf l}$ to these
expressions one may now isolate the required monomials:
\begin{equation}
D_{2{\bf l}+{\bf m}}({\bf x})
= (-1)^{|{\bf l}+{\bf m}|} \partial_{\bf a}^{\bf l}
\left[\prod_\alpha \left(\frac{a_\alpha^2}{2}
\partial_\alpha \right)^{m_\alpha}
\phi_{|{\bf l}+{\bf m}|}({\bf x}) \right]
\label{5.14}
\end{equation}
where the elements of ${\bf m}$ are all either 0 or 1, and as before
$|{\bf l}+{\bf m}| = \sum_\gamma (l_\alpha+m_\alpha)$. Following the
derivation of (\ref{5.12}), one therefore obtains the monomial
expansions
\begin{equation}
D_{2{\bf l}+{\bf m}}({\bf x})
= {\sum_{\bf k}}' D^{(2{\bf l}+{\bf m})}_{2{\bf k}+{\bf m}}(\lambda)
\prod_\alpha x_\alpha^{2k_\alpha + m_\alpha},
\label{5.15}
\end{equation}
in which the sum continues to be restricted to $|{\bf k}| \leq |{\bf
l}|+1$ and
\begin{eqnarray}
D^{(2{\bf l}+{\bf m})}_{2{\bf k}+{\bf m}}(\lambda)
&=& \frac{v({\bf a})}{(|{\bf l}|-|{\bf k}|+1)!}
{\sum_{\bf p}}' A_{{\bf k}+{\bf m}+{\bf p}}(\lambda)
\nonumber \\
&&\times\ \prod_\alpha C^{(k_\alpha l_\alpha m_\alpha)}_{p_\alpha}
a_\alpha^{2(l_\alpha+p_\alpha+m_\alpha)},\ \ \ \
\label{5.16}
\end{eqnarray}
where the sum is again over all $0 \leq p_\alpha \leq l_\alpha$.
Equation (\ref{5.12}) is now clearly a special case of (\ref{5.16}) in
which ${\bf m}$ vanishes. If one desires these coefficients for all
monomials of degree at most $N$, then the basic integrals $A_{\bf n}$
will be required for all ${\bf n}$ such that $0 \leq |{\bf n}| \leq
N+1$.

It is straightforward to check that (\ref{5.15}) and (\ref{5.16})
reproduce the partial results in Sec.\ \ref{sec:int_compute} if one
sets $q=1$.

\subsection{Recursion relations for the \emph{A} integrals}
\label{sec:A_recursion}

We finally reduce the computation of the $A$ integrals to algebraic
recursion relations, given only the pair
\begin{eqnarray}
A_{000} &=& \frac{2 F(\varphi,k)}{\sqrt{a_1^2 - a_3^2}}
\nonumber \\
A_{100} &=& \frac{2 E(\varphi,k)}{(a_2^2-a_1^2) \sqrt{a_1^2 - a_3^2}}
- \frac{F(\varphi,k)}{a_2^2-a_1^2}
\label{5.17}
\end{eqnarray}
[see (\ref{3.18})--(\ref{3.22})].

The relations are based on the following three identities. First, by
integrating the identity
\begin{equation}
\partial_t \frac{1}{\prod_\alpha (a_\alpha^2 +t)^{k_\alpha+\frac{1}{2}}}
= -\frac{1}{\prod_\alpha (a_\alpha^2 +t)^{k_\alpha+\frac{1}{2}}}
\sum_\beta \frac{k_\beta + \frac{1}{2}}{a_\beta^2 + t},
\label{5.18}
\end{equation}
over the range $\lambda \leq t < \infty$, one obtains the relation
\begin{equation}
\sum_\beta \left(k_\beta + \frac{1}{2} \right)
A_{{\bf k} + {\bf \hat e}_\beta}({\bf a},\lambda)
= \frac{1}{\prod_\alpha
(a_\alpha^2 + \lambda)^{k_\alpha + \frac{1}{2}}}.
\label{5.19}
\end{equation}

Second, it is easy to verify the homogeneity relation
\begin{equation}
A_{\bf k}(\kappa {\bf a},\kappa^2 \lambda)
= \kappa^{-2(k_1+k_2+k_3)-1} A_{\bf k}({\bf a},\lambda).
\label{5.20}
\end{equation}
for arbitrary scale factor $\kappa > 0$. By taking the derivative of
both sides with respect to $\kappa$ and setting $\kappa = 1$, one
obtains the Euler relation
\begin{eqnarray}
&&\sum_\beta (a_\beta^2 + \lambda)
\left(k_\beta + \frac{1}{2} \right)
A_{{\bf k} + {\bf \hat e}_\beta}({\bf a},\lambda)
\nonumber \\
&&\ \ \ \ \ \ \ = \left(|{\bf k}| + \frac{1}{2} \right)
A_{\bf k}({\bf a},\lambda).
\label{5.21}
\end{eqnarray}

By combining (\ref{5.19}) and (\ref{5.21}) one obtains for each
$\gamma$,
\begin{eqnarray}
&& \sum_{\alpha (\neq \gamma)} (a_\gamma^2 - a_\alpha^2)
\left(k_\alpha + \frac{1}{2}\right)
A_{{\bf k} + {\bf \hat e}_\alpha}({\bf a},\lambda)
\label{5.22} \\
&&=\ \frac{a_\gamma^2 + \lambda}
{\prod_\alpha (a_\alpha^2 + \lambda)^{k_\alpha + \frac{1}{2}}}
- \left(|{\bf k}| + \frac{1}{2} \right)
A_{\bf k}({\bf a},\lambda).
\nonumber
\end{eqnarray}

Third, by noting the trivial identity
\begin{equation}
a_\alpha^2 - a_\beta^2 = (a_\alpha^2 + t) - (a_\beta^2 + t)
\label{5.23}
\end{equation}
one obtains for any $\alpha \neq \beta$,
\begin{equation}
A_{\bf k}({\bf a},\lambda) = \frac{1}{a_\alpha^2-a_\beta^2}
\left[A_{{\bf k} - {\bf \hat e}_\alpha}({\bf a},\lambda)
- A_{{\bf k} - {\bf \hat e}_\beta}({\bf a},\lambda) \right].
\label{5.24}
\end{equation}

Equation (\ref{5.22}) provides a relation between $A_{\bf k}$ and any
two of its ``forward'' neighbors, while (\ref{5.24}) provides a
relation between $A_{\bf k}$ and any two of its ``backward'' neighbors.
It is easy to see that by applying these two relations appropriately
one may recursively generate any $A_{\bf k}$ from the pair (\ref{5.17})
alone. In numerical implementations care must be taken, however. The
recursion (\ref{5.24}) will likely be unstable if $|a_\alpha-a_\beta|$
is too small. In this case an approximate form for $A_{\bf k}$ will
need to be computed for large enough $|{\bf k}|$, and the recursion
relations iterated \emph{backward} to smaller values \cite{numrec}.

\section{Cases of degeneracy}
\label{sec:degenerate}

Great simplifications occur when two or more of the semi-major axes are
identical. The cases of most interest are spheroids where $a_1 = a_2
\equiv a$, and either $a_3 > a$ (prolate spheroid) or $a_3 < a$ (oblate
spheroid). The third case, $a_1=a_2=a_3 \equiv a$, corresponds to the
trivial case of the sphere.

\subsection{Sphere case}
\label{sec:spherecase}

For the case of the sphere one finds the fully analytic result
\begin{eqnarray}
A_K(a,\lambda) &=& \int_\lambda^\infty
\frac{dt}{(a^2 + t)^{K + \frac{3}{2}}}
\nonumber \\
&=& \frac{1}{K + \frac{1}{2}}
\frac{1}{(a^2 + \lambda)^{K + \frac{1}{2}}}
\label{6.1}
\end{eqnarray}
where $K = |{\bf k}|$. From (\ref{3.25}), for exterior points one
obtains
\begin{equation}
\lambda = |{\bf x}|^2 - a^2,\ |{\bf x}| \geq a.
\label{6.2}
\end{equation}

\subsection{Prolate and oblate spheroids}
\label{sec:pro_ob_spheroids}

For the more interesting case of spheroids, the basic integral
\begin{equation}
A({\bf a},\lambda) = \int_\lambda^\infty
\frac{dt}{(a^2 + t) \sqrt{a_3^2 + t}}
\label{6.3}
\end{equation}
is elementary, and for the prolate case one obtains
\begin{equation}
A({\bf a},\lambda) = \frac{1}{\sqrt{a_3^2 - a^2}}
\ln\left(\frac{1 + \sqrt{\frac{a_3^2-a^2}{a_3^2+\lambda}}}
{1 - \sqrt{\frac{a_3^2-a^2}{a_3^2+\lambda}}} \right),
\label{6.4}
\end{equation}
while for the oblate case one obtains
\begin{equation}
A({\bf a},\lambda) = \frac{2}{\sqrt{a^2 - a_3^2}}
\arctan \sqrt{\frac{a^2-a_3^2}{a_3^2+\lambda}}.
\label{6.5}
\end{equation}
In either case, the higher order integrals
\begin{equation}
A_{k k_3}({\bf a},\lambda) = \int_\lambda^\infty
\frac{dt}{(a^2+t)^{k+1}(a_3^2+t)^{k_3+\frac{1}{2}}},
\label{6.6}
\end{equation}
where $k = k_1+k_2$, may now be generated by iterating (\ref{5.22}),
which may now be put in the form
\begin{widetext}
\begin{eqnarray}
A_{k+1,k_3}({\bf a},\lambda) &=& \frac{1}{(k+1)(a_3^2-a^2)}
\left[\frac{1}{(a^2+\lambda)^{k+1} (a_3^2+\lambda)^{k_3-\frac{1}{2}}}
- \left(k+k_3+\frac{1}{2} \right) A_{kk_3}({\bf a},\lambda) \right]
\nonumber \\
A_{k,k_3+1}({\bf a},\lambda) &=& \frac{1}{(k_3+\frac{1}{2})(a^2-a_3^2)}
\left[\frac{1}{(a^2+\lambda)^k (a_3^2+\lambda)^{k_3+\frac{1}{2}}}
- \left(k+k_3+\frac{1}{2} \right) A_{kk_3}({\bf a},\lambda) \right].
\label{6.7}
\end{eqnarray}
Here, the first and second lines follow by setting $\gamma=3$ and
$\gamma = 1$ or 2, respectively.  Note that care should be taken for
small $|a_3-a|$ since the terms in brackets are then small as well, so
that the overall result remains finite.
\end{widetext}

Finally, equation (\ref{3.25}) for $\lambda$ is quadratic, with
solution
\begin{eqnarray}
\lambda &=& \sqrt{\frac{1}{4}(a_3^2 + a^2 - |{\bf x}|^2)^2
+ a^2 a_3^2 \left(\sum_\alpha \frac{x_\alpha^2}{a_\alpha^2} - 1 \right)}
\nonumber \\
&&-\ \frac{1}{2}(a_3^2 + a^2 - |{\bf x}|^2),
\label{6.8}
\end{eqnarray}
which is consistently positive for ${\bf x}$ outside $V_s$.

\section{Formal solution to the scattering problem}
\label{sec:formal_scatt}

Now that a complete formalism for the evaluation of the necessary
integrals has been presented, one may turn finally to the solution of
the scattering problem (\ref{2.9}) using the high contrast formulation
(\ref{2.24}). The main work in the application of the mean field
approach is the computation of the electric field internal to the
scatterer, which involves the evaluation of the ${\cal D}^{\bf m}({\bf
x})$ [equation (\ref{2.46})] inside the ellipsoid. For this purpose,
one may therefore set $\lambda \equiv 0$ in all of the formulas derived
in Secs.\ \ref{sec:solid_ellipsoid} and (\ref{sec:degenerate}).

\subsection{Noninductive solutions}
\label{sec:noninduct}

Before proceeding further, it is worth further clarifying the nature of
the projection in (\ref{2.24}), which removes the noninductive part of
the electric field. This is most easily done directly from the integral
equation (\ref{2.9}), where, for simplicity, we consider the case of
homogeneous background as well as scatterer. Using translation
invariance of $g = g(|{\bf x}-{\bf x}'|)$ [equation (\ref{2.6})], and
integrating by parts, as in (\ref{2.23}), one obtains
\begin{eqnarray}
{\bf E}({\bf x}) &=& {\bf E}_b({\bf x}) +
Q \int_{V_s} d^3x' g({\bf x},{\bf x}') {\bf E}({\bf x}')
\nonumber \\
&&\ - \frac{Q}{\kappa_b^2} \nabla
\int_{\partial V_s} d^2r' g({\bf x},{\bf r}')
{\bf \hat n}({\bf r}') \cdot {\bf E}({\bf r}'),\ \ \ \
\label{7.1}
\end{eqnarray}
where $Q=\kappa^2 - \kappa_b^2$ and we have used $\nabla \cdot {\bf E}
= 0$. The second line is a pure gradient, and hence part of $\nabla
\Phi$. In (\ref{2.22}) we restricted the class of solutions to those
with vanishing normal component. However, it is clear from this term
that a contribution with ${\bf \hat n} \cdot {\bf E} = O(\kappa_b^2)$
[which provides the leading finite $\kappa_b$ correction to the
boundary condition (\ref{2.22})], although contributing negligibly to
the inductive part of ${\bf E}$---which is contained entirely the first
line of (\ref{7.1})---does make a finite contribution to $\Phi$.

To explore this further, let us seek purely noninductive solutions
${\bf E} = -\nabla \Phi$, which requires also $\nabla^2 \Phi = 0$.
This separation is consistent only to leading order in
$\kappa_b^2/\kappa^2$, where (\ref{7.1}) reduces to
\begin{equation}
{\bf E}_b({\bf x}) = -\frac{\kappa^2}{\kappa_b^2}
\nabla \int_{\partial V_s} d^2r'
\frac{{\bf \hat n}({\bf r}') \cdot \nabla \Phi({\bf r}')}
{4\pi |{\bf x}-{\bf r}'|}
\label{7.2}
\end{equation}
Thus, the background field ${\bf E}_b = -\nabla \Phi_b$ must
also be noninductive, and one identifies
\begin{eqnarray}
\Phi_b({\bf x}) &=& \frac{\kappa^2}{\kappa_b^2}
\int_{\partial V_s} d^2r'
\frac{{\bf \hat n}({\bf r}') \cdot \nabla' \Phi({\bf r}')}
{4\pi |{\bf x}-{\bf r}'|}
\nonumber \\
&=& -\frac{\kappa^2}{\kappa_b^2}
\nabla \cdot \int_{V_s} d^3x'
\frac{\nabla' \Phi({\bf x}')}{4\pi |{\bf x}-{\bf x}'|}.
\label{7.3}
\end{eqnarray}
Note that a solution to the Laplace equation is uniquely specified
by a Neumann boundary condition, so there is no contradiction
in the fact that the full $\Phi$ appears in the second line, whereas
only its boundary normal derivative appears in the first.

If $\Phi$ is a polynomial, the last integral is precisely of the type
considered in Sec.\ \ref{sec:solid_ellipsoid}. The two gradients
imply that $\Phi_b$ is a polynomial of the same order. In particular,
the linear form $\Phi_b({\bf x}) = x_\alpha$ leads to
\begin{eqnarray}
\Phi({\bf x}) &=& \chi_\alpha x_\alpha
\nonumber \\
\chi_\alpha &\equiv& \frac{\kappa_b^2}{\kappa^2}
\frac{2\pi}{v({\bf a}) A_{{\bf \hat e}_\alpha}({\bf a},0)}
\label{7.4}
\end{eqnarray}
This is equivalent the well known solution for an ellipsoid in a uniform
static applied electric field ${\bf E}_0$:
\begin{equation}
{\bf E}_\mathrm{int} = \sum_\alpha {\bf \hat e}_\alpha
\chi_\alpha E_{0,\alpha}.
\label{7.5}
\end{equation}

Higher order polynomials may be handled in a similar fashion. For example,
the form
$\Phi_b = x_\alpha x_\beta$ ($\alpha \neq \beta$), leads to
\begin{eqnarray}
\Phi({\bf x}) &=& \chi_{\alpha \beta} x_\alpha x_\beta
\nonumber \\
\chi_{\alpha\beta} &=& \frac{\kappa_b^2}{\kappa^2}
\frac{\pi}{v({\bf a}) (a_\alpha^2 + a_\beta^2)
A_{{\bf \hat e}_\alpha + {\bf \hat e}_\beta}({\bf a},0)}.
\label{7.6}
\end{eqnarray}
A somewhat messier calculation allows one to compute the response to
$\Phi_b = x_\alpha^2 - x_\beta^2$, ($\alpha \neq \beta$) in the form
$\Phi =  c_0 + \sum_\alpha c_\gamma x_\gamma^2$, with coefficients
obeying the constraint $\sum_\gamma c_\gamma = 0$.  Progressively
higher order polynomial forms for $\Phi_b$ may be reduced to
progressively higher order matrix inversions for the polynomial
coefficients of $\Phi$ \cite{foot:ellipseharm}.

It is clear from (\ref{7.3}) that all such solutions generate very
small internal fields, $|{\bf E}_\mathrm{int}|/|{\bf E}_b| =
O(\kappa_b^2/\kappa^2)$. However, via the last term on the right hand
side of (\ref{7.1}), the external field is of the same order as the
background field, but, once again, is invisible to a purely inductive
measurement.

One may also compute corrections to these solutions, in powers
of $\kappa_b^2/\kappa^2$, arising from the terms in (\ref{7.1})
left out of (\ref{7.3}). For metallic targets, however, these
corrections are negligible, and one concludes that the noninductive
modes respond quasistatically in the frequency regime of interest
here. The physical origin of this result is that these modes are
controlled by the induced polarization charges on the surface of
the target, which respond essentially instantaneously to the
applied field. Current flow, required to generate magnetic fields
that would influence an inductive measurement, is negligible.

\subsection{Perturbation theory for inductive excitations}
\label{sec:leadinduct}

Consider now solutions to the internal field equation (\ref{2.24}).
Simple dimensional analysis shows that the Coulomb integral is of order
\begin{equation}
\frac{k \mu \sigma L_s^2}{c}
= \frac{L_s^2}{2\pi \delta_s^2}
\equiv \frac{\eta_s^2}{2\pi},
\label{7.7}
\end{equation}
where $L_s$ is target characteristic diameter, and $\delta_s(\omega)
= c/\sqrt{2\pi \sigma \mu \omega}$ is the target skin depth \cite{Jackson}.
The parameter
\begin{equation}
\eta_s = \frac{L_s}{\delta_s}
= \frac{\pi L_s}{5\ \mathrm{cm}}
\left(\frac{\mu}{\mu_0} \right)^{1/2}
\left(\frac{\sigma}{10^7\ \mathrm{S/m}} \right)^{1/2}
\left(\frac{f}{100\ \mathrm{Hz}} \right)^{1/2}
\label{7.8}
\end{equation}
[written here in MKS units; compare (\ref{2.17})] is small at low
frequencies, and is the formal mean field expansion parameter. The
dimensional quantities inserted here typical of compact metallic
targets that might be of interest.  To zeroth order, for $\eta_s
\ll 1$, equation (\ref{2.19}) reduces simply to ${\bf A} = {\bf A}_b$:
the target is transparent to magnetic excitations [as opposed to its
essentially complete opaqueness (\ref{7.3}) to polarizing fields].
In particular, the zeroth order solutions
\begin{equation}
{\bf E}_\mathrm{int} = \hat {\cal P}_\sigma {\bf E}_b
= {\bf Z}_{lmp}^{({\bf a};i)}({\bf x}),
\label{7.9}
\end{equation}
may be parameterized by the basis functions defined by (\ref{2.39}),
and rescaled according to (\ref{2.32}). Here, the $\nabla \Phi_b$ part
of ${\bf E}_b$, which has been projected out of (\ref{7.9}), generates,
via Sec.\ \ref{sec:noninduct}, $O(\kappa_b^2/\kappa^2)$ polarization
corrections to ${\bf E}_\mathrm{int}$.

One may conveniently compute corrections iteratively:
\begin{equation}
{\bf E}({\bf x}) = \sum_{n=0}^\infty
\left(\frac{ik \mu \sigma}{c} \right)^n
\Delta {\bf E}_n({\bf x}),
\label{7.10}
\end{equation}
in which the terms satisfy the recursion relation
\begin{equation}
\Delta {\bf E}_{n+1}({\bf x}) =
\hat {\cal P}_\sigma \int_{V_s} d^3x'
\frac{\Delta {\bf E}_n({\bf x})}{|{\bf x}-{\bf x}'|},
\label{7.11}
\end{equation}
beginning with $\Delta {\bf E}_0 = \hat {\cal P}_\sigma {\bf E}_b$. The
series (\ref{7.10}) is an explicit expansion in powers of the small
parameter $k\mu \sigma/c = \eta_s^2/2\pi L_s^2$.  Since the basis
functions (\ref{7.9}) are polynomials of degree $N = l+2p+i-1$, and
application of the Coulomb integral adds two to the degree of the
numerator (application of the projection operator does not change
this), the $n$th term in the series (\ref{7.10}) is of degree $N+2n$.

Consider, for example, the linear form
\begin{equation}
\Delta {\bf E}_0
= \frac{a_\beta}{a_\alpha} x_\alpha {\bf \hat e}_\beta
- \frac{a_\alpha}{a_\beta} x_\beta {\bf \hat e}_\alpha,\
(\alpha \neq \beta),
\label{7.12}
\end{equation}
which circulates in the $\alpha \beta$-plane, and indeed lies in the
space of functions defined by (\ref{2.22}). These are, in fact, three
independent linear combinations of the three $l=1$, $p=0$ harmonic basis
functions ${\bf Z}_{lmp}^{({\bf a};1)}$. This form corresponds to a
constant magnetic field
\begin{equation}
{\bf H}_b = \frac{1}{ik\mu} \nabla \times {\bf E}_b
= \frac{1}{ik\mu}
\frac{a_\alpha^2 + a_\beta^2}{a_\alpha a_\beta}
{\bf \hat e}_\gamma,
\label{7.13}
\end{equation}
where $(\alpha \beta \gamma)$ is taken to be a cyclic permutation of
$(123)$, and the magnetic field is therefore orthogonal to the $\alpha
\beta$-plane. This will be an adequate model of the background field if
the transmitter is sufficiently far from the target.

\begin{figure}

\includegraphics[width=\columnwidth]
{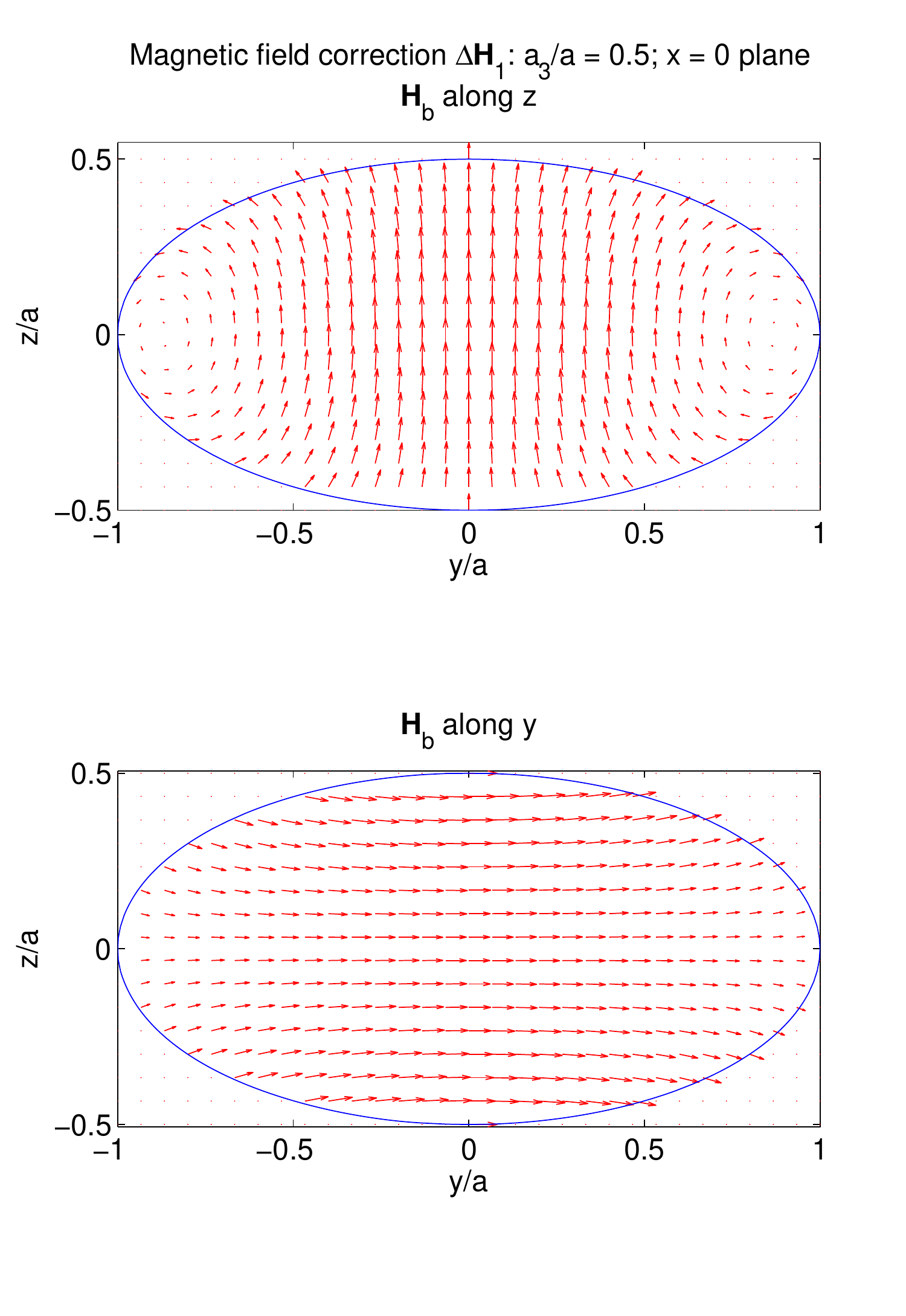}

\caption{(Color online) Leading magnetic field correction (\ref{7.15})
in the $yz$-plane for a spheroid, $a_1 = a_2 \equiv a$, using aspect ratio
$a_3/a = 0.5$. \textbf{Upper plot:} background field (\ref{7.13}) taken
along the $z$-axis. \textbf{Lower plot:} background field taken along
the $y$-axis. In both cases, for $x=0$ the field also lies in the
$yz$-plane. For background field along $x$ (not shown), the field
in this plane is along $x$ as well.}
\label{fig:h1}
\end{figure}

Applying the results of Sec.\ \ref{sec:solid_ellipsoid} [specifically,
(\ref{5.1}) and the first line of (\ref{5.13}) with $n=1$], one
obtains the leading correction
\begin{eqnarray}
\Delta {\bf E}_1 &=& a_\alpha a_\beta v({\bf a})
\nonumber \\
&\times& \hat {\cal P}_\sigma
\left\{x_\alpha {\bf \hat e}_\beta
\left[A_{{\bf \hat e}_\alpha}({\bf a},0)
- \sum_\nu A_{{\bf \hat e}_\nu
+ {\bf \hat e}_\alpha}({\bf a},0) x_\nu^2 \right] \right.
\nonumber \\
&-& \left. x_\beta {\bf \hat e}_\alpha
\left[A_{{\bf \hat e}_\beta}({\bf a},0)
- \sum_\nu A_{{\bf \hat e}_\nu
+ {\bf \hat e}_\beta}({\bf a},0) x_\nu^2 \right] \right\}.
\nonumber \\
\label{7.14}
\end{eqnarray}
The projection operator (especially applied to the cubic terms) is very
messy, and its result will not be displayed here. However, since it
subtracts only a gradient, the curl of (\ref{7.14}) directly produces
the leading correction to the magnetic field:
\begin{eqnarray}
\Delta {\bf H}_1 &=& \frac{1}{ik\mu} \nabla \times \Delta {\bf E}_1
\nonumber \\
&=& \frac{a_\alpha a_\beta v({\bf a})}{ik\mu}
\Big({\bf \hat e}_\gamma \big\{A_{{\bf \hat e}_\alpha}({\bf a},0)
+ A_{{\bf \hat e}_\beta}({\bf a},0)
\nonumber \\
&&-\ [A_{{\bf \hat e}_\alpha + {\bf \hat e}_\beta}({\bf a},0)
+ 3 A_{2{\bf \hat e}_\alpha}({\bf a},0)] x_\alpha^2
\nonumber \\
&&-\ [A_{{\bf \hat e}_\alpha + {\bf \hat e}_\beta}({\bf a},0)
+ 3 A_{2{\bf \hat e}_\beta}({\bf a},0)] x_\beta^2
\nonumber \\
&&-\ [A_{{\bf \hat e}_\alpha + {\bf \hat e}_\gamma}({\bf a},0)
+ A_{{\bf \hat e}_\beta + {\bf \hat e}_\gamma}({\bf a},0)] x_\gamma^2 \big\}
\nonumber \\
&&+\ 2 x_\gamma
[A_{{\bf \hat e}_\alpha + {\bf \hat e}_\gamma}({\bf a},0) x_\alpha
{\bf \hat e}_\alpha
\nonumber \\
&&\ \ \ \ \ \ +\ A_{{\bf \hat e}_\beta
+ {\bf \hat e}_\gamma}({\bf a},0) x_\beta
{\bf \hat e}_\beta] \Big),
\label{7.15}
\end{eqnarray}
in which, again, $(\alpha \beta \gamma)$ is cyclic permutation of
$(123)$. It is easily verified that this correction is divergence-free,
as required. Example field patterns are shown in Fig.\ \ref{fig:h1}
for the case of an oblate spheroid (discus-shape), where the required
$A$-coefficients are computed in closed form using the results in Sec.\
\ref{sec:degenerate}.

It is clear that the complexity of the polynomial form $\Delta {\bf
E}_n$ increases rapidly with $n$, and a numerical implementation is
required to keep track of all the terms. In later sections we will show
theoretical results, and comparisons to experiment, using all 232 basis
functions with $l+2p \leq 7$, where convergence is found even when
$\eta_s$ is not small.

\subsection{External field}
\label{sec:ext_field}

It is the field external to the scatterer that is relevant to
target detection. Equation (\ref{2.19}) allows computation of
(the inductive part of) the external field once the internal
field is known \cite{foot:noninductext}. Specifically, from the
series (\ref{7.10}), one obtains
\begin{equation}
{\bf E}^{(\mathrm{ind})}_\mathrm{ext}({\bf x})
\equiv ik {\bf A}({\bf x})
= \sum_{n=0}^\infty \left(\frac{ik\mu \sigma}{c} \right)^n
\Delta {\bf E}^{(\mathrm{ind})}_n({\bf x})
\label{7.16}
\end{equation}
in which $\Delta {\bf E}_0^{(\mathrm{ind})} = ik {\bf A}_b$ is
the inductive part of the background field (both inside and
outside the target), and
\begin{equation}
\Delta {\bf E}^{(\mathrm{ind})}_n({\bf x})
= \int_{V_s} d^3x' \frac{\Delta {\bf E}_{n-1}({\bf x}')}
{|{\bf x}-{\bf x}'|},\ {\bf x} \notin V_s,\ n \geq 1.
\label{7.17}
\end{equation}

If the result for $\Delta {\bf E}_{n-1}$ is a polynomial, as
described in Sec.\ \ref{sec:leadinduct}, then the results of Sec.\
\ref{sec:solid_ellipsoid} directly produce the external field
in the form of sum of products of polynomials with elliptic
functions. The latter are now functions of position ${\bf x}$
via the nonzero value of $\lambda({\bf x})$---see (\ref{3.25}).

For example, the linear form (\ref{7.12}) for the background
field produces a leading correction
\begin{eqnarray}
\Delta {\bf E}^{(\mathrm{ind})}_1 &=& a_\alpha a_\beta v({\bf a})
\nonumber \\
&\times& \left\{x_\alpha {\bf \hat e}_\beta
\left[A_{{\bf \hat e}_\alpha}({\bf a},\lambda)
- \sum_\nu A_{{\bf \hat e}_\nu
+ {\bf \hat e}_\alpha}({\bf a},\lambda) x_\nu^2 \right] \right.
\nonumber \\
&-& \left. x_\beta {\bf \hat e}_\alpha
\left[A_{{\bf \hat e}_\beta}({\bf a},\lambda)
- \sum_\nu A_{{\bf \hat e}_\nu
+ {\bf \hat e}_\beta}({\bf a},\lambda) x_\nu^2 \right] \right\},
\nonumber \\
\label{7.18}
\end{eqnarray}
which is identical in form to (\ref{7.14}), but with the
projection operator omitted, and now including nonzero $\lambda$.

\begin{figure*}

\includegraphics[width=5.0in]{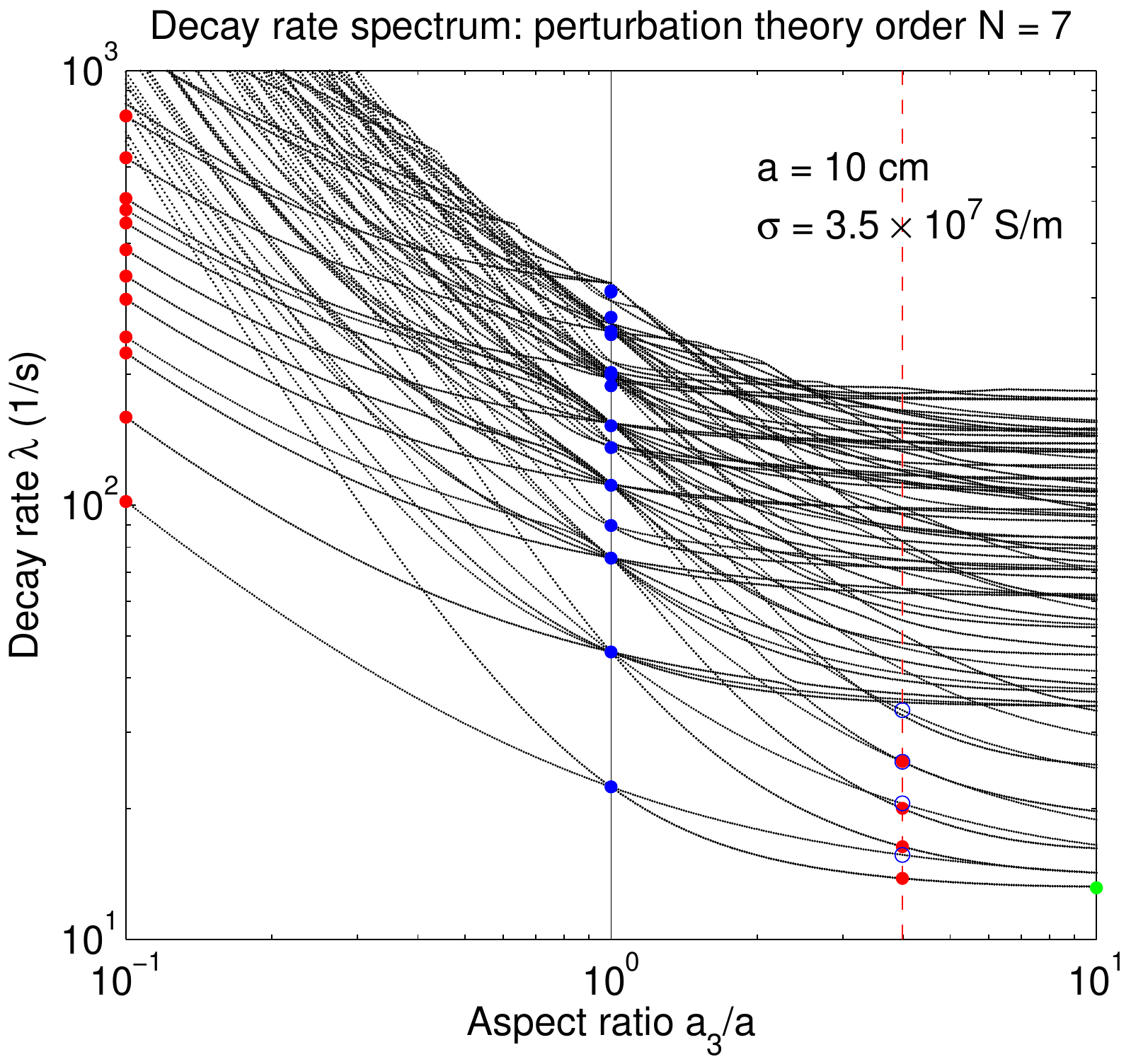}

\caption{(Color online) Spheroid decay rate spectrum vs.\ aspect
ratio $\alpha = a_3/a$, with conductivity $\sigma$ and radius $a$
fixed as indicated. Unit permeability, $\mu = \mu_b = 1$, is used
as well. The first 125 decay rates (out of a total of 232 that
are computed using all 232 basis functions with order $N = l + 2p
\leq 7$) are plotted for each $0.1 \leq \alpha \leq 10$. Apparent
sharp bends in the curves at the top of the spectrum are artifacts
of this truncation. Blue dots at $\alpha = 1$ show exact
analytic results for the sphere, with degeneracy effects from
enhanced symmetry evident. Red dots at $\alpha = 4$ mark the four
(doubly degenerate) vertically circulating modes shown in Fig.\
\ref{fig:vertmodes}; blue circles mark the four azimuthally
circulating modes shown in Fig.\ \ref{fig:azmodes}. The green
dot at the right is the analytic result for the lowest mode for
an infinite cylinder, $\alpha \to \infty$. More slowly increasing
branches at the left, $\alpha \ll 1$, correspond to modes with
current patterns circulating in the $xy$-plane. The red dots
at $\alpha = 0.1$ mark the first twelve such modes shown in Fig.\
\ref{fig:horizmodes}.}

\label{fig:decayrates}
\end{figure*}

\begin{figure*}

\includegraphics[width=5.0in]{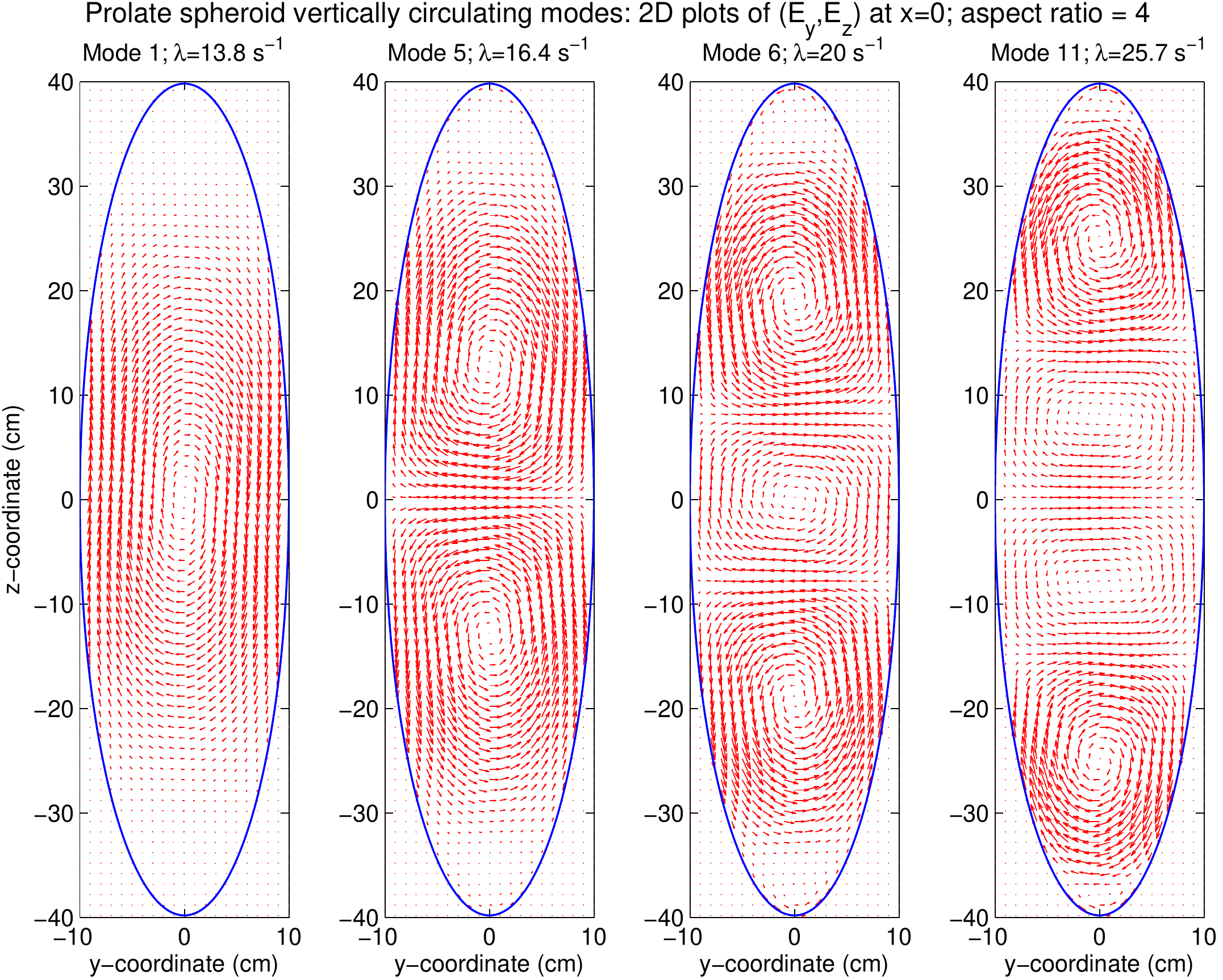}

\caption{(Color online) Four lowest order vertically circulating
mode shapes, and corresponding decay rates (red dots at $a_3/a
= 4$ in Fig.\ \ref{fig:decayrates}), for a $10 \times 10 \times 40$
cm radius aluminum prolate spheroid: $(E_y,E_z)$ are plotted in the
$x = 0$ plane (where $E_x$ vanishes identically).}

\label{fig:vertmodes}
\end{figure*}

\begin{figure*}

\includegraphics[width=5.5in]{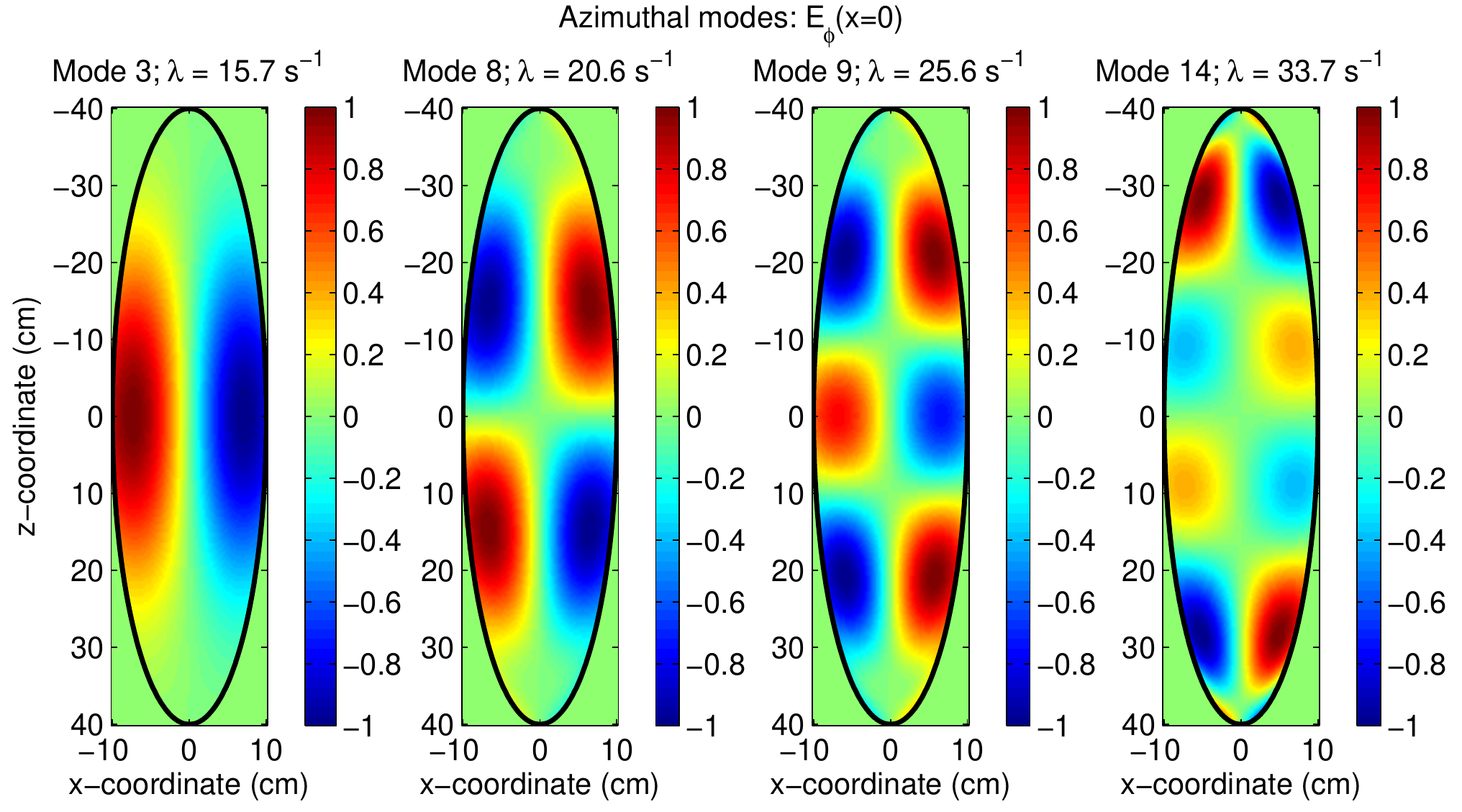}

\caption{(Color online) Four lowest order azimuthally circulating
mode shapes, and corresponding decay rates (blue circles at
$a_3/a = 4$ in Fig.\ \ref{fig:decayrates}), for a $10 \times 10
\times 40$ cm radius aluminum prolate spheroid. The azimuthal
component of the electric field $E_\phi$ (or $E_x$), normalized by
its maximum magnitude, is plotted in the $x=0$ plane; all other
components vanish. Currents in the lowest mode (far left) circulate
in the same direction at all heights. As the mode order increases
(from left to right), the number of nodal surfaces, across which
the circulation direction of the current changes sign, increases.}

\label{fig:azmodes}
\end{figure*}

\subsection{Far-field asymptotics: multipole expansion}
\label{sec:multipole}

Computation of the external field greatly simplifies if one is far from
the target---greater than a few maximum target radii, say---but not so
far that non-quasistatic corrections to the Green function (\ref{2.4})
become important.

First, for large $|{\bf x}|/L_s$ the cubic equation (\ref{3.25}) for
$\lambda$ has the expansion
\begin{equation}
\lambda({\bf x}) = |{\bf x}|^2
- \sum_\alpha a_\alpha^2 \hat x_\alpha^2
+ O(1/|{\bf x}|^2),
\label{7.19}
\end{equation}
in which ${\bf \hat x} = {\bf x}/|{\bf x}|$ is the unit vector with
components $\hat x_\alpha = x_\alpha/|{\bf x}|$. For large $\lambda$
one finds in turn,
\begin{widetext}
\begin{eqnarray}
A_{\bf k}({\bf a},\lambda) &=& \int_\lambda^\infty
\frac{dt}{t^{|{\bf k}|+3/2}}
\frac{1}{\prod_\alpha (1 + a_\alpha^2/t)^{k_\alpha+1/2}}
= \int_\lambda^\infty \frac{dt}{t^{|{\bf k}|+3/2}}
\left[1 - \frac{1}{2t} \sum_\nu (2k_\alpha + 1)
a_\alpha^2 + O(1/t^2) \right]
\nonumber \\
&=& \frac{2}{2|{\bf k}|+1} \frac{1}{\lambda^{|{\bf k}|+1/2}}
- \frac{\sum_\alpha (2k_\alpha + 1) a_\alpha^2}{2|{\bf k}|+3}
\frac{1}{\lambda^{|{\bf k}|+3/2}}
+ O\left(\frac{1}{\lambda^{|{\bf k}|+5/2}} \right),
\label{7.20}
\end{eqnarray}
with, as before, $|{\bf k}| = \sum_\alpha k_\alpha$. Substituting
(\ref{7.19}), one obtains explicitly
\begin{equation}
A_{\bf k}({\bf a},\lambda) = \frac{2}{2|{\bf k}|+1}
\frac{1}{|{\bf x}|^{2|{\bf k}|+1}}
+\left[\sum_\alpha a_\alpha^2 \hat x_\alpha^2
- \frac{\sum_\alpha (2k_\alpha + 1)
a_\alpha^2}{2|{\bf k}|+3} \right]
\frac{1}{|{\bf x}|^{2|{\bf k}|+3}}
+ O\left(\frac{1}{|{\bf x}|^{2|{\bf k}|+5}} \right).
\label{7.21}
\end{equation}
\end{widetext}

As an example, leading behavior of the external field correction
(\ref{7.18}) takes the simple magnetic dipole form
\begin{equation}
\Delta {\bf E}^{(\mathrm{ind})}_1
= \frac{4 a_\alpha a_\beta v({\bf a})}{15|{\bf x}|^2}
\left(\hat x_\alpha {\bf \hat e}_\beta
- \hat x_\beta {\bf \hat e}_\alpha \right).
\label{7.22}
\end{equation}
Higher order terms in (\ref{7.16}) will generally all have a leading
$1/|{\bf x}|^2$ term, but with more complicated angular dependence.
This may be formalized via a vector multipole expansion using the
vector harmonics (\ref{2.35}) \cite{Jackson}.

\begin{figure*}

\includegraphics[width=7.0in]{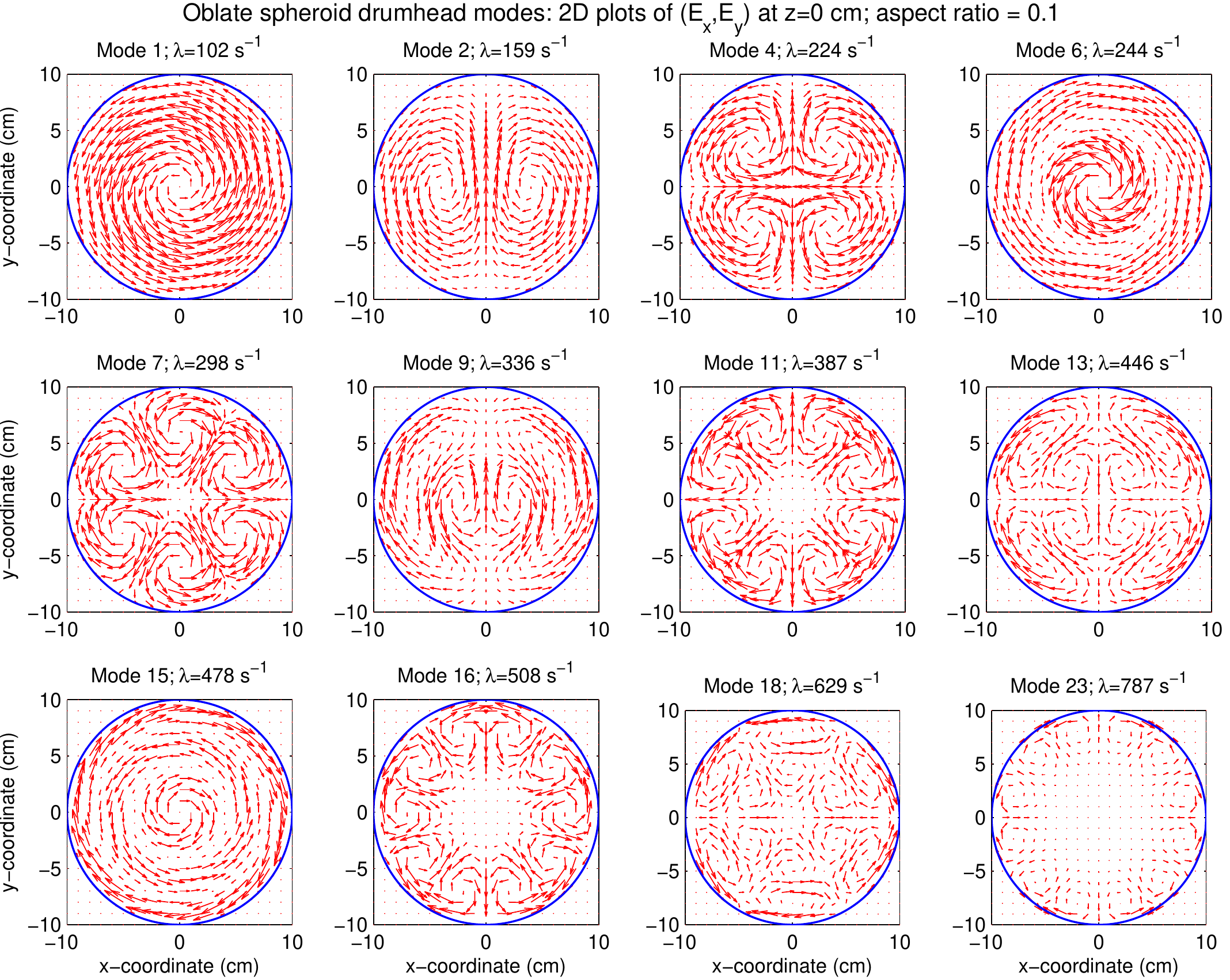}

\caption{(Color online) Twelve lowest order horizontal circulating
``drumhead'' mode shapes, and corresponding decay rates (red dots at
$a_3/a = 0.1$ in Fig.\ \ref{fig:decayrates}), for a $20 \times 20
\times 2$ cm radius aluminum prolate spheroid: $(E_x,E_y)$ are plotted
in the $z = 0$ plane (where $E_x$ vanishes). Modes 1,6,15 have
azimuthal symmetry (angular momentum index $m = 0$) and are
non-degenerate. The remainder, with $m > 0$, are doubly degenerate,
with the second mode shape obtained via a $\pi/2m$ rotation about the
$z$-axis (modes 2,9 have $m=1$; modes 4,13 have $m=2$; modes 7,18 have
$m=3$; mode 11 has $m=4$; mode 16 has $m=5$; mode 23 has $m=6$).  For
fixed $m$, the higher order modes have increasingly complex radial
structure, with the circulation direction changing sign with radius
(see modes 6,9,13,15,18).}

\label{fig:horizmodes}
\end{figure*}

\section{Time domain response}
\label{sec:tdem_response}

Having presented the general theory, and some leading order
perturbation results in the previous sections, we now turn to more
sophisticated applications of the theory that require numerical
implementation of high order expansions in $\eta_s$. Specifically,
theoretical predictions for time-domain EM measurements will now
presented (see Secs.\ \ref{sec:tdem} and \ref{sec:td_eigen}) and
compared to experimental data. As described previously, time domain
measurements are those of the freely decaying response of a target
after termination of an applied pulse.

Any time-domain response may, of course, be written as the Fourier
superposition of a spectrum of frequency domain responses, each of
which may be individually computed using the previous theory. However,
as will be described below, the formulation in terms of a superposition
of freely decaying modes, equation (\ref{1.1}), is numerically more
efficient because these need only be computed once for a given target,
avoiding recomputation of the perturbation series for each of a
continuous set of frequencies. In fact, these modes can be used to
directly solve the frequency domain problem as well.

A more important point is that a rapidly terminated pulse (over tens of
microseconds, in the experiments to be analyzed further below) has a
spectrum that includes some very high frequencies (e.g., tens of kHz),
for which $\eta$ is a far larger than can be handled at any achievable
order in perturbation theory. However, this part of the spectrum mainly
excites very rapidly decaying modes that disappear from the later-time
domain response. Thus, the mode approach allows accurate prediction of
the signal at later time even when it fails at earlier time. In fact,
the very large number of modes appearing at early-time make it a poor
representation of the response. A complementary description in terms of
the inward diffusion of screening surface currents \cite{W2003,W2004}
should be used instead. It will be seen below that a merging of the two
theories provides a complete description of the signal.

\subsection{Mode computation}
\label{sec:modecomp}

The freely decaying mode computation is implemented via the generalized
eigenvalue equation (\ref{2.43}). Specifically, the modes are written
as a superposition (\ref{2.42}) of the truncated set of vector harmonic
modes (\ref{2.39}), rescaled according to (\ref{2.32}), with $l + 2p
\leq N$ for some chosen upper limit $N$. Results will be shown for
$N=7$ (which yields a total of 232 basis functions---116 for each type
$i=1,2$), which will be seen to suffice for accurate comparison with
experiment. Since ${\bf Z}_{lmp}^{({\bf a};i)}({\bf x})$ is a
polynomial of degree $l+2p+i-1$, the ${\bf x}'$ integral defining the
$H$-matrix (\ref{2.29}) produces, via the results of Sec.\
\ref{sec:solid_ellipsoid}, a polynomial of degree $l+2p+i+1$. The final
${\bf x}$ integral [which, it should be recalled, automatically
implements the projection operator $\hat {\cal P}_\sigma$ in
(\ref{2.24}) and (\ref{2.41})], in both (\ref{2.29}) and the definition
of the $O$-matrix (\ref{2.28}) (with uniform $\sigma$ here) is then a
pure polynomial integral, and is trivial.

Figure \ref{fig:decayrates} show results for the decay rate spectrum of
a range of spheroids ($a_1 = a_2 \equiv a$). The lowest 125 decay rates
(out of a total of 232 computed at this at order $N=7$) are plotted for
each of 201 aspect ratios $\alpha = a_3/a$ in the range $0.1 \leq
\alpha \leq 10$. Even though these are plotted for a particular choices
of radius and conductivity (as well as permeability), in the high
contrast limit the combination $\lambda \mu \sigma a^2$ is independent
of all three. Hence these results may be trivially rescaled to obtain
results for any spheroid with the same geometry. The mode
eigenfunctions scale trivially with $a$ as well. Note as well that the
azimuthal symmetry (which is exactly preserved in by the perturbation
theory at fixed $N$) means that the $z$ angular momentum index $m$ is a
``good quantum number'', making the matrices $O$ and $H$ block
diagonal. The eigenvales for $\pm m$ are also degenerate, so that many
of the lines in Fig.\ \ref{fig:decayrates} actually represent pairs of
modes.

Exact results for the sphere, at $\alpha = 1$, are shown by cyan dots,
and indicate the accuracy of the method for rather large effective
values of $\eta_s = O(10)$ [obtained by substituting $a$ for $L_s$ and
$\lambda$ for $f$ in (\ref{7.8})]. The total angular momentum index $l$
is now also a good quantum number, and the vast increase in degeneracy
is evident.

The infinite right circular cylinder of radius $a$ corresponds to
$\alpha \to \infty$. In this limit, translation invariance implies that
the $z$-dependence of the modes is given by $e^{ikz}$ for arbitrary
wavenumber $k$. The analytic result for one of the $k=0$ modes (with
current traveling up one side of the cylinder and down the other), is
shown as the green dot, and is seen to have converged even at $\alpha =
10$.

A few mode shapes for the prolate spheroid with $\alpha = 4$ are
illustrated in Figs.\ \ref{fig:vertmodes} and \ref{fig:azmodes}. The
four modes in Fig.\ \ref{fig:vertmodes}, consisting of an increasing
number of vortices circulating in the $yz$-plane and indicated by the
red dots in Fig.\ \ref{fig:decayrates}, are doubly degenerate (with the
second mode obtained by $90^\circ$ rotation about the $z$-axis). One
may think of these modes as approximating those of an infinite cylinder
with $k \approx n \pi/a_3$, $n = 1,2,3,4$.

The modes shown in Fig.\ \ref{fig:azmodes}, indicated by the blue
circles in Fig.\ \ref{fig:decayrates}, are non-degenerate. Here the
current always circulates in the $xy$-plane, but the flow direction
oscillates with $z$. Again, they may be thought of in terms of those of
an infinite cylinder with $k \approx n \pi/a_3$, $n = 1,2,3,4$.

A set of twelve modes for an oblate spheroid, with aspect ratio
$\alpha = 0.1$ are shown in Fig.\ \ref{fig:horizmodes}. As seen
Fig.\ \ref{fig:decayrates}, since the radius $a$ is fixed, the
mode decay rates increase as they become vertically ``squeezed''
by decreasing $a_3$. We denote these ``drumhead modes'' because
a plot of the vertical component of the magnetic field  $H_z =
\partial_x E_y - \partial_y E_z$ would look very similar to the
surface height pattern of a vibrating drumhead.  Of course, the latter
oscillate at fixed frequency, rather than relax exponentially, but
there are parallels in the physics of the underlying mode patterns.

There are two independent time scales that operate to determine the
scaling of $\lambda$ with $a_3$ for a particular mode pattern, and the
interplay between them basically determines the shapes of the curves in
Fig.\ \ref{fig:decayrates}.

For a circulating current vortex that is very thin (the dimension $a_3$
for the modes in Fig.\ \ref{fig:horizmodes}) compared to its horizontal
extent (the dimension $a_v$ of one of the individual vortices in the
mode current pattern), the decay is dominated the decay of stored
magnetic energy via Joule heating. Thus, the dissipated power scales as
$P = I^2 R \sim (J a_v a_3)^2/(\sigma a_3) = J^2 a_v^2 a_3/\sigma$,
where $J$ is the characteristic current density. The magnetic energy is
estimated as $U_H = \mu B^2 V_\mathrm{eff}$, in which $B \sim I/c a_v =
J a_3/c$ is the magnetic field within a circulating current field and
$V_\mathrm{eff} \sim a_v^3$ is the effective volume over which it is
supported (mostly outside the target when $a_3/a_v \ll 1$), then scales
as $U_H \sim J^2 a^3 a_3^2/c^2$ \cite{foot:ring}. The ratio $\lambda
\sim P/U_H \sim c^2/\sigma \mu a a_3 = c^2/\sigma \mu a^2 \alpha$, as
claimed. For the modes in Fig.\ \ref{fig:decayrates}, $a_v/a$ remains
roughly fixed as $a_3/a \to 0$, and this produces the observed $\lambda
\sim 1/\alpha$ scaling for $\alpha \ll 1$.

On the other hand, if two oppositely oriented currents streams lie very
close to each other, the decay is dominated by transverse diffusion of
the streams into each other, which causes them to cancel out. The
diffusion time scale is $\tau_D \sim D d^2$, where $d$ is the current
stream separation,
\begin{equation}
D = \frac{c^2}{4\pi \sigma \mu}
\label{8.1}
\end{equation}
is the diffusion constant, and one estimates $\lambda \sim 1/\tau_D
\sim c^2/\mu \sigma d^2$. For modes with alternating current sheets in
the vertical dimension one has $d \propto a_3$ (e.g., the mode
geometries pictured in Figs.\ \ref{fig:vertmodes} or \ref{fig:azmodes},
where $d/a_3 \approx n$ for $n=1,2,3,4$, but squeezed in the vertical
dimension), and this leads to the $\lambda \sim 1/\alpha^2$ scaling
seen for the steeper curves on the left side of Fig.\
\ref{fig:decayrates}.

\begin{figure}

\includegraphics[width=0.9\columnwidth]{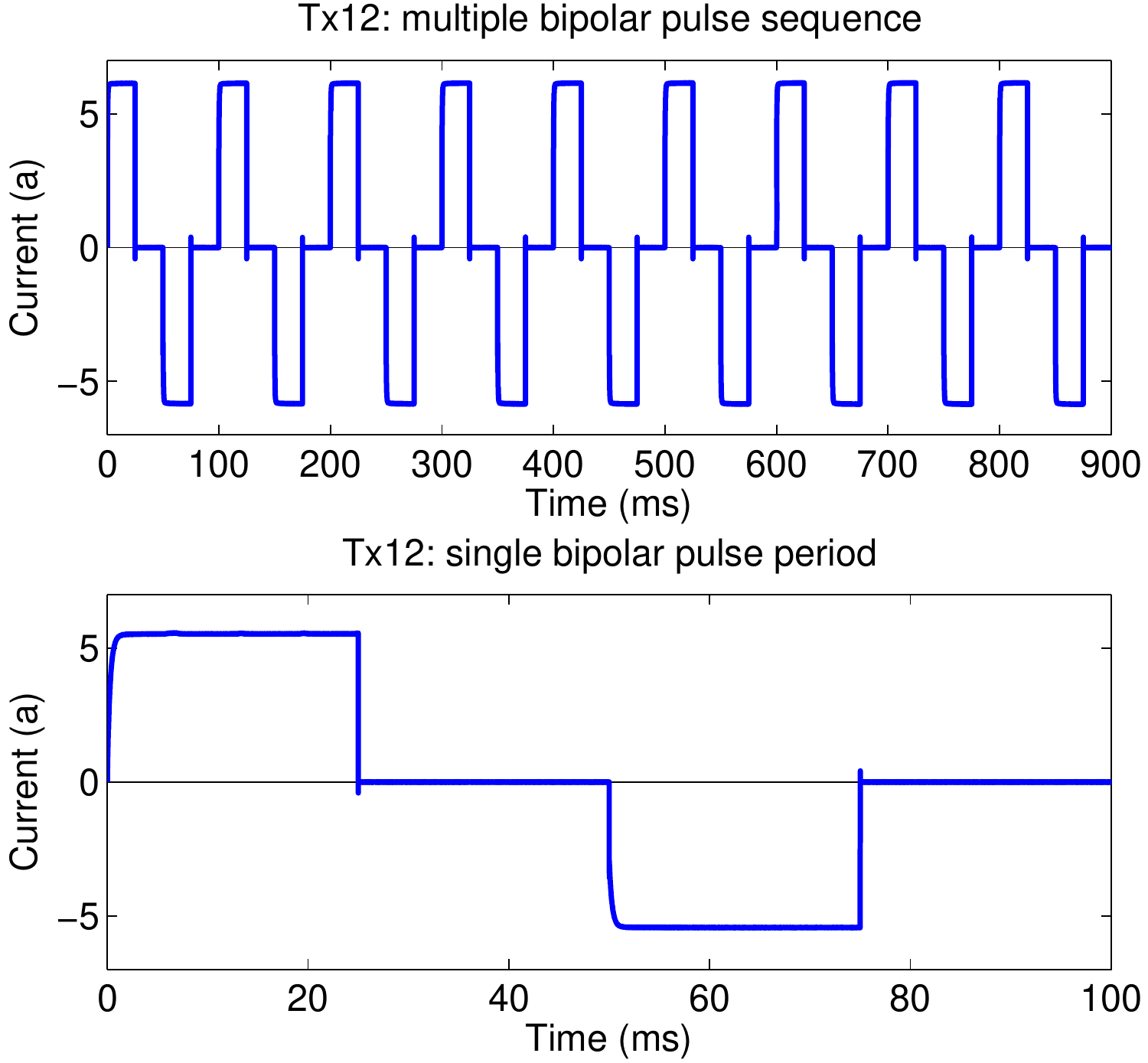}

\caption{(Color online) Transmitter pulse train generated by the
NRL-TEMTADS platform. The pulse train (with maximum current
approximately 5.7 a), is bipolar, meaning that it consists of identical
pulses of alternating sign. The average current is then zero, which
helps average out certain transients. The leading edge of each pulse is
actually a sequence of three exponential relaxations, with time
constants of 2.5 $\mu$s, 0.33 ms, and 4 ms. The trailing edge is a
linear offramp over a 10 $\mu$s interval. The total length of both the
pulse and the following quiescent detection interval is 25 ms.}

\label{fig:pulse}
\end{figure}

\begin{figure*}

\includegraphics[width=3.3in,clip]{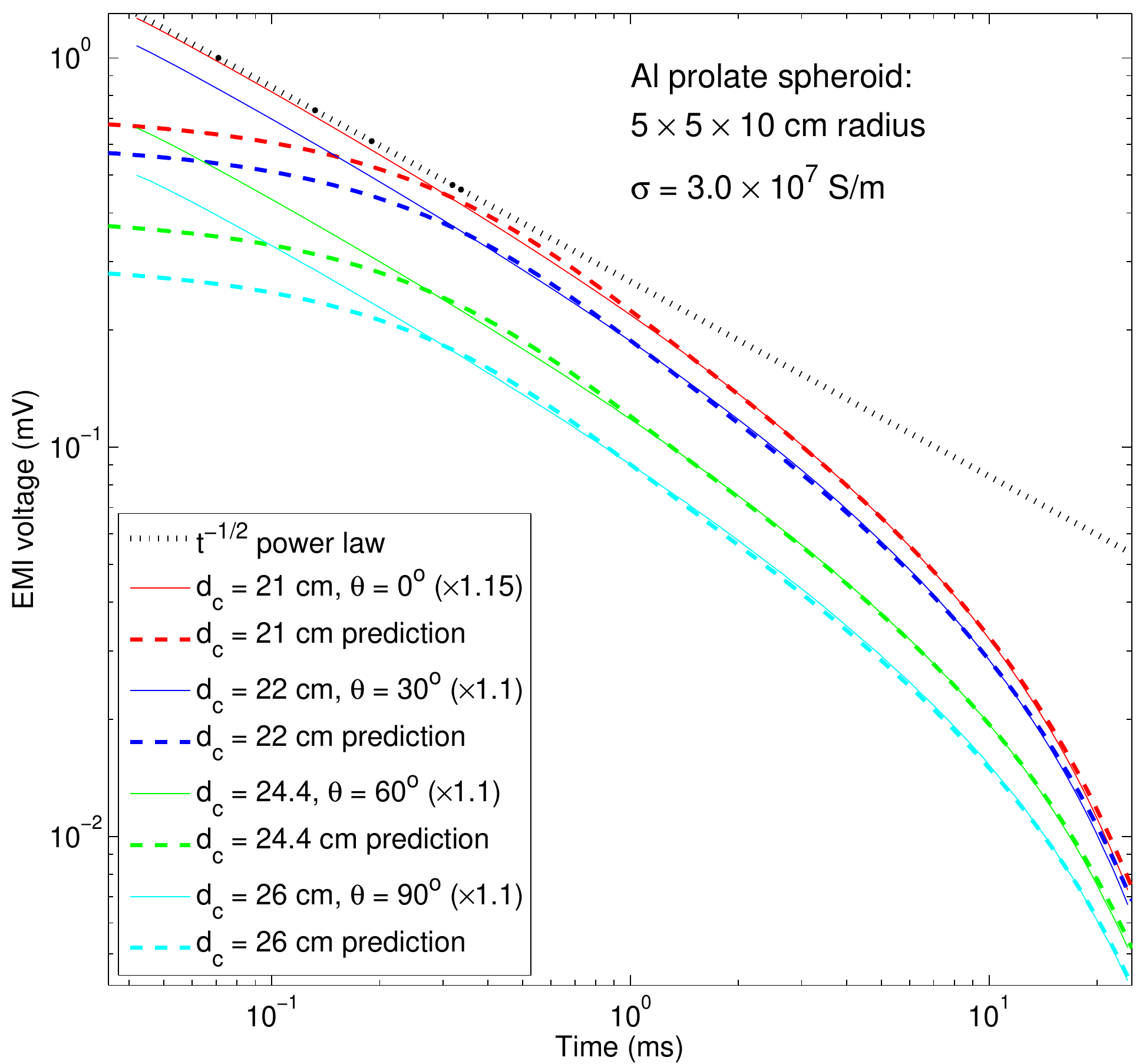}
\quad
\includegraphics[width=3.3in,clip]{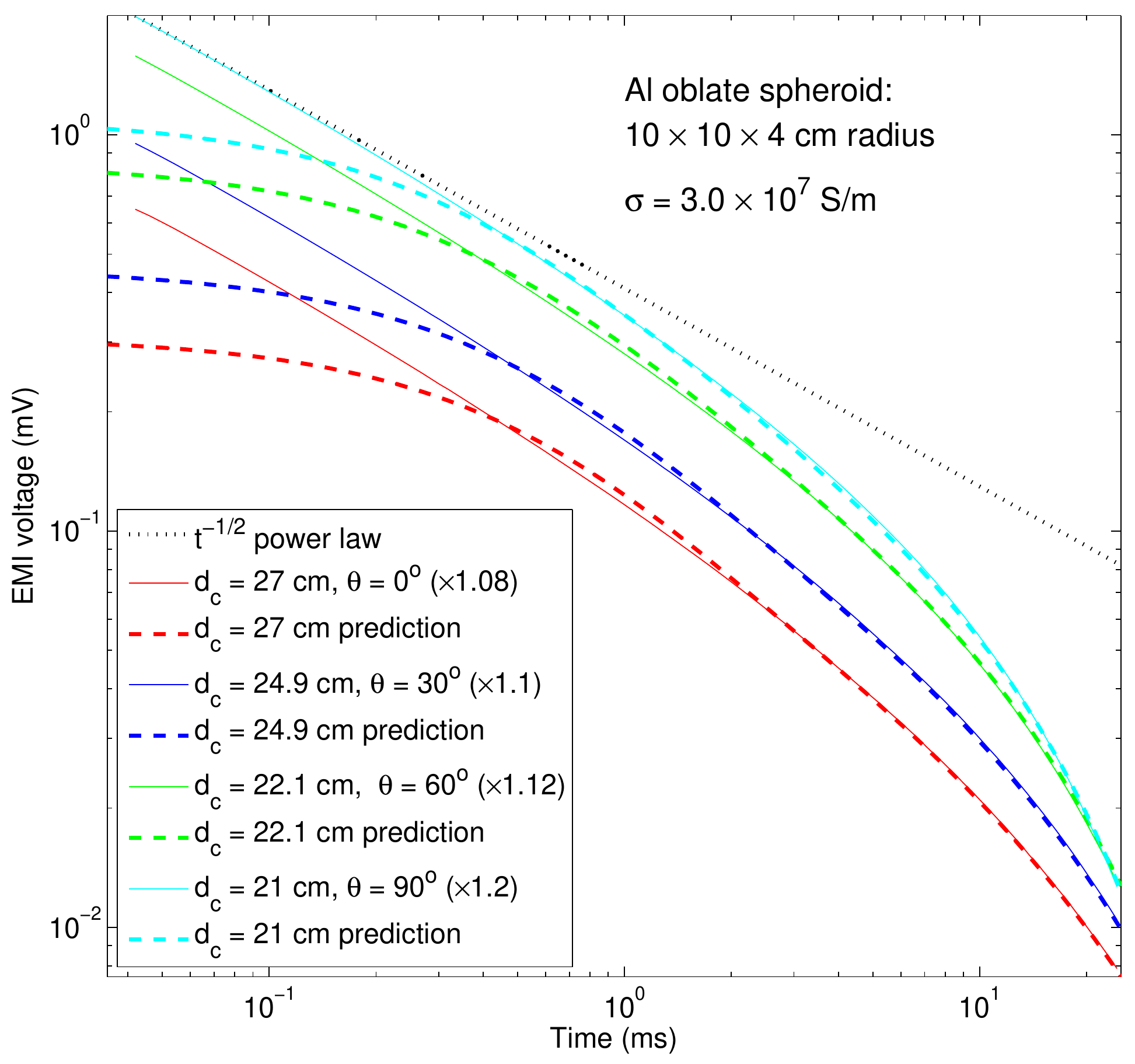}

\caption{(Color online) Comparisons of data (solid lines; taken by the
NRL-TEMTADS platform \cite{TEMTADS,foot:temtads}) and \emph{ab initio}
theoretical predictions (dashed lines) for a $5 \times 5 \times 10$ cm
radius aluminum prolate spheroid (left) and a $10 \times 10 \times 4$
cm radius aluminum oblate spheroid (right), at various depths-to-center
$d_c$, with various axial tilt angles $\theta$. The target is centered
under the transmitter-receiver pair. Plotted are the measured receiver
voltages beginning immediately after the pulse termination times seen
in Fig.\ \ref{fig:pulse}, averaged over about ten cycles. The
multipliers indicated in each legend entry reflect a $\sim 10$\%
variability in the transmitter current, as well as residual coil
characterization \cite{foot:temtads} and target positioning errors, and
are applied to the data to optimize the fit. Straight dashed lines show
the predicted $1/\sqrt{t}$ early time divergence \cite{W2003}. This
first principles agreement, over nearly two decades in both time and
voltage, is remarkable.}

\label{fig:datacomp}
\end{figure*}

\subsection{Comparisons with experimental data}
\label{sec:datacomp}

\subsubsection{Mode excitation and voltage amplitude computation}
\label{subsec:excite}

Having explored some details of the mode physics, we now illustrate,
through comparisons with experimental data on artificial aluminum
spheroids, how the mode theory is used to compute measurable quantities
through incorporation of a detailed model of the measurement platform.

The first step is to relate the excitation coefficients $A_n$ in the
free decay series (\ref{1.1}) to the source excitation pulse, encoded
in ${\bf S}$ on the right hand side of the wave equation (\ref{2.3}).
In the time domain, this equation reads
\begin{equation}
\frac{1}{D({\bf x})} \partial_t {\bf E}({\bf x},t)
+ \nabla \times \nabla \times {\bf E}({\bf x},t)
= -\frac{4\pi \mu}{c^2} \partial_t {\bf j}_S({\bf x},t).
\label{8.2}
\end{equation}
One seeks the solution in the form
\begin{equation}
{\bf E}({\bf x},t) = \sum_n {\cal A}_n(t) {\bf e}^{(n)}({\bf x})
\label{8.3}
\end{equation}
generalizing (\ref{1.1}). Substituting this into (\ref{8.2}), and using
the mode defining relation (\ref{2.40}) and the orthogonality condition
(\ref{2.45}), one obtains the equation of motion for the individual
amplitudes:
\begin{equation}
(\partial_t + \lambda_n) {\cal A}_n = -\partial_t j_n
\label{8.4}
\end{equation}
where
\begin{equation}
j_n(t) = \int d^3x {\bf e}^{(n)^*}({\bf x})
\cdot {\bf j}_S({\bf x},t)
\label{8.5}
\end{equation}
is the inner product of the source current with the mode eigenfunction.
For a compact transmitter loop ${\cal C}_T$ with $N_T$ windings
carrying current $I_T(t)$ (see Fig.\ \ref{fig:pulse} for an example),
this reduces to the line integral \cite{foot:lineint}
\begin{eqnarray}
j_n(t) &=& N_T a_n I_0(t)
\nonumber \\
a_n &\equiv& \oint_{{\cal C}_T} {\bf e}^{(n)*}({\bf x})
\cdot d{\bf l}
\label{8.6}
\end{eqnarray}
The solution to (\ref{8.4}) is
\begin{eqnarray}
{\cal A}_n(t) &=& N_T a_n I_n(t)
\nonumber \\
I_n(t) &\equiv& -\int_{-\infty}^t dt'
e^{-\lambda_n(t-t')} \partial_{t'} I_0(t')
\nonumber \\
&=& I_n(t_p) e^{-\lambda_n (t-t_p)},
\label{8.7}
\end{eqnarray}
where the second line is valid during any quiescent interval following
the termination of a pulse at time $t_p$. Note that for a very sharp
pulse termination, on a time-scale shorter than $1/\lambda_n$, one may
approximate $-\partial_{t'} I_0(t') \simeq \Delta I_0 \delta(t-t_p)$,
where $\Delta I_0$ is the down-step size (e.g., about 5 a in Fig.\
\ref{fig:pulse}). The termination then contributes $\Delta I_0$ to
$I_n(t_p)$. In any case, it follows that the required coefficient in
(\ref{1.1}) is given by
\begin{equation}
A_n = {\cal A}_n(t_p) = N_T a_n I_n(t_p)
\label{8.8}
\end{equation}

For a compact receiver loop ${\cal C}_R$ with $N_R$ windings, the
voltage is given by the line integral
\begin{equation}
V(t) = N_R \oint_{{\cal C}_R} {\bf E}({\bf x},t) \cdot d{\bf l}.
\label{8.9}
\end{equation}
Substituting (\ref{8.3}), one obtains the voltage series (\ref{1.2})
with
\begin{eqnarray}
V_n = A_n N_R b_n = N_T N_R a_n b_n I_n(t_p)
\label{8.10}
\end{eqnarray}
where
\begin{eqnarray}
b_n = \oint_{{\cal C}_R} {\bf e}^{(n)}({\bf x}) \cdot d{\bf l}.
\label{8.11}
\end{eqnarray}
is the corresponding receiver loop line integral.

This completes the specification of the measured voltage in terms of
modal quantities given the transmitter and receiver loop
characteristics. Note that computation of $a_n$ and $b_n$ requires
evaluation of the external electric field. For this purpose, the right
hand side of (\ref{2.41}) is evaluated using the previously computed
internal forms (\ref{2.42}) for the mode eigenfunctions. This
evaluation requires the various quantities worked out in Sec.\
\ref{sec:solid_ellipsoid} for positive values of the parameter
$\lambda({\bf x})$.  The far field asymptotic forms described in
(\ref{sec:multipole}) may be used at sufficient target standoff. The
line integrals defining $a_n,b_n$ are then performed numerically
\cite{numrec} through evaluations at a discrete set of points along the
loops ${\cal C}_T,{\cal C}_R$ \cite{foot:rotate}. The pulse wave forms
$I_0(t)$ are typically given by a sequence of relaxing exponentials and
linear ramps, for which $I_n(t)$ may be evaluated analytically
\cite{foot:multipulse}.

All of these algorithms have been implemented numerically to produce
the comparisons now described. Given the precomputation of the mode
properties (which requires 10--20 minutes for a given target on a
standard workstation), computation of the voltage series (\ref{1.2}) is
found to take only about 1 s. This speed is critical to efficient
solution inverse problems which underlie, for example, the UXO
discrimination problem. For the latter, properties of an unknown target
are estimated by searching over different candidate targets to find the
one that produces the best fitting voltage curve predictions
\cite{foot:inverse}.

\begin{figure}

\includegraphics[width=\columnwidth]
{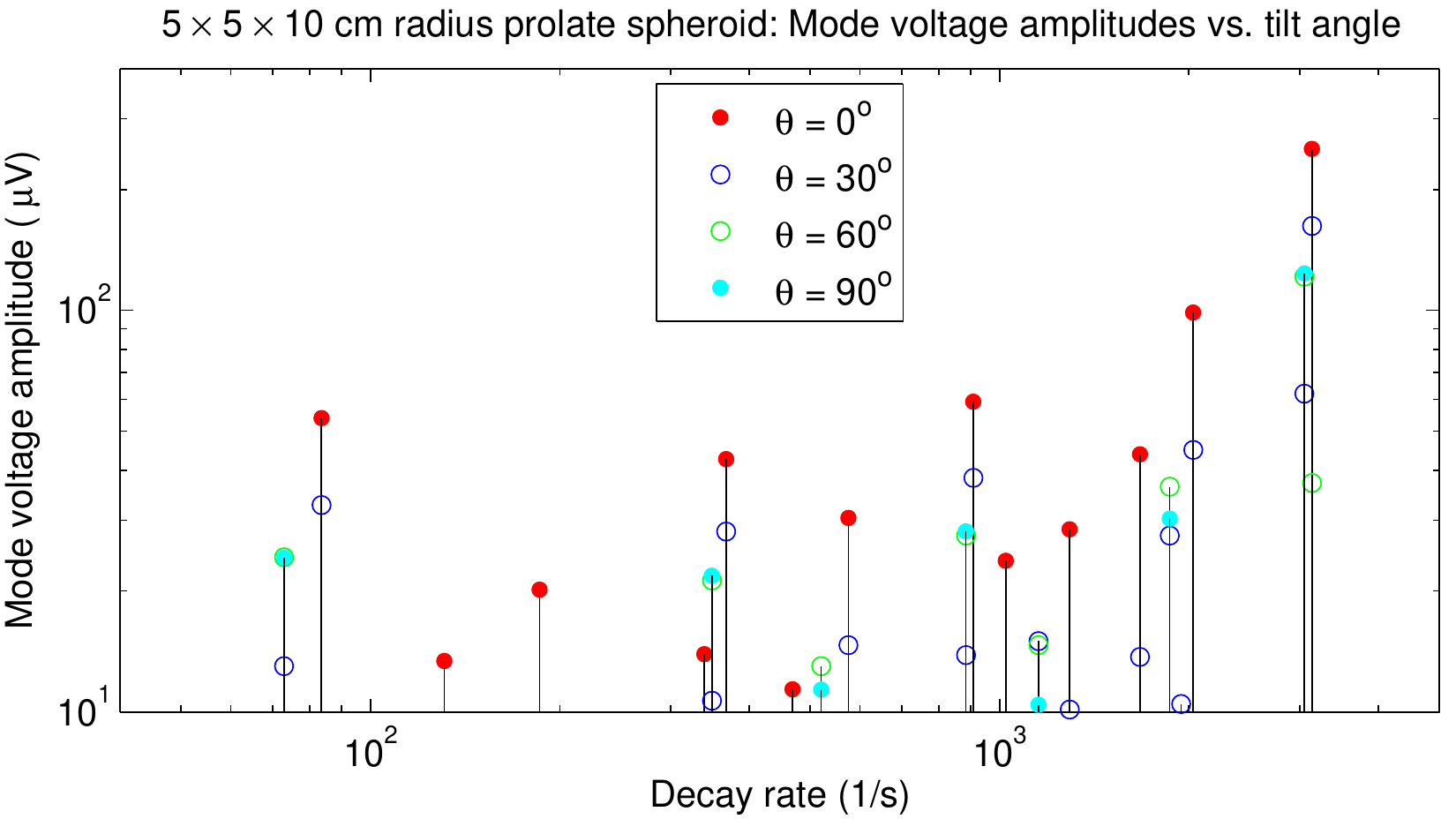}

\smallskip

\includegraphics[width=\columnwidth]
{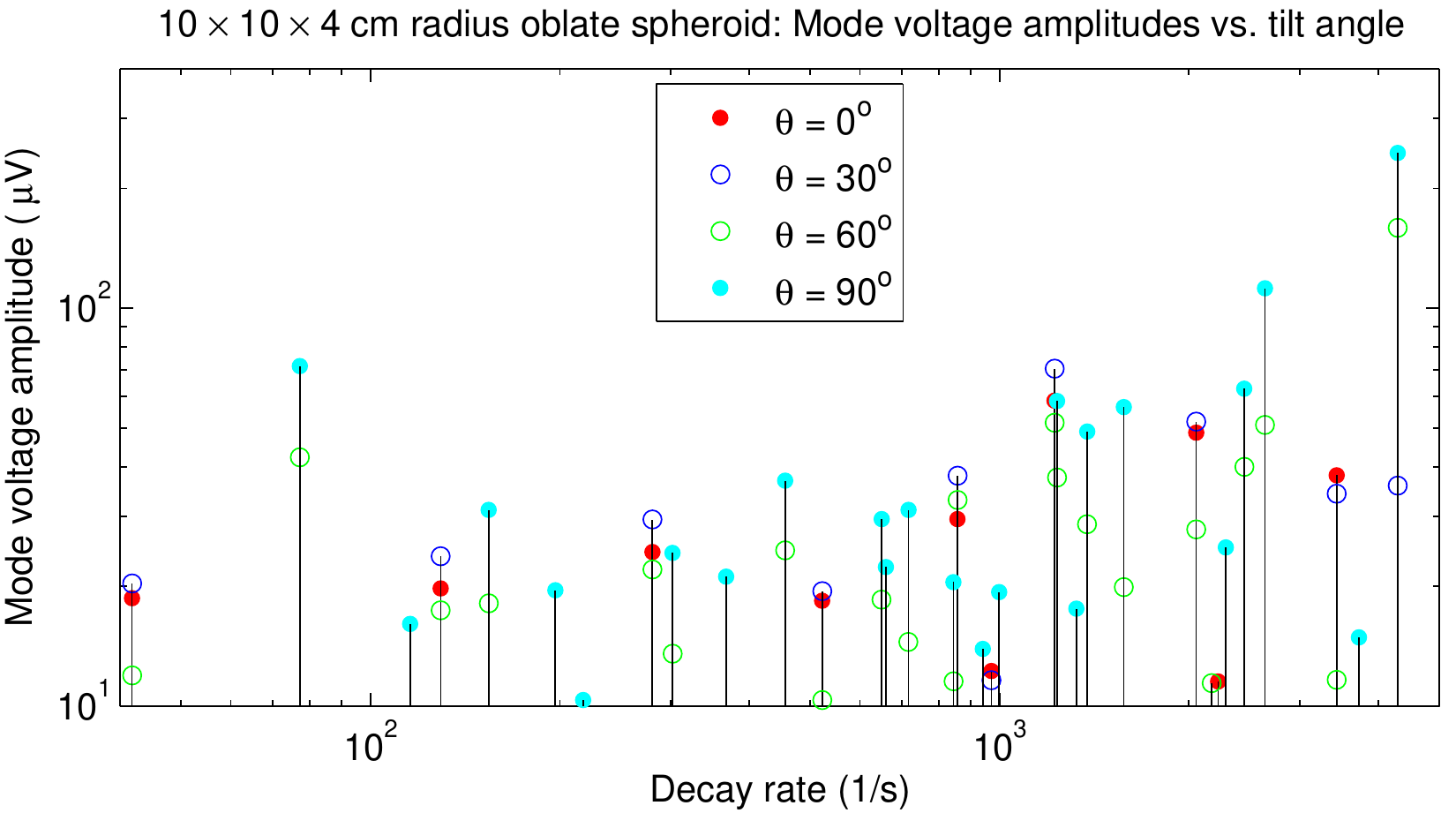}

\caption{Plot of the largest individual mode voltage amplitudes $V_n$
in the series (\ref{1.2}), plotted vs.\ mode decay rate $\lambda_n$,
corresponding to the time-domain curves in Fig.\ \ref{fig:datacomp}
(\textbf{top:} prolate spheroid; \textbf{bottom:} oblate spheroid). Of
particular note is the complete complementarity of the excited modes
for vertical ($\theta = 0^\circ$; red dots) and horizontal ($\theta =
90^\circ$; blue dots) targets, which is primarily responsible for the
decay rate of the curves at later time. A mixture of the two sets is
excited at intermediate tilt angles (blue and green circles). It is the
the failure of the amplitudes to decrease with increasing decay rate
that is responsible for the diverging signals at early time.}

\label{fig:vamps}
\end{figure}

\subsubsection{Comparison with NRL-TEMTADS measurements}
\label{subsec:temtads}

In Fig.\ \ref{fig:datacomp} we show comparisons between the theory and
data taken on artificial aluminum spheroids using the Naval Research
Labs TEMTADS platform \cite{TEMTADS}. The platform consists of a $5
\times 5$ horizontal array of 25 independent, well calibrated, high
dynamic range concentric transmitter and receiver coils. The coils are
square ($35 \times 35$ cm, with $N_T = 35$ windings for the
transmitters; $25 \times 25$ cm, with $N_R = 16$ windings, for the
receivers) and laid out flat in the same plane with 40 cm between
centers in each direction \cite{foot:temtads}. The pulse waveform is
described in Fig.\ \ref{fig:pulse}. For illustrative purposes, only the
strongest signal, from the coil under which the target was centered, is
shown in Fig.\ \ref{fig:datacomp} \cite{foot:inverse}.

The data quality is seen to be very high, with no visible noise or
distortion over more than two decades dynamic range of voltage and
time. The model predictions (dashed lines) are seen to very accurately
reproduce the data, except at very early time where, as explained
earlier, the absence of the contribution of more complex modes in the
series (\ref{1.2}) with decay times faster than about 0.25 ms
($\lambda_n \agt 4000$ s$^{-1}$) cause them to fall below the data
curves (solid lines). In comparison, the fundamental modes here are a
few tens of inverse seconds, (as can read off Fig.\
\ref{fig:decayrates} for $a_3/a = 2,0.4$ after applying the $1/\sigma
a^2$ scaling), and hence have decay times comparable to the extent of
the 25 ms measurement window. However, as also noted previously, the
early time portion of the data follows the predicted $1/\sqrt{t}$
divergence (ultimately cut off only by the finite 10 $\mu$s pulse
off-ramp width, which lies invisibly below the TEMTADS measurement
window) \cite{W2003}. Significantly, the $N=7$th order mean field
prediction succeeds in overlapping this regime, so that by simply
substituting a $V_\mathrm{pred}(t*)\sqrt{t^*/t}$ tail to the early time
prediction (below an optimally chosen crossover time $t^* \sim 0.25$
ms), one obtains a model fit to the that is accurate (at the $\sim 5$\%
level) over the entire dynamic range of the data.

The different curves in Fig.\ \ref{fig:datacomp} correspond to a
combination of different target depths and orientations. The changing
depth-to-center $d_c$ has week effect on the shape of the curves,
mainly changing the overall amplitude (with a roughly $1/d_c^6$ dipolar
dependence). This is a reflection of the fact that, at these standoffs
and for a centered target, the applied magnetic field in the target
region is fairly uniform and vertical.

The effect of orientation is more interesting, visibly changing the
steepness of the curves at later time, with the oblate spheroid (right
panel) showing a stronger effect of this type than the prolate spheroid
(left panel). The reason for this (as quantified in Fig.\
\ref{fig:vamps}, which shows a plot of the individual mode amplitudes)
is that for a vertical symmetry axis ($\theta = 0$), it is horizontally
circulating modes of the type shown in Figs.\ \ref{fig:azmodes} and
\ref{fig:horizmodes} that are most strongly excited [consistent with
the line integrals (\ref{8.6}), (\ref{8.11})], whereas for a horizontal
symmetry axis it is the vertically circulating modes, of the type shown
in Fig.\ \ref{fig:vertmodes}). Looking at Fig.\ \ref{fig:decayrates}
(and also the upper panel in Fig.\ \ref{fig:vamps}), one sees that for
$a_3/a > 1$ the leading vertical mode (curve containing the lowest red
dot) decays slightly more slowly than the leading horizontal mode
(curve containing the lowest blue circle). This explains why the red
curve ($\theta = 0^\circ$) in the left panel of Fig.\
\ref{fig:datacomp} is slightly steeper at late time than the cyan curve
($\theta = 90^\circ$). Conversely, for $a_3/a < 1$, it is the
horizontal modes that are more slowly decaying (see also the lower
panel in Fig.\ \ref{fig:vamps}), and the decay rate gap is much larger
(diverging as $a_3/a \to 0$). This explains why, in the right panel,
the cyan curve is significantly steeper at late time than the red
curve.

There is no distinction between the shapes of the curves at early time,
other than the overall amplitude of the $1/\sqrt{t}$ divergence.
However, the behavior of the amplitudes in Fig.\ \ref{fig:vamps} with
increasing decay rate explains the origin of this divergence in the
mode picture. As shown in Ref.\ \cite{W2003}, immediately after pulse
termination, the currents form a very thin sheet on the target surface.
The delta-function-like feature requires a superposition of an
essentially infinite number of modes (cut off only by the ultimately
finite pulse off-ramp rate), and one indeed sees in Fig.\
\ref{fig:vamps} that the mode amplitudes, if anything, are actually
growing with increasing decay rate. In fact, as can be seen explicitly
in the exact solution for the sphere, there is an infinite subset of
modes (corresponding there to a fixed subset angular momentum indices
$l,m$ that depend on the geometry of the background exciting field)
which have asymptotically constant excitation $V_0$, and make a voltage
contribution
\begin{equation}
V(t) \approx V_0 \sum_{p=p_0}^\infty e^{-\lambda_0 p^2 t}
\approx V_0 \sqrt{\frac{\pi}{4\lambda_0 t}},\ t \to 0,
\label{8.12}
\end{equation}
where $\lambda_0 \sim 1/\mu \sigma a^2$ describes the asymptotic
behavior of the decay rates for large enough $p \geq p_0$.
Equivalently, one expects, for a general target with a sufficiently
regular surface, that there is a density of states $\rho_e(\lambda)
\sim 1/\sqrt{\lambda}$ [a small fraction of the total, which increases
as $\rho_\mathrm{tot}(\lambda) \sim \sqrt{\lambda}$], with constant
excitation for large $\lambda$, which indeed leads to $V(t) \sim \int
\rho_e(\lambda) e^{-\lambda t} d\lambda  \sim 1/\sqrt{t}$ for $t \to
0$.

It is observations such as those above, connecting the geometry of the
decay curves to the geometry of the target, that are critical to a
workable inversion scheme \cite{foot:inverse}. It is also clear how
data from multiple sensors, or from different platform positions, which
see different effective orientations of the target, can greatly aid in
such an effort.

As a final comment, the quality of the fits points to an interesting
implication regarding the accuracy of the higher order mode
contributions. It is apparent from the exact sphere comparison (blue
dots) in Fig.\ \ref{fig:decayrates} that the decay rates, and
presumably the mode shapes, can be trusted quantitatively only,
perhaps, for the first few dozen modes (recall, also, that this figure
shows only the first 125 out of 232 computed modes). However, it is
clear that their summed contribution to the induced voltage measurement
is quantitatively extremely accurate. This points to the conclusion
that the overall effect of a group of modes with similar decay rates
(in a density of states sense) depends only on the part of the mode
Hilbert space that they cover, not on the detailed partitioning of that
subspace between individual modes. One can imagine a very complex
applied field, generated by an intricate set of transmitter coils
surrounding the entire target that is tuned to excite a single high
order mode, for which the prediction will badly fail. However, for the
relatively uniform fields of interest here, the conclusion appears
valid.

\section{Generalizations of the theory}
\label{sec:generalizations}

In this final section we describe various generalizations of the basic
theory. In Sec.\ \ref{sec:inhomo_perm} we consider permeable targets,
$\mu \neq \mu_b$. In Sec.\ \ref{sec:himagcontrast} we consider
simplifications in the high contrast limit $\mu/\mu_b \gg 1$, relevant
to ferrous targets where $\mu/\mu_b = O(10^2)$. The high magnetic
contrast limit parallels in many ways that of the high conductivity
contrast limit (e.g., it enforces a vanishing surface normal ${\bf \hat
n} \cdot {\bf H}$ of the internal magnetic field). However, there are
some surprising subtleties in the external field computation, which is
shown to vanish when $\mu_b/\mu \to 0$. A computation of the leading
$O(\mu_b/\mu)$ dependence of ${\bf \hat n} \cdot {\bf H}$ is then
required, and we show how to accomplish this within the Chandrasekhar
formalism. In Sec.\ \ref{sec:magmodecomp} we consider the computation
of the freely decaying eigenmodes. Unlike in the nonmagnetic case
(Sec.\ \ref{sec:td_eigen}) where the eigenmodes follow trivially from
the diagonalization of the Coulomb integral operator (\ref{2.41}), in
the magnetic case the frequency dependence enters the operator in a
more complicated way, and a sequential search must be performed to find
the decay rates. In Sec.\ \ref{sec:gentarggeom} we consider more
realistic target geometries, including hollow targets and multiple
targets. Finally, in Sec.\ \ref{sec:inhomobgperm} we consider the
effects of background permeability variations. We have shown that
background conductivity variations do not impact an induction
measurement, but even a very small background permeability has a strong
impact.

All of these generalizations require significantly more work to
implement, and their applications to experimental data will therefore
be described elsewhere.

\subsection{Generalization to inhomogeneous permeability}
\label{sec:inhomo_perm}

For inhomogeneous permeability, (\ref{2.3}) is replaced by,
\begin{eqnarray}
\mu_b \nabla \times \left(\frac{1}{\mu_b}
\nabla \times {\bf E}_b \right)
- \kappa_b^2 {\bf E}_b &=& {\bf S}
\nonumber \\
\mu \nabla \times \left(\frac{1}{\mu}
\nabla \times {\bf E} \right)
- \kappa^2 {\bf E} &=& {\bf S}.
\label{9.1}
\end{eqnarray}
Subtracting the first equation from the second, one obtains
\begin{eqnarray}
\nabla \times \left[\frac{1}{\mu_b}
\nabla \times ({\bf A}-{\bf A}_b) \right]
&=& \frac{1}{ik\mu_b} \left[\kappa^2 {\bf E}
- \kappa_b^2 {\bf E}_b \right]
\nonumber \\
&&+\ \left(\nabla \frac{\mu}{\mu_b} \right) \times {\bf H},
\label{9.2}
\end{eqnarray}
in which the magnetic field has been introduced via ${\bf H} = {\bf
B}/\mu = (ik\mu)^{-1} \nabla \times {\bf E}$, and we have again
represented the electric field in the form (\ref{2.10}), (\ref{2.11})
in terms of a Coulomb gauge vector potentials and the gradient of a
scalar potential. We now define the background quasistatic (symmetric)
tensor Green function ${\bf G}_A({\bf x},{\bf x}')$ by
\begin{eqnarray}
\nabla \times \left[\frac{1}{\mu_b}
\nabla \times {\bf \hat G}_A({\bf x},{\bf x}') \right]
&=& {\bf \hat P}_T \delta({\bf x}-{\bf x}')
\nonumber \\
\nabla \cdot {\bf \hat G}_A({\bf x},{\bf x}') &=& 0,
\label{9.3}
\end{eqnarray}
in which ${\bf \hat P}_T = \delta({\bf x}-{\bf x}') \openone + \nabla
\nabla (4\pi |{\bf x}-{\bf x}'|)^{-1}$ is the transverse
(divergence-free) projection of the delta function. One obtains the
formal solution
\begin{eqnarray}
{\bf A}({\bf x}) - {\bf A}_b({\bf x})
&=& \int d^3x' {\bf \hat G}_A({\bf x},{\bf x}')
\cdot \left\{\left(\nabla \frac{\mu}{\mu_b} \right)
\times {\bf H}({\bf x}') \right.
\nonumber \\
&+& \left. \frac{1}{ik\mu_b} \left[\kappa^2 {\bf E}({\bf x}')
- \kappa_b^2 {\bf E}_b({\bf x}') \right] \right\}.
\label{9.4}
\end{eqnarray}
The Green function ${\bf \hat G}_A$ accounts explicitly for variations
in the background permeability, and in the non-magnetic limit
(\ref{9.4}) reduces to (\ref{2.18}). Analytic forms for ${\bf \hat
G}_A$ also exist, e.g., for horizontally stratified backgrounds. Since
$\mu/\mu_b$ is typically discontinuous at the target boundary it is
convenient to eliminate the resulting surface term by integrating the
${\bf H}$ term on the right hand side by parts. One obtains,
\begin{widetext}
\begin{equation}
{\bf A}({\bf x}) - {\bf A}_b({\bf x}) = \int d^3x'
\left\{\frac{4\pi}{c} {\bf \hat G}_A({\bf x},{\bf x}')
\cdot [\sigma {\bf E}({\bf x}') - \sigma_b {\bf E}_b({\bf x}')]
+ \left(\frac{\mu}{\mu_b} - 1 \right)
[\nabla' \times {\bf \hat G}_A({\bf x},{\bf x}')]
\cdot {\bf H}({\bf x}') \right\}
\label{9.5}
\end{equation}
\end{widetext}
in which the curl operation acts on the second index of ${\bf \hat
G}_A$. So far, no approximations have been made, but in the high
contrast limit one may drop the $\epsilon_b {\bf E}_b$ term and
restrict the integral to the target volume $V_s$. For uniform $\mu_b$
one may replace
\begin{equation}
{\bf \hat G}_A({\bf x},{\bf x}')
\to \openone \frac{\mu_b}{4\pi |{\bf x}-{\bf x}'|}.
\label{9.6}
\end{equation}
The transverse projection terms do not contribute because the right
hand side of (\ref{9.2}), and all terms derived from it in (\ref{9.4})
and (\ref{9.5}), are divergence free. With these simplifications,
(\ref{9.5}) now reduces to
\begin{eqnarray}
{\bf A}({\bf x}) - {\bf A}_b({\bf x}) &=& \frac{4\pi \mu_b}{c}
\int_{V_s} d^3x' \frac{\sigma({\bf x}') {\bf E}({\bf x}')}
{4\pi |{\bf x}-{\bf x}'|}
\label{9.7} \\
&+& \nabla \times
\int_{V_s} d^3x' \frac{[\mu({\bf x}')-\mu_b]
{\bf H}({\bf x}')}{4\pi |{\bf x}-{\bf x}'|}.
\nonumber
\end{eqnarray}

\subsubsection{Coupled integral equations for {\bf E} and {\bf H}}
\label{subsec:coupled_eqns}

One may now, in principle, use (\ref{9.5}) or (\ref{9.7}) as the basis
for a low frequency perturbation theory. However, upon substituting for
${\bf H}$, the appearance of the curl of ${\bf E}$, in addition to
${\bf E}$ itself, on the right hand side is found to decrease numerical
stability, and it is preferable to use a more symmetric approach in
which ${\bf E}$ and ${\bf H}$ are treated on an equal footing.

One may obtain a second equation coupling the two fields by taking the
curl of both sides of (\ref{9.5}). However, the left hand side then
produces the combination $\mu {\bf H} - \mu_b {\bf H}_b$ rather than
${\bf H} - {\bf H}_b$, which turns out to be less convenient. In order
to obtain the latter, we construct an alternative to (\ref{9.1}) by
formulating the Maxwell equations in terms of ${\bf H}$ instead of
${\bf E}$. One may combine the Maxwell equations for ${\bf H}$ and
${\bf H}_b$ in the form
\begin{eqnarray}
\nabla \times ({\bf H} - {\bf H}_b) &=&
- ik (\epsilon {\bf E} - \epsilon_b {\bf E}_b)
\nonumber \\
\nabla \cdot [\mu_b ({\bf H} - {\bf H}_b)]
&=& -\nabla \cdot [(\mu-\mu_b){\bf H}],
\label{9.8}
\end{eqnarray}
in which the source term has been canceled and the right hand side of
the second equation vanishes outside of the target volume $V_s$. In
order to formulate these as an integral equation, define the magnetic
field tensor Green function by
\begin{equation}
{\bf \hat G}_H({\bf x},{\bf x}') = \frac{1}{\mu_b}
\nabla \times {\bf \hat G}_A({\bf x},{\bf x}'),
\label{9.9}
\end{equation}
which obeys
\begin{eqnarray}
\nabla \times {\bf \hat G}_H({\bf x},{\bf x}')
&=& {\bf \hat P}_T \delta({\bf x}-{\bf x}')
\nonumber \\
\nabla \cdot [\mu_b {\bf \hat G}_H({\bf x},{\bf x}')] &=& 0,
\label{9.10}
\end{eqnarray}
and represents a generalization of the Biot-Savart law. Define also a
background scalar Green function $g_H$ satisfying
\begin{equation}
-\nabla \cdot [\mu_b \nabla g_H({\bf x},{\bf x}')]
= \delta({\bf x}-{\bf x}').
\label{9.11}
\end{equation}
Together these can be used to construct a formal solution to
(\ref{9.8}) in the form
\begin{widetext}
\begin{equation}
{\bf H}({\bf x}) - {\bf H}_b({\bf x})
=  \int d^3x' \left\{\frac{4\pi}{c}{\bf \hat G}_H({\bf x},{\bf x}')
\cdot [\sigma {\bf E}({\bf x}') - \sigma_b {\bf E}_b({\bf x}')]
- (\mu-\mu_b) \nabla [\nabla' g_H({\bf x},{\bf x}')]
\cdot {\bf H}({\bf x}') \right\},
\label{9.12}
\end{equation}
\end{widetext}
whose structure may be compared to that of (\ref{9.5}). Once again, in
the high contrast limit one may drop the $\sigma_b {\bf E}_b$ term and
restrict the integral to $V_s$. For uniform background one obtains
\begin{equation}
g_H({\bf x},{\bf x}') = \frac{1}{4\pi \mu_b |{\bf x}-{\bf x}'|}
\label{9.13}
\end{equation}
and (\ref{9.12}) reduces to
\begin{eqnarray}
{\bf H}({\bf x}) - {\bf H}_b({\bf x}) &=& \frac{4\pi}{c}
\nabla \times \int_{V_s} d^3x' \frac{\sigma({\bf x}') {\bf E}({\bf x}')}
{4\pi |{\bf x}-{\bf x}'|}
\label{9.14} \\
&+& \frac{1}{\mu_b} \nabla \nabla \cdot
\int_{V_s} d^3x' \frac{[\mu({\bf x}')-\mu_b]
{\bf H}({\bf x}')}{4\pi |{\bf x}-{\bf x}'|},
\nonumber
\end{eqnarray}
whose structure may be compared to (\ref{9.7}).

Equations (\ref{9.5}) and (\ref{9.12}), or their homogeneous background
counterparts (\ref{9.7}) and (\ref{9.14}), are the basic results of
this section. They provide closed integral equations for ${\bf E}, {\bf
H}$ that generalizes the non-magnetic form (\ref{2.19}). Defining the
coefficients
\begin{eqnarray}
Q_{EE} &=& 4\pi ik \mu_b/c, \ \
Q_{EH} = ik
\nonumber \\
Q_{HH} &=& 1/\mu_b,\ \
Q_{HE} = 4\pi/c
\label{9.15}
\end{eqnarray}
and the integral equations may be written in the block form
\begin{equation}
\left[\begin{array}{c} ik{\bf A} \\
{\bf H} \end{array} \right]
= \left[\begin{array}{c} ik{\bf A}_b \\
{\bf H}_b \end{array} \right]
+ \left[\begin{array}{cc}
Q_{EE} \hat {\cal K}_1 & Q_{EH} \hat {\cal K}_2 \\
Q_{HE} \hat {\cal K}_2 & Q_{HH} \hat {\cal K}_3 \\
\end{array} \right]
\left[\begin{array}{c} \sigma {\bf E} \\
(\mu-\mu_b) {\bf H} \end{array} \right]
\label{9.16}
\end{equation}
in which the operators
\begin{eqnarray}
\hat {\cal K}_1[{\bf F}] &=& \int_{V_s} d^3x'
\frac{{\bf F}({\bf x}')}{4\pi|{\bf x}-{\bf x}'|}
\nonumber \\
\hat {\cal K}_2[{\bf F}]
&=& \nabla \times \hat {\cal K}_1[{\bf F}]
\nonumber \\
\hat {\cal K}_3[{\bf F}]
&=& \nabla \nabla \cdot \hat {\cal K}_1[{\bf F}]
\label{9.17}
\end{eqnarray}
represent the basic Coulomb integral operators.  These may all be shown
to be symmetric as well.

\subsubsection{Basis function expansion}
\label{subsec:basisexpand}

The basis function expansion solution to (\ref{9.7}) and (\ref{9.14})
requires now separate field expansions [compare (\ref{2.25})]
\begin{eqnarray}
\sigma({\bf x}) {\bf E}({\bf x})
&=& \sum_M a_M {\bf Z}^E_M({\bf x})
\nonumber \\
\mu({\bf x}) {\bf H}({\bf x})
&=& \sum_M b_M {\bf Z}^H_M({\bf x}).
\label{9.18}
\end{eqnarray}
The electric field basis functions ${\bf Z}^E_m$ continue to obey the
divergence free and Neumann boundary conditions (\ref{2.26}). The
magnetic field basis functions ${\bf Z}^H_m$ obey only the first
condition (except in the limit $\mu/\mu_b \to \infty$---see below)
\cite{foot:Hbasis}. For homogeneous ellipsoids, one may continue use
$\tilde {\bf Z}^{(1)}_{lmp}$ [see (\ref{2.39})], but simply drop the
$(1-x^2)$ factor and use $\tilde {\bf Z}^{(2)}_{lmp} = \nabla \times
[x^{l+2p} {\bf X}_{lm}]$ in place of ${\bf Z}^{(2)}_{lmp}$ [both still
rescaled via (\ref{2.32})].

Define now the coefficients
\begin{eqnarray}
a_{b,L} &=& \int_{V_s} d^3x {\bf Z}^{E*}_L({\bf x})
\cdot {\bf E}_b({\bf x})
\nonumber \\
b_{b,L} &=& \int_{V_s} d^3x {\bf Z}^{H*}_L({\bf x})
\cdot {\bf H}_b({\bf x}),
\label{9.19}
\end{eqnarray}
and the matrix elements
\begin{eqnarray}
O^E_{LM} &=& \int_{V_s} d^3x \frac{{\bf Z}^{E*}_L({\bf x})
\cdot {\bf Z}^E_M({\bf x})}{\sigma({\bf x})}
\nonumber \\
O^H_{LM} &=& \int_{V_s} d^3x \frac{{\bf Z}^{H*}_L({\bf x})
\cdot {\bf Z}^H_M({\bf x})}{\mu({\bf x})}
\nonumber \\
R_{LM} &=& \int_{V_s} d^3x {\bf Z}_L^{E*}
\cdot \hat {\cal K}_1 [{\bf Z}_M^E]
\nonumber \\
S_{LM} &=& \int_{V_s} d^3x {\bf Z}^{H*}_L
\cdot \hat {\cal K}_3 [(1-\mu_b/\mu) {\bf Z}^H_M]
\nonumber \\
U_{LM} &=& \int_{V_s} d^3x {\bf Z}^{E*}_L \cdot
\hat {\cal K}_2 [(1-\mu_b/\mu){\bf Z}^H_M]
\nonumber \\
V_{LM} &=& \int_{V_s} d^3x {\bf Z}^{H*}_M
\cdot \hat {\cal K}_2[{\bf Z}^E_L({\bf x}')].
\label{9.20}
\end{eqnarray}
In terms of these, equation (\ref{9.16}) reduces to the super-matrix
equation [compare (\ref{2.27})--(\ref{2.30}) for the nonmagnetic case]:
\begin{eqnarray}
&&\left(\begin{array}{cc}
{\bf O}^E & {\bf 0} \\
{\bf 0} & {\bf O}^H
\end{array} \right)
\left(\begin{array}{c} {\bf a} \\
{\bf b} \end{array} \right)
\label{9.21} \\
&&\hskip0.5in =\ \left(\begin{array}{c} {\bf a}_b \\
{\bf b}_b \end{array} \right)
+ \left(\begin{array}{cc}
Q_{EE} {\bf R} & Q_{EH} {\bf U} \\
Q_{HE} {\bf V} & Q_{HH} {\bf S}
\end{array} \right)
\left(\begin{array}{c} {\bf a} \\
{\bf b} \end{array} \right),
\nonumber
\end{eqnarray}
in which ${\bf O}^E$, ${\bf O}^H$, ${\bf R}$ are self-adjoint. For a
homogeneous target, ${\bf \hat S}$ is self adjoint as well, and ${\bf
U}^\dagger = (1-\mu_b/\mu) {\bf V}$.  The block matrix on the right
hand side of (\ref{9.21}) may then be made self adjoint by reexpressing
the equations in terms of $\sqrt{Q_{HE} (1-\mu_b/\mu)} {\bf a}$ and
$\sqrt{Q_{EH}} {\bf b}$. This is important for numerical purposes.

Since the basis functions for ellipsoids remain polynomials, for the
homogeneous target case the Coulomb integrals entering (\ref{9.20}) may
all be performed analytically using the techniques described in Secs.\
\ref{sec:int_compute}, \ref{sec:solid_ellipsoid} and
\ref{sec:degenerate}.

\subsection{High magnetic contrast limit}
\label{sec:himagcontrast}

For ferrous (e.g., steel) targets one typically finds very large
permeability contrast $\mu/\mu_b = O(10^2)$. Although this is far
smaller than the $O(10^7)$ conductivity contrast, it will often be the
case that 1\% accuracy is more than sufficient, and it is then
advantageous to seek simplifications in the $\mu/\mu_b \to \infty$
limit.

Estimating the terms on the right hand side of (\ref{9.7}), one sees
that the ratio of the ${\bf E}$ term to the ${\bf H}$ term is of order
\begin{equation}
\kappa a \frac{\mu_b}{\mu}
= \sqrt{\frac{\lambda}{\lambda_c}} \frac{\mu_b}{\mu},
\label{9.22}
\end{equation}
in which the the curl operation in $\hat {\cal K}_2$ is approximated by
the inverse target size $1/a$, and $\lambda_c = 4\pi \sigma \mu a^2$ is
a target characteristic decay rate scale. Therefore, for modes with
decay rates $\lambda/\lambda_c < (\mu/\mu_b)^2$ the $Q_{EH} \hat {\cal
K}_2$ term dominates \cite{foot:lambdac}.

Estimating the terms on the right hand side of (\ref{9.14}) requires
more care. Nominally, the ratio of the ${\bf E}$ term to the ${\bf H}$
term is also given by (\ref{9.22}). However, precisely as in the high
contrast limit for the ${\bf E}$ field, the nominally diverging ${\bf
H}$ term actually forces the boundary normal component ${\bf H} \cdot
{\bf \hat n}$ to scale with the factor (\ref{9.22}), and the resulting
term [which takes the form of a gradient, precisely as does the
$1/\kappa_b^2$ term in (\ref{2.9})] serves to cancel the boundary
normal component arising from the ${\bf E}$ term.

In the simultaneous high conductivity and high magnetic contrast limit,
(\ref{9.16}) may therefore written in the remarkably symmetric form
\begin{eqnarray}
{\bf E} &=& {\bf E}_b + Q_{EH} \hat {\cal K}_2 [\mu {\bf H}]
- \nabla \Phi_E
\nonumber \\
{\bf H} &=& {\bf H}_b + Q_{HE} \hat {\cal K}_2 [\sigma {\bf E}]
- \nabla \Phi_H
\label{9.23}
\end{eqnarray}
in which \emph{both} $\Phi_E$ and $\Phi_H$ are determined by the
condition that the boundary normal components of their respective
fields vanish. More explicitly, from (\ref{9.14}) one identifies
\begin{eqnarray}
\Phi_H &=& -Q_{HH} \nabla \cdot
\hat {\cal K}_1[(\mu-\mu_b){\bf H}]
\nonumber \\
&=& \int_{\partial V_s} d^2r'
\frac{[\mu({\bf r}')/\mu_b - 1]
{\bf \hat n}({\bf r}') \cdot {\bf H}({\bf r}')}
{4\pi|{\bf x}-{\bf x}'|}
\nonumber \\
&&+\ \int d^3x' \frac{\nabla' \cdot {\bf H}({\bf x}')}
{4\pi|{\bf x}-{\bf x}'|}.
\label{9.24}
\end{eqnarray}
Note that the second term vanishes identically for a homogeneous
target. The surface integral term clearly diverges with $\mu/\mu_b$
unless the surface normal ${\bf \hat n}({\bf r}') \cdot {\bf H}({\bf
r}') = O(\mu_b/\mu)$ vanishes. This leads to a finite result for
$\nabla \Phi_H$ that appears to depend on the subleading $O(\mu_b/\mu)$
dependence of ${\bf H}$. However, self-consistently, this term must
simply act to cancel the leading order surface normal component ${\bf
H}_b + Q_{HE} \hat {\cal K}_2[\sigma {\bf E}]$. This leads to a
relation for $\Phi_H$ that depends only on the leading form for ${\bf
E}$. This derivation is entirely equivalent to the high conductivity
contrast limit derived in Sec.\ \ref{sec:highconlim} in which ${\bf E}$
was separated into inductive and gradient parts.

\subsubsection{Basis function expansion}
\label{subsec:magbasisexpand}

Since both ${\bf E}$ and ${\bf H}$ now obey the same boundary
condition, one solves (\ref{9.23}) using \emph{identical} divergence
free basis functions ${\bf Z}^H_M = {\bf Z}^E_M \equiv {\bf Z}_M$. The
identity (\ref{2.23}) for $\sigma {\bf E}$, and the analogous one for
$\mu {\bf H}$, implies that the gradients are orthogonal to the ${\bf
Z}_M$ and the basis function expansion (\ref{9.18}) now yields the
off-diagonal form
\begin{eqnarray}
&&\left(\begin{array}{cc}
{\bf O} & {\bf 0} \\
{\bf 0} & {\bf O}
\end{array} \right)
\left(\begin{array}{c} {\bf a} \\
{\bf b} \end{array} \right)
\label{9.25} \\
&&\hskip0.35in =\ \left(\begin{array}{c} {\bf a}_b \\
{\bf b}_b \end{array} \right)
+ \left(\begin{array}{cc}
{\bf 0} & Q_{EH} {\bf V} \\
Q_{HE} {\bf V} & {\bf 0}
\end{array} \right)
\left(\begin{array}{c} {\bf a} \\
{\bf b} \end{array} \right),
\nonumber
\end{eqnarray}
in which we note that ${\bf V}^\dagger = {\bf V}$ is now self adjoint.
This may be reduced to the single equation for ${\bf a}$:
\begin{equation}
{\bf O} {\bf a} = {\bf a}_b
+ Q_{EH} {\bf V} {\bf O}^{-1} {\bf b}_b
+  Q_{EH} Q_{HE} {\bf V} {\bf O}^{-1} {\bf V} {\bf a}.
\label{9.26}
\end{equation}

The solution to the eigenvalue problem, in which the background fields
vanish, may be obtained by first solving the generalized eigenvalue
problem for ${\bf V}$:
\begin{equation}
{\bf V}{\bm \alpha}_n = \eta_n {\bf O} {\bm \alpha}_n,
\label{9.27}
\end{equation}
from which one identifies the electric and magnetic eigenvectors
\begin{eqnarray}
{\bf a}_n &=& {\bm \alpha}_n
\nonumber \\
{\bf b}_n &=& \eta_n Q_{HE} {\bm \alpha}_n
= \frac{1}{Q_{EH}\eta_n} {\bf a}_n,
\label{9.28}
\end{eqnarray}
with the consistency condition
\begin{equation}
Q_{EH} Q_{HE} \eta_n^2 = 1.
\label{9.29}
\end{equation}
Using (\ref{9.15}) this determines the decay rates $\lambda_n =
i\omega_n$
\begin{equation}
\lambda_n = \frac{c^2}{4\pi \eta_n^2}.
\label{9.30}
\end{equation}

For ellipsoids, the volume $V_s$ is symmetric under inversion ${\bf x}
\to -{\bf x}$. It is easy to check that $\hat {\cal K}_1$ is then
\emph{even} under inversion, while $\hat {\cal K}_2$ is \emph{odd}.
This implies that, even for arbitrary $\epsilon_b/\epsilon, \mu_b/\mu$,
if ${\bf e}^{(n)}({\bf x}), {\bf h}^{(n)}({\bf x})$ form an eigenmode,
so does ${\bf e}^{(n)}(-{\bf x}),-{\bf h}^{(n)}(-{\bf x})$. The same
symmetry implies that one can always choose solutions with a definite
parity \cite{foot:parity}, and this implies that ${\bf e}^{(n)}$, ${\bf
h}^{(n)}$ have opposite parity. However, in the present case, where the
two are multiples of each other, it follows that \emph{both} ${\bf
e}^{(n)}, \pm {\bf h}^{(n)}$ are eigenmodes, i.e., the $\eta_n$ must
come in oppositely signed pairs, and hence that the $\lambda_n$ are
each doubly degenerate (beyond any other degeneracies arising, e.g.,
from rotation invariance for spheroidal targets).

\subsubsection{External field}
\label{subsec:extfield}

There is a subtle problem that arises when one attempts to compute the
external field in the high magnetic contrast limit: we will show that
the inductive part of ${\bf E}-{\bf E}_b$ vanishes identically.
Specifically, whenever the boundary normal component of ${\bf H}$
vanishes, $\hat {\cal K}_2[\mu {\bf H}]$ becomes a perfect gradient.
The $\hat {\cal K}_1[\sigma {\bf E}]$ term in (\ref{9.16}) produces an
$O(\mu_b/\mu)$ inductive contribution, as will the \emph{leading
correction} to ${\bf H}$. Therefore, the total inductive part of the
external field is of order $\mu_b/\mu$, and its accurate computation
requires the leading correction to ${\bf H}$. However, it is actually
only the correction to the boundary normal component that one requires,
and we will see that this may be extracted directly from $\Phi_H$ in
(\ref{9.23}). We will show that the leading correction to $\lambda_n$
also follows only from this normal component.

To see that the EMI contribution of the ${\bf H}$ field term comes
directly from its surface normal component, note that one may write
\begin{equation}
\nabla \frac{1}{4\pi |{\bf x}|} = \nabla \times {\bf A}_\mathrm{mon}
\label{9.31}
\end{equation}
where ${\bf A}_\mathrm{mon}$ is the vector potential associated with a
monopole field ${\bf B}_\mathrm{mon} = -{\bf x}/|{\bf x}|^2$. One
choice is \cite{foot:monoflux}
\begin{eqnarray}
{\bf A}_\mathrm{mon}({\bf x})
&=& \frac{\tan(\theta/2)}{|{\bf x}|} \hat {\bm \phi}
= \frac{1-\cos(\theta)}{\sin(\theta)}
\frac{1}{4\pi |{\bf x}|} \hat {\bm \phi}
\nonumber \\
&=& \frac{{\bf \hat z} \times {\bf x}}{4\pi (|{\bf x}|+z)|{\bf x}|}.
\label{9.32}
\end{eqnarray}
Other choices are related to this one by a gauge transformation. Using
this form, one obtains for any vector field ${\bf F}$
\begin{widetext}
\begin{eqnarray}
\left(\nabla \frac{1}{4\pi|{\bf x}-{\bf x}'|} \right)
\times {\bf F}({\bf x}') &=&
[\nabla \times {\bf A}_\mathrm{mon}({\bf x}-{\bf x}')]
\times {\bf F}({\bf x}')
\nonumber \\
&=& [{\bf F}({\bf x}') \cdot \nabla]
{\bf A}_\mathrm{mon}({\bf x}-{\bf x}')
- \nabla [{\bf A}_0({\bf x}-{\bf x}')
\cdot {\bf F}({\bf x}')]
\nonumber \\
&=& -\nabla' \cdot [{\bf F}({\bf x}')
{\bf A}_\mathrm{mon}({\bf x}-{\bf x}')]
- \nabla [{\bf A}_\mathrm{mon}({\bf x}-{\bf x}')
\cdot {\bf F}({\bf x}')],
\label{9.33}
\end{eqnarray}
where, in the last line, we have used  $\nabla {\bf A}_\mathrm{mon} =
-\nabla' {\bf A}_\mathrm{mon}$ and $\nabla' \cdot {\bf B} = 0$. The
divergence dot product in the first term acts on the ${\bf B}$ index.
Inserting this result into in (\ref{9.16}), one obtains, using ${\bf F}
= (\mu - \mu_b){\bf H}$:
\begin{eqnarray}
\nabla \times \int_{V_s} d^3 x'
\frac{[\mu({\bf x}')-\mu_b]{\bf H}({\bf x}')}
{4\pi|{\bf x}-{\bf x}'|}
&=& - \int_{\partial V_s} d^2r' [\mu({\bf x}')-\mu_b]
{\bf A}_\mathrm{mon}({\bf x}-{\bf r}')
{\bf \hat n}({\bf r}') \cdot {\bf H}({\bf r}')
\nonumber \\
&&-\ \nabla \int_{V_s} d^3x' [\mu({\bf x}')-\mu_b]
{\bf A}_\mathrm{mon}({\bf x}-{\bf x}')
\cdot {\bf H}({\bf x}').
\label{9.34}
\end{eqnarray}
\end{widetext}
The second term is a perfect gradient, and therefore does not
contribute to an inductive measurement. The first term depends only on
the magnetic field surface normal, and therefore vanishes, as claimed,
to leading order in $\mu_b/\mu$.

\subsubsection{Leading surface normal component computation}
\label{subsec:lead_ndoth}

Specializing, for simplicity, to a homogeneous target, the full fields
${\bf E},{\bf H}$ may in principle be obtained through a perturbation
expansion in $g \equiv \mu_b/\mu$:
\begin{equation}
{\bf H}({\bf x}) = {\bf H}_0({\bf x})
+ g {\bf H}_1({\bf x}) + \ldots,
\label{9.35}
\end{equation}
and similarly for ${\bf E}$. With this definition, the leading order
inductive part of the external electric field is given by
\begin{eqnarray}
{\bf E}_\mathrm{ind}({\bf x})
&=& {\bf E}_{b,\mathrm{ind}}({\bf x})
+ Q_{EE} \sigma {\cal K}_1[{\bf E}_0]({\bf x})
\label{9.36} \\
&-& Q_{EH} \mu_b \int_{\partial V_s} d^2r'
{\bf A}_\mathrm{mon}({\bf x}-{\bf r}')
{\bf \hat n} ({\bf r}') \cdot {\bf H}_1({\bf r}').
\nonumber
\end{eqnarray}
On the other hand, from (\ref{9.24}) one obtains
\begin{equation}
\Phi_H({\bf x}) = \int_{\partial V_s} d^2r'
\frac{{\bf \hat n}({\bf r}') \cdot {\bf H}_1({\bf r}')}
{4\pi |{\bf x}-{\bf r}'|},
\label{9.37}
\end{equation}
which establishes a relation between $\Phi_H$ and the leading nonzero
surface field normal component and also shows that $\Phi$ obeys
Laplace's equation
\begin{equation}
\nabla^2 \Phi_H = 0,\ {\bf x} \notin \partial V_s.
\label{9.38}
\end{equation}
On the other hand, the vanishing of ${\bf \hat n} \cdot {\bf H}_0$ on
$\partial V_s$ leads, from the second line of (\ref{9.23}), to a
leading order Neumann-type boundary condition
\begin{equation}
{\bf \hat n}({\bf r}) \cdot \nabla \Phi_H({\bf r})
= {\bf \hat n}({\bf r}) \cdot {\bf H}_b({\bf r})
+ Q_{HE} {\bf \hat n}({\bf r}) \cdot
\hat {\cal K}_2[\sigma {\bf E}_0]({\bf r}),
\label{9.39}
\end{equation}
for ${\bf r} \in \partial V_s$.

Together, (\ref{9.37}) and (\ref{9.38}) uniquely define $\Phi_H$.
Inverting (\ref{9.37}) for ${\bf \hat n} \cdot {\bf H}_1$ and inserting
the result in (\ref{9.36}) then finally produces the desired inductive
component of the external field.

\subsubsection{Solution via Chandrasekhar approach}
\label{subsec:chandrasoln}

The computation defined by Sec.\ \ref{subsec:lead_ndoth} can be
conveniently implemented for homogeneous ellipsoids using the
Chandrasekhar methods of Sec.\ \ref{sec:chandra_theory}. We outline the
approach here.

We assume that ${\bf E}_0$ has already been determined, approximated as
a polynomial of some degree $N$, by solving the leading order equation
(\ref{9.25}). The result, along with a polynomial approximation for the
background field, can then be used to compute the right hand side of
(\ref{9.39})
\begin{eqnarray}
f^H({\bf x}) &\equiv& Q_{HE} \sigma {\bf n}({\bf x})
\cdot \nabla \times \int_{V_s} d^3x'
\frac{{\bf E}_0({\bf x}')}{4\pi|{\bf x}-{\bf x}'|}
\nonumber \\
&&+\ {\bf n}({\bf x}) \cdot {\bf H}_b({\bf x})
\nonumber \\
&=& \sum_{\bf k} f^H_{\bf k} x_1^{k_1} x_2^{k_2} x_3^{k_3}
\label{9.40}
\end{eqnarray}
in the form of a polynomial of degree $N+2$. Here ${\bf n} =
\sum_\alpha {\bf \hat e}_\alpha x_\alpha/a_\alpha^2$ is the
un-normalized unit vector. Now, since $\Phi_H$ obeys the Laplace
equation, it must have a spherical harmonic expansion
\begin{equation}
\Phi_H({\bf x}) = \sum_{l,m} \phi_{H,lm} x^l Y_{lm}(\theta,\phi),
\label{9.41}
\end{equation}
in which we restrict $l \leq N+2$. Solving for the coefficients
$\phi_{H,lm}$ requires one to compare the normal derivative of
(\ref{9.41}) to (\ref{9.40}) on the boundary. Even though $x$ depends
on $\theta,\phi$ on the boundary of an ellipsoid, the spherical
harmonic expansion still provides a unique specification for a function
on $\partial V_s$, and we therefore need to express (\ref{9.39}) in
this form.

To achieve this, we use the fact that $x^l Y_{lm}$ is a polynomial of
degree $l$ (see Sec.\ \ref{subsec:basisexpand}), and defining $\xi^2 =
\sum_\alpha (x_\alpha/a_\alpha)^2$ ($=1$ on the boundary), we can more
generally express
\begin{equation}
\xi^p x^l Y_{lm}(\theta,\phi) = \sum_{\bf k}
{\cal Y}_{lmp;{\bf k}} x_1^{k_1} x_2^{k_2} x_3^{k_3}
\label{9.42}
\end{equation}
as a polynomial with some known set of coefficients ${\cal Y}_{lmp;{\bf
k}}$. Here, the sum is restricted by the condition $k_1+k_2+k_3 =
l+2p$. Since the spherical harmonics are complete, this relationship is
invertible:
\begin{equation}
x_1^{k_1} x_2^{k_2} x_3^{k_3} = \sum_{l,m,p}
[{\cal Y}^{-1}]_{{\bf k};lmp}
\xi^{2p} x^l Y_{lm}(\theta,\phi),
\label{9.43}
\end{equation}
where the sum is restricted by the same condition.

Inserting (\ref{9.42}) (for $p=0$) into (\ref{9.41}) and taking the
normal derivative one obtains
\begin{eqnarray}
{\bf n}({\bf x}) \cdot \nabla \phi_H({\bf x})
&=& \sum_{\bf k} \phi^H_{\bf k}
x_1^{k_1} x_2^{k_2} x_3^{k_3}
\nonumber \\
\phi^H_{\bf k} &=& \left(\sum_\alpha
\frac{k_\alpha}{a_\alpha^2} \right)
\sum_{l,m} \phi_{H,lm} {\cal Y}_{lm0;{\bf k}}\ \ \ \ \ \
\label{9.44}
\end{eqnarray}
Substituting (\ref{9.43}) into (\ref{9.44}) and setting $\xi = 1$ the
boundary condition takes the form
\begin{equation}
\bar f^H_{lm} = \sum_{l',m'} \Phi_{lm;l'm'} \phi_{H,l'm'},
\label{9.45}
\end{equation}
where $\bar f^H_{lm} = \sum_p \bar f^H_{lmp}$ and similarly for $\bar
H_{b,lm}$ and $\bar \phi_{H,lm}$, and
\begin{equation}
\Phi_{lm;l'm'} = \sum_{\bf k} \left(\sum_\alpha
\frac{k_\alpha}{a_\alpha^2} \right)
\sum_{l,m} {\cal Y}_{l'm'0;{\bf k}}
[{\cal Y}^{-1}]_{{\bf k};lm,\frac{l'-l}{2}}.
\label{9.46}
\end{equation}
The sum over ${\bf k}$ is restricted by $k_1 + k_2 + k_3 = l'$, and
this also explains the value of the last index on ${\cal Y}^{-1}$. It
follows that $l'- l \geq 0$ is even, and the matrix ${\bm \Phi}$ has an
upper triangular-like structure. Therefore, if $f({\bf x})$ is a
polynomial of degree $N + 2$, then $\bar f_{lm}$ is nonzero only for $l
\leq N + 2$, and it follows as well that one requires only $l' \leq N +
2$: $\phi_{H,l'm'}$ is nonzero only for $l \leq N+2$, and only a finite
sub-block of ${\bm \Phi}$ needs to be inverted. In fact, the upper
triangular structure allows the inversion of this sub-block to be
reduced to a sequence of inversions of $(2l + 1) \times (2l + 1)$
blocks in descending order, $l = N+2,N+1,N,\ldots,3,2,1$.

The next step is to use (\ref{9.37}) to determine ${\bf n}({\bf r})
\cdot {\bf H}_1({\bf r})$ in terms of $\Phi_H$. Expanding
\begin{equation}
{\bf n}({\bf r}) \cdot {\bf H}_1({\bf r})
= \sum_{l,m} h^{(1)}_{lm} r^l Y_{lm}(\theta,\phi),
\label{9.47}
\end{equation}
restricting again $l \leq N+2$, the problem then reduces to computing
the expansion coefficients $h^{(1)}_{lm}$ in terms of the
$\phi^H_{lm}$. However, the Chandrasekhar method makes this very
straightforward. Let $p({\bf x})$ be a monomial of degree $n$, then
\begin{eqnarray}
\phi_0({\bf x}) &=& \int_{\partial V_s} d^2r'
\frac{p({\bf r}')}{|{\bf n}({\bf r}')| |{\bf x} - {\bf r}'|}
\nonumber \\
&=& \int d^3x' \frac{p({\bf x}')}{|{\bf x} - {\bf x}'|}
\delta(1 - \xi^{\prime 2})
\nonumber \\
&=& \partial_\mu|_{\mu=1} \int d^3x'
\frac{p({\bf x}')}{|{\bf x} - {\bf x}'|}
\theta(\mu - \xi^{\prime 2})
\nonumber \\
&=& \partial_\mu|_{\mu=1} \mu^{\frac{n+2}{2}}
\phi({\bf x}/\sqrt{\mu})
\nonumber \\
&=& \frac{1}{2}\left[n+2
- {\bf x} \cdot \nabla \right] \phi({\bf x})
\label{9.48}
\end{eqnarray}
where
\begin{equation}
\phi({\bf x}) = \int_{V_s} d^3x'
\frac{p({\bf x}')}{|{\bf x}-{\bf x}'|}.
\label{9.49}
\end{equation}
The last integral is of the type computed in Sec.\
\ref{sec:solid_ellipsoid}. Note that although $\phi$ has degree $n+2$,
the operator acting on $\phi$ in (\ref{9.48}) cancels all terms of this
degree, and $\phi_0$ is actually of degree $n$. Thus, inserting, via
(\ref{9.42}), the polynomial representation of (\ref{9.47}) into
(\ref{9.37}) and applying the identity (\ref{9.48}) one obtains a
linear relationship of the form
\begin{equation}
\phi_{H,lm} = \sum_{l',m'} {\cal H}_{lm;l'm'} h^{(1)}_{l'm'},
\label{9.50}
\end{equation}
in which ${\cal H}_{lm;l'm'}$ are known coefficients that follow from
the matrix ${\cal Y}$ and the results of Sec.\
\ref{sec:solid_ellipsoid}. Here, both $l,l' \leq N+2$, so the
relationship can be inverted to obtain the $h^{(1)}_{lm}$.

Finally, the external field can be computed via (\ref{9.36}). However,
the $(|{\bf x}|+z)^{-1}$ singularity in ${\bf A}_\mathrm{mono}$ means
that this integral is not in Chandrasekhar form. It is therefore better
to work from the original form
\begin{equation}
{\bf E}_\mathrm{ind}({\bf x}) = {\bf E}_{b,\mathrm{ind}}
+ Q_{EE} \sigma \hat {\cal K}_1[{\bf E}_0]
+ Q_{EH} \mu_b \hat {\cal K}_2[{\bf H}_1],
\label{9.51}
\end{equation}
[where, in an abuse of notation, ${\bf E}_\mathrm{ind}$ here differs
from that in (\ref{9.36}) by various gradients] in which ${\bf H}_1$ is
\emph{any} field with consistent surface normal values. One choice is
\begin{eqnarray}
{\bf H}_1({\bf x}) &=& \sum_{{\bf k},\alpha}
{\bf \hat e}_\alpha h^{(1)}_{{\bf k},\alpha}
x_1^{k_1} x_2^{k_2} x_3^{k_3}
\nonumber \\
h^{(1)}_{{\bf k},\alpha} &=& \frac{a_\alpha^2}{3}
h^{(1)}_{{\bf k} + {\bf \hat e}_\alpha},
\label{9.52}
\end{eqnarray}
where $h^{(1)}_{\bf k} = \sum_{l,m} {\cal Y}_{{\bf k};lm0}
h^{(1)}_{lm}$ are the coefficients in the equivalent monomial expansion
form of (\ref{9.47}).

\begin{figure}
\includegraphics[width=\columnwidth]{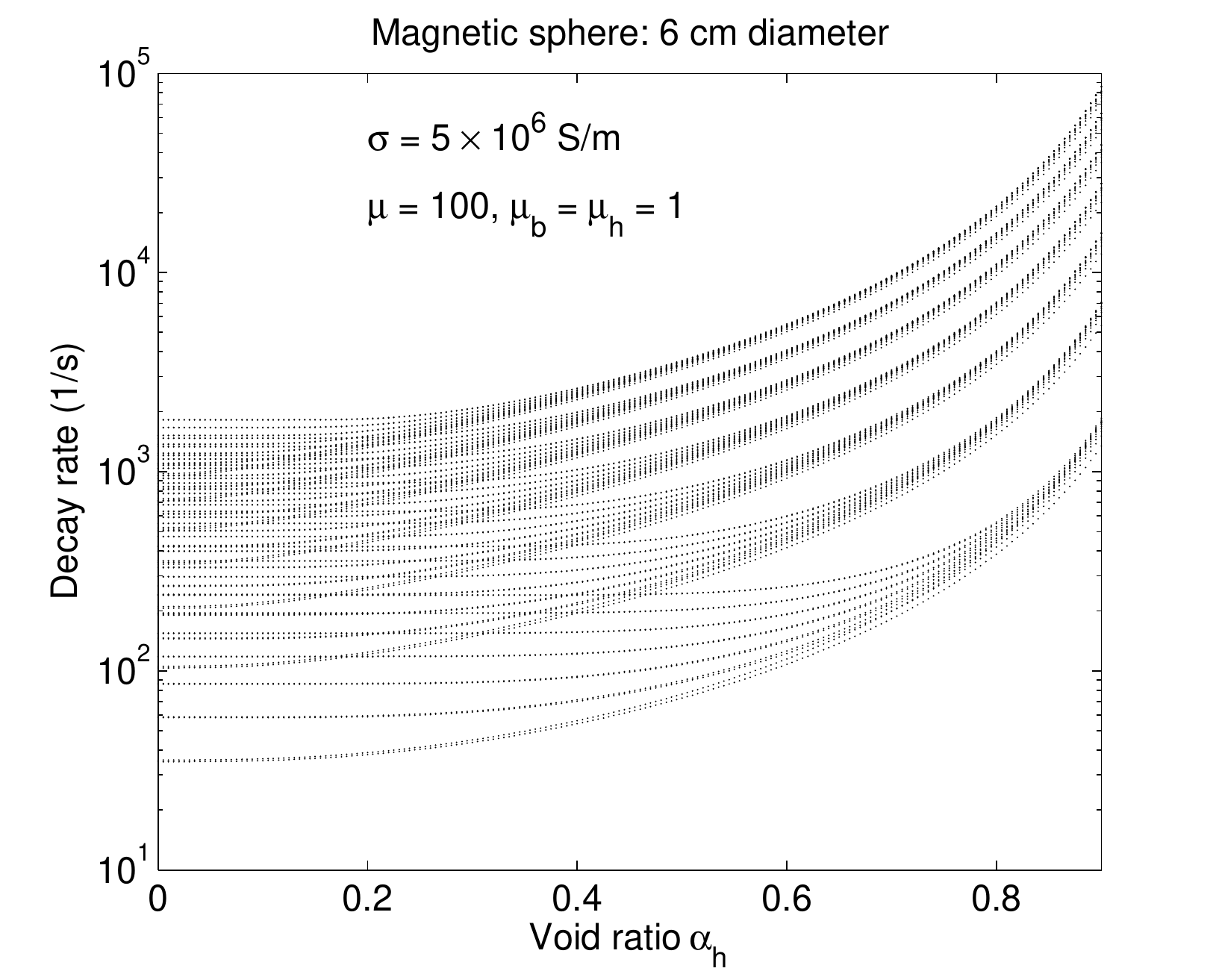}
\includegraphics[width=\columnwidth]{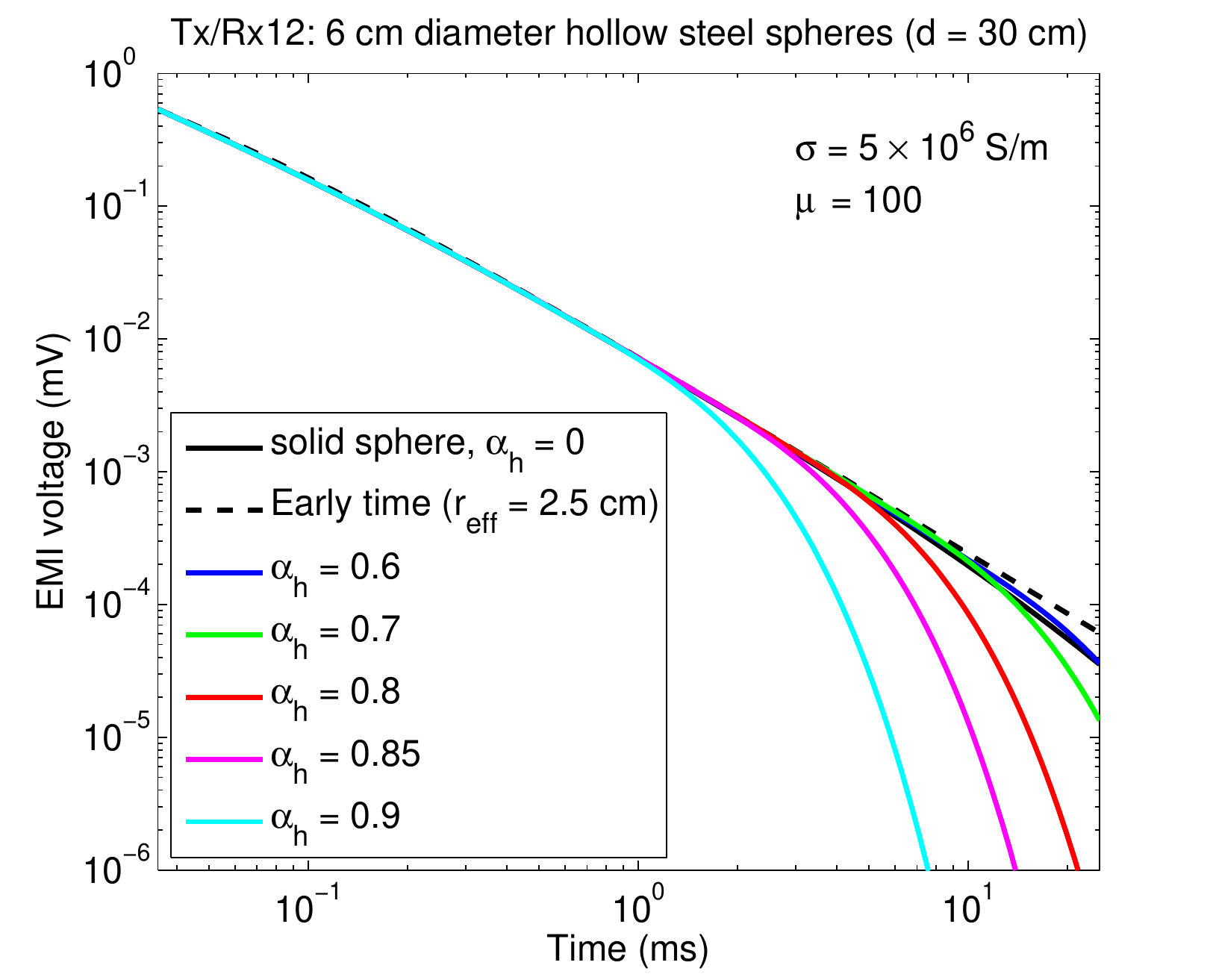}

\caption{(Color online) \textbf{Top:} Exact analytic results for the
decay rate spectrum of hollow steel spheres ($\mu = 100$, $\sigma = 5
\times 10^6$ S/m) as a function of the void ratio $\alpha_h =
r_\mathrm{void}/r_\mathrm{sphere}$. The decay rates increase rapidly as
$\alpha_h \to 1$. \textbf{Bottom:} Exact TDEM voltage predictions using
the NRL-TEMTADS platform for a sequence of hollow steel spheres
centered 20 cm below the transmitter, including the solid sphere
($\alpha_h = 0$, solid black line). Since the surface geometry of all
targets is identical, the same early time curve (dashed line) fits all
data sets. However, the multi-exponential regime begins earlier for
thinner-shelled targets.}

\label{fig:hollowsphere}
\end{figure}

\begin{figure*}
\includegraphics[width=5.5in]
{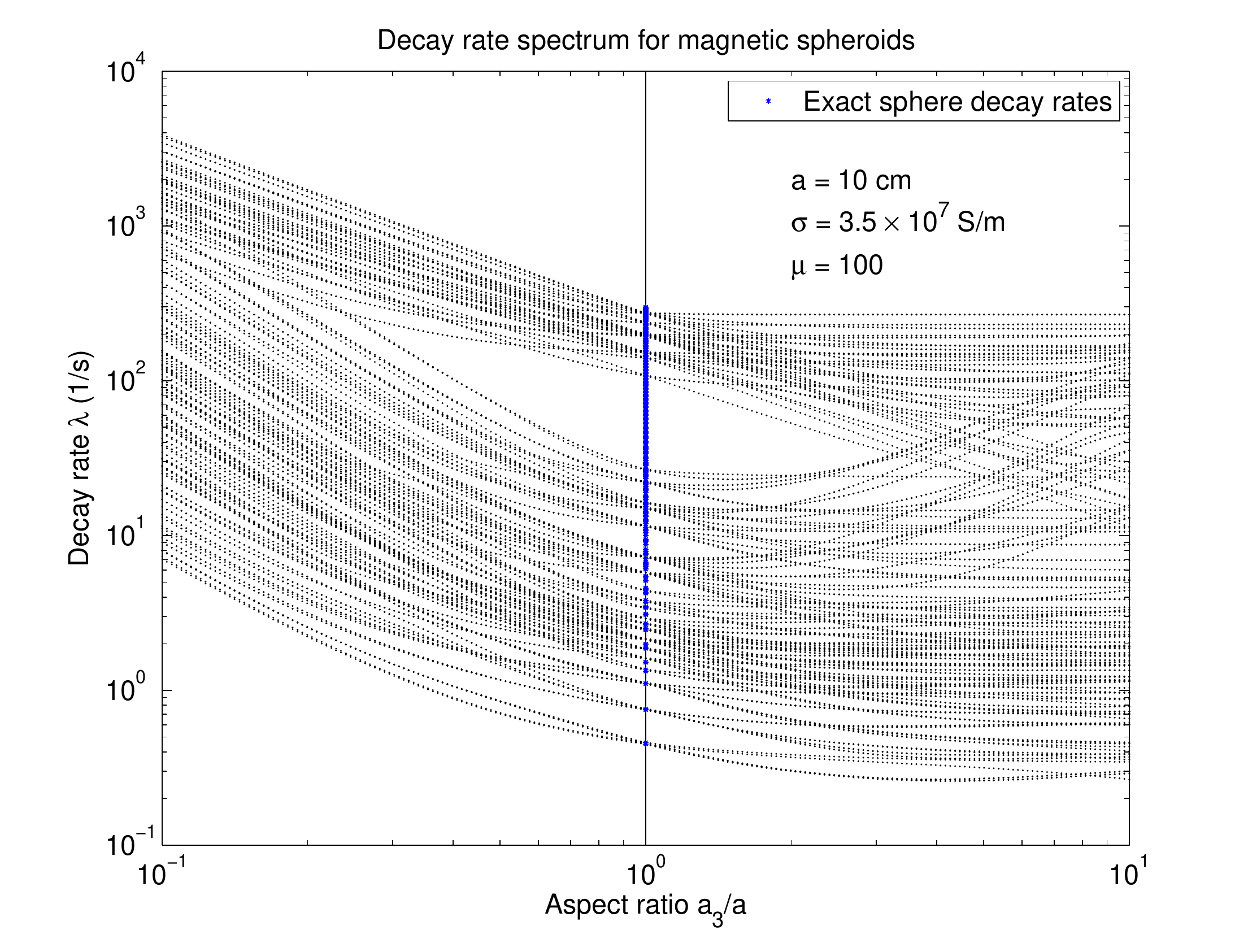}

\caption{Decay rate spectrum vs.\ aspect ratio $\alpha = a_3/a$ for
strongly magnetic/ferrous spheroids ($\mu = 100$). All 232 decay rates
computed using 232 basis functions each for the electric and magnetic
fields with order $N = l + 2p \leq 7$ are plotted for 201 values in the
range $0.1 \leq \alpha \leq 10$. Blue dots at $\alpha = 1$ show exact
analytic results for the sphere. The corresponding results for
nonmagnetic spheroids are shown in Fig.\ \ref{fig:decayrates}. As
described in the text, the large value of $\mu$ effectively divides the
modes into two sets, one with much higher decay rates. The apparent
mode gap near the center of the figure would be filled in with
increasing $N$ (as would smaller gaps that contain unmatched sphere
modes).}

\label{fig:decayrates_mag}
\end{figure*}

\subsection{Freely decaying mode computation}
\label{sec:magmodecomp}

The freely decaying modes once again correspond to solutions to the
homogeneous equations (\ref{9.16}) or (\ref{9.21}) in which the
background terms vanish. Solutions will exist only for special values
of the frequency $\omega = -i\lambda$. For nonmagnetic targets (see
Sec.\ \ref{sec:td_eigen}) the frequency enters the eigenvalue equation
(\ref{2.41}) as a trivial multiplier (in the coefficient $Q_{EE}$), and
the modes are straightforward solutions to the generalized eigenvalue
equation (\ref{2.43}).

For magnetic targets the frequency enters the $Q$-parameters
(\ref{9.15}) in a way that cannot simply be factored out of the $2
\times 2$ block operator. Rather, one obtains a form
\begin{equation}
\hat {\cal O} {\bf A} = \hat {\cal H}(\lambda) {\bf A}
\label{9.53}
\end{equation}
in which ${\bf A} = ({\bf a},{\bf b})^T$ and $\hat {\cal O}$, $\hat
{\cal H}$ represent the $2 \times 2$ operators in (\ref{9.21}). Thus,
the modes correspond to special values $\lambda_n$ in which $\hat {\cal
H}(\lambda_n)$ has a \emph{unit} (generalized) eigenvalue. To compute
these one therefore proceeds as follows. First, solve the generalized
eigenvalue problem
\begin{equation}
\hat {\cal O} {\bf A}_n(\lambda) = \eta_n(\lambda)
\hat {\cal H}(\lambda) {\bf A}_n(\lambda)
\label{9.54}
\end{equation}
for each fixed $\lambda$. The $|\eta_n(\lambda)|$ are increasing
functions of $\lambda$, and beginning with small $\lambda$, there will
be sequence of increasing values $\lambda_n$ for which
$\eta_n(\lambda_n) = 1$ \cite{foot:negeval}. These are the sought after
decay rates, and the corresponding ${\bf A}_n(\lambda_n)$ represent the
mode shapes. This search requires on the order of ten repeated
diagonalizations to find an individual $\lambda_n$, hence thousands to
find the entire spectrum. The task is therefore quite numerically
intensive (though at least the subblock matrices ${\bf R}$, ${\bf S}$,
${\bf U}$, ${\bf V}$ do not need to be recomputed at each step).
Spectra resulting from such a computation, using the ferrous value $\mu
= 100$, are shown in Fig.\ \ref{fig:decayrates_mag}.

There is an interesting feature in Fig.\ \ref{fig:decayrates_mag} that
deserves comment. Unlike for $\mu=\mu_b$ (Fig.\ \ref{fig:decayrates})
where the larger decay rate values are found to progressively fill in
as the basis function order $N$ increases, the $\mu/\mu_b = 100$ case
shown in the figure apparently displays two classes of modes that are
widely separated in decay rate (at least near the center of the plot).
As $N$ increases the gap is expected to gradually fill in (as indicated
by the exact sphere results).

The pattern occurs because of the large value of $\mu/\mu_b$, and is
explained by examining the mode shapes themselves. As discussed in
Sec.\ \ref{sec:himagcontrast}, as $\mu/\mu_b \to \infty$ the magnetic
field boundary normal vanishes, ${\bf H} \cdot {\bf \hat n} \to 0$. In
fact, this convergence is conditional. For given finite $\mu/\mu_b$
there will be more slowly decaying modes for which ${\bf H} \cdot {\bf
\hat n} = O(\mu_b/\mu)$ is indeed small (compared to the transverse
component $|{\bf \hat n} \times {\bf H}|$), and the decay rate
converges to a finite value in the limit $\mu_b/\mu \to 0$. However,
there are also magnetically polarized modes for which ${\bf H} \cdot
{\bf \hat n}$ is not small compared to $|{\bf \hat n} \times {\bf H}|$
(e.g., for which ${\bf H}$ is fairly uniform inside the target), and
these decay much more rapidly. The mode shapes of the higher and lower
groups of modes in Fig.\ \ref{fig:decayrates_mag} indeed exhibit
precisely this difference. The interesting non-monotonic behavior of
some of the mode curves with increasing aspect ratio probably
originates from similar effects (and would go away at high enough order
$N$).

It should be noted that this division is directly analogous to the
distinct inductive and polarization/non-inductive responses discussed
in Secs.\ \ref{sec:tdem} and \ref{sec:noninduct} that occur for large
conductivity contrast, $\sigma/\sigma_b \gg 1$, though the division
here is much less extreme because $\mu/\mu_b \ll \sigma/\sigma_b$
\cite{foot:earlyt}.

Note, finally, that the reduction/truncation of the magnetic field
basis functions described in Sec.\ \ref{subsec:magbasisexpand} to those
with strictly vanishing surface normal eliminates these magnetic
polarization modes at the outset. The corresponding off-diagonal
reduction (\ref{9.25}) of the matrix equation also restores a simple
dependence of the decay rates on the spectrum of ${\cal H}$, namely the
inverse quadratic relation (\ref{9.30}). The approximate validity of
this relation can also be used to speed up the previously described
$\lambda_n$ search for large but finite $\mu/\mu_b$.

\subsection{More general target geometries}
\label{sec:gentarggeom}

\subsubsection{Hollow targets}
\label{subsec:hollowtarg}

The theory presented here is valid even if $V_s$ is not simply
connected, or even connected at all. However, the Chandrasekhar theory
is specific to solid ellipsoids (Sec.\ \ref{sec:solid_ellipsoid}).
Unfortunately, many of the most interesting targets, especially UXO,
are hollow (UXO shell thicknesses are in the neighborhood of 10\% of
the radius).  At early time currents reside only near the outer surface
of the target and the hollow portion plays no role \cite{W2003,W2004}.
However, at intermediate time and beyond, when the currents penetrate
throughout the target, the dynamics is very different. In particular,
the leading decay rates scale inversely with the shell thickness (see
the upper panel of Fig.\ \ref{fig:hollowsphere}; analogous also to the
scaling with small aspect ratio seen on the left hand side of Fig.\
\ref{fig:decayrates}). The result is a strong deviation from the solid
sphere result that begins earlier and earlier as the shell thickness
decreases (lower panel of Fig.\ \ref{fig:hollowsphere}). Accurate
modeling in this regime therefore requires extension of the formalism
to hollow targets.

There are main two issues. The first is the generalization of the
polynomial basis functions described in Sec.\ \ref{subsec:ellpsdbasis}.
The rescaling (\ref{2.32}) from the sphere to the ellipsoid works only
if the hollow portion is concentric and geometrically similar to the
surrounding target, i.e., with axes defined by ${\bf a}_h = \alpha_h
{\bf a}$, $0 \geq \alpha_h < 1$. The underlying $i=1$ sphere basis
functions [first line of (\ref{2.39})] then remain unchanged, while the
$i=2$ basis functions are replaced by
\begin{equation}
{\bf Z}_{lmp}^{(2)} = \nabla \times
[(x^2-\alpha_h^2)(1-x^2)x^{l+2p} {\bf X}_{lm}].
\label{9.55}
\end{equation}
The factor $(x^2-\alpha_h^2)(1-x^2)$ enforces vanishing of the normal
component ${\bf Z}_{lmp}^{(2)}$ at both boundaries. The
concentric-similar assumption is almost certainly adequate for most
UXO.

The second issue is the evaluation of the Coulomb integral matrix
elements (\ref{2.29}) or (\ref{9.20}). Given a polynomial $p({\bf x})$,
the basic Coulomb integral may be written as the difference of two
solid ellipsoid integrals
\begin{eqnarray}
P({\bf x}) &=& \hat {\cal K}_1[p]({\bf x})
\label{9.56} \\
&=& \int_{V_s({\bf a})} d^3x'
\frac{p({\bf x}')}{4\pi |{\bf x}-{\bf x}'|}
- \int_{V_s({\bf a}_h)} d^3x'
\frac{p({\bf x}')}{4\pi |{\bf x}-{\bf x}'|}.
\nonumber
\end{eqnarray}
Both may be evaluated analytically using the methods of Sec.\
\ref{sec:solid_ellipsoid} for arbitrary ${\bf a}_h$ (and arbitrary
center placement).

The problem is that, since ${\bf x}$ is external to the $V_s({\bf
a}_h)$, the second integral is not a polynomial, but a sum of
polynomials multiplied by elliptic functions [themselves depending on
the nonzero function $\lambda({\bf x})$ defined by (\ref{3.25})]. The
second volume integral entering the matrix elements, namely of $P({\bf
x})$, or its derivatives, multiplied by a second polynomial, then
requires a numerical evaluation. This is not a major barrier to
implementing the more general theory, but it does require some extra
numerical effort that will be part of future work.

\subsubsection{Multiple targets}
\label{subsec:multitarget}

The formalism may be also be extended, with a similar level of effort,
to multiple targets. The individual target basis functions are
unchanged from those of an isolated target. The difficulty again comes
from the second volume integral entering the matrix elements: the
mutual induction manifests in the integrals such as
\begin{equation}
R_{12} = \int_{V_s^2} d^3x p_2({\bf x})
\int_{V_s^1} d^3x' \frac{p_1({\bf x}')}
{4\pi |{\bf x}-{\bf x}'|},
\label{9.57}
\end{equation}
in which the external field in a second target volume $V_s^2$,
generated by a polynomial function $p_1$ in the first target volume
$V_s^1$, is integrated against a polynomial function $p_2$ in the
second target. Once again, even for simple targets where this external
field can be expressed in terms of elliptic functions, the second
integral requires in general a numerical evaluation. For well separated
targets, the mutual induction will be weak, and a local gradient
expansion about the center of $V_s^2$ will provide accurate analytic
results. However, for closely spaced, strongly interacting targets,
which is the physically more interesting regime, such an approximation
will break down.

\subsection{Effects of inhomogeneous background permeability}
\label{sec:inhomobgperm}

Beyond the general results (\ref{9.5}) and (\ref{9.12}), the results
presented in this section are specialized to the limit of homogeneous
background permeability. The problem is that, unlike background
conductivity variations, which do not affect the inductive response as
long as the conductivity contrast remains high, even small soil
permeability variations (around unity) will impact the induced voltage
via the $\mu_b$-dependence of the magnetic Green function (\ref{9.3}).

The biggest impact is the direct reflection from the ground surface. In
the absence of an air-ground permeability contrast the soil is
essentially transparent to the low frequency signals considered here.
However, even for very small air-ground contrast (say, 1\% or less),
the direct reflection can mask the signals (which remain essentially
unchanged, at the 1\% level) from smaller and/or more deeply buried
targets. The ground response is much flatter in frequency than that of
the target, and so the latter can still be distinguished (though even
this assumption can break down if the permeability is
frequency-dependent, e.g., if the soil magnetic impurities have
nontrivial magnetodynamics). If the background permeability has slow
enough spatial variation, the problem can also be ameliorated through a
target-absent background subtraction.  If the permeability is known and
the ground is flat, this subtraction could also be computed
theoretically (essentially from the corresponding plane interface
Fresnel coefficient). Again, for 1\% level accuracy, the target
response may be computed assuming homogeneous ground permeability, and
simply added to the ground reflection.

For larger permeability contrast, and/or highly variable soils (e.g.,
strongly magnetic volcanic soils \cite{foot:kahoolawe}) the target
response becomes much more difficult to discern. Given accurate prior
knowledge of $\mu_b({\bf x})$, there is no barrier in principle to
computing an adequate approximation to the magnetic Green function
(\ref{9.3}) in the target neighborhood (e.g., as a polynomial-modified
Coulomb singularity; analytic forms are also available for horizontally
stratified media) and using this to solve for the internal field mode
shapes. Intuitively, one expects adjustments in the target current
patterns due to mutual induction with neighboring magnetic impurity
concentrations. More difficult are the background and external fields,
which will experience $\mu_b$ over a larger volume and hence require a
more comprehensive model to compute.

In conclusion, there appears to be some interesting future work to be
performed modeling the basic phenomenology of the effects of
magnetically variable backgrounds, but in the presence of high,
unpredictable variability \cite{foot:kahoolawe}, the background
subtraction problem has no simple solution---the number of unknown
parameters is too large to constrain using limited survey data confined
to above-ground measurements.

\acknowledgments

This material is based upon work supported by SERDP, through the US
Army Corps of Engineers, Humphreys Engineer Center Support Activity
under Contract No.\ W912HQ-09-C-0024. The author thanks D. Steinhurst
for discussion and for providing experimental data. The author has also
greatly benefitted from discussions with E. M. Lavely, M. Blohm, A.
Becker, T. Smith and F. Morrison.


\begin{thebibliography}{}
\bibitem{foot:magsoils} Even relatively weak soil permeability (1\% or
    less) can have a very large effect. The air-soil magnetic contrast
    leads to reflections that can mask the weaker signals from smaller,
    more deeply buried targets. This becomes an issue, especially, in
    certain volcanic areas \cite{foot:kahoolawe}. In Sec.\
    \ref{sec:inhomobgperm} we briefly discuss how one may attempt to
    incorporate such effects into the model, but such an approach can
    be useful only if the variability is sufficiently predictable and
    the required magnetostatic model sufficiently simple.
\bibitem{foot:ebprime} Clearly, for the highly conducting targets
    of interest, $\sigma/\omega$ will always dominate $\epsilon'$,
    but for very dry soils, or in air, $\epsilon_b'$ could dominate
    $\sigma_b/\omega$. However, the background response will still be
    quasi-static, and it is easily verified that this more general
    case may be accounted for with the simple replacement $\sigma_b
    \to \sigma_b - i\omega \epsilon_b'/4 \pi$ in the development of
    Sec.\ \ref{sec:gf_form}. This has no effect on the final
    result, which relies only on the formal limit $\kappa_b \to 0$.
\bibitem{foot:bdycond} This is, of course, the same conclusion
    as would be obtained from the usual condition of continuity
    of $\epsilon {\bf \hat n} \cdot {\bf E}$ across the bounday
    in the limit where $|\epsilon|/|\epsilon_b| \gg 1$.
\bibitem{foot:ortho} The matrix ${\bf O}$ would be the identity matrix
    ${\bf I}$ if the basis functions are additionally orthornormal with
    respect to the density $\kappa^2$. However, this is not required
    and may be inconvenient.
\bibitem{foot:ellipseharm} The integral relation (\ref{7.3}) may
    actually be diagonalized using ellipsoidal harmonics, but the
    harmonics themselves are very complicated \cite{MF}, and their
    computation is not necessarily easier than the matrix inversion
    route based on the Chandrasekhar approach laid out in Secs.\
    \ref{sec:chandra_theory}--\ref{sec:solid_ellipsoid} \cite{Chandra}.
\bibitem{foot:noninductext} The non-inductive part may in principle
    be computed as well using (\ref{2.20}). For homogeneous target
    and background, where the right hand side of (\ref{2.20})
    is supported on the target boundary, this may be formulated as a
    Laplace-Neumann problem with boundary values provided by the
    internal field. However, as discussed in the Introduction, this
    case is not physically realistic, and $\sigma_b$ is seldom
    sufficently constrained to enable accurate computation of
    $\Phi_\mathrm{ext}$.
\bibitem{foot:ring} This is the same estimate as would used to
    determine the rate of decay of current flowing in circular loop.
\bibitem{foot:multipulse} For larger targets, $1/\lambda_n$ for
    lower order modes be much larger (100 ms or more) than the
    interval between periodic pulses (typically tens of ms), and
    the integral (\ref{8.7}) will encompass several pulses. A
    typical measurement averages over many pulses, and the resulting
    $I_n(t_p)$ is a geometric series that may still be summed
    analytically.
\bibitem{foot:inverse} Applications of the present theory to
    target inversion problems, where, for example, multi-sensor
    data providing multiple ``looks'' at the target is extremely
    useful, lies beyond the scope of the present article, and
    will be presented elsewhere.
\bibitem{foot:temtads} The transmitter coil windings are
    actually spread out over 7.8 cm in height. However,
    modeling them as 1D ideal loops centered at 3.9 cm above
    their base suffices for targets at reasonable standoff (say,
    centered 20 cm or more below the plane of the loop). The
    receiver coils are compact in height, and lie at the bases
    of the transmitter coils, so their centers lie 3.9 cm below the
    effective center of the latter. Due to the rapid $\sim 1/d^6$
    decay of the signals with target depth, these details turn
    out to be quantitatively important: prior to incorporating
    them, the comparisons in Fig.\ \ref{fig:datacomp} contained
    10--20\% systematic biases that could not be explained by any
    known measurement uncertainty (e.g., the known $\sim 10\%$
    pulse amplitude variability cited in the caption to Fig.\
    \ref{fig:datacomp}) or calibration error.
\bibitem{foot:lambdac} One typically finds that for slowest decay rate
    $\lambda_1$ one has $\kappa a \simeq \pi$, which yields $\lambda_1
    \simeq \lambda_c/\pi^2$. thus $\lambda_c$ is actually an order of
    magnitude larger than the fundamental decay rate. Thus, according
    to (\ref{9.22}), for $\mu/\mu_b = O(10^2)$, one will need
    $\lambda/\lambda_1 = O(10^5)$ for the $Q_{EE} \hat {\cal K}_1$ term
    to dominate.
\bibitem{foot:monoflux} If one integrates $-{\bf \hat n} \cdot \nabla
    \frac{1}{|{\bf x}|}$ over some surface $S$ the result is the
    monopole flux through that surface. Using Ampere's law, the same
    result is obtain by integrating the tangential component of ${\bf
    A}$ around the contour $C$ enclosing $S$. The form (\ref{9.31}) may
    then be derived by taking $C$ to be the circle at fixed polar angle
    $\theta$, which then equates $2\pi |{\bf A}_0| \sin(\theta)$ to the
    enclosed flux $[1-\cos(\theta)]/2$.
\bibitem{foot:parity} This follows very generally from the formal
    observation that the parity operator commutes with ${\cal H}$,
    irrespective of the value of $g$. Therefore the eigenvectors may
    always be chosen to have definite parity.
\end{thebibliography}
\end{document}